\documentclass[a4paper,11pt]{article}
\usepackage[DIV12]{typearea}
\usepackage{longtable}
\usepackage{booktabs}
\usepackage{amsmath}
\usepackage{amssymb}
\usepackage{bbm}
\usepackage[utf8]{inputenc}
\usepackage{pdflscape}
\usepackage{xspace}
\usepackage[caption=false]{subfig}
\usepackage{nicefrac}
\usepackage{graphicx}
\usepackage{cite}  
\usepackage{nccfoots}
\usepackage{cancel}
\usepackage[small]{caption2}
\usepackage{multirow}
\usepackage{scalerel}
\usepackage{eufrak}

\usepackage[all,cmtip]{xy}
\newlength{\xywd}
\newcommand{\xyrightarrow}[2][]{%
  \sbox{0}{$\scriptstyle#1$}%
  \xywd=\wd0
  \sbox{0}{$\scriptstyle#2$}%
  \ifdim\wd0>\xywd \xywd=\wd0 \fi
  \xymatrix@C\dimexpr\xywd+1em\relax{{}\ar[r]^{#2}_{#1}&{}}%
}

\DeclareMathOperator{\re}{Re}
\DeclareMathOperator{\im}{Im}

\newcommand{\subind}[1]{{{\ensuremath\scriptscriptstyle{(\hspace{-0.7pt}#1\hspace{-0.7pt})}}}}

\newcommand{\rep}[1]{\ensuremath\boldsymbol{#1}}

\newcommand{\x}{\ensuremath\times}
\newcommand{\Z}[1]{\ensuremath{\mathbbm{Z}_{#1}}} 
\newcommand{\SO}[1]{\ensuremath{\mathrm{SO}(#1)}}

\newcommand{\SL}[1]{\ensuremath{\mathrm{SL}(#1)}}
\newcommand{\GL}[1]{\ensuremath{\mathrm{GL}(#1)}}
\newcommand{\U}[1]{\ensuremath{\mathrm{U}(#1)}}

\newcommand{\I}{\mathrm{i}}
\newcommand{\Id}{\mathbbm{1}}

\newcommand{\CP}{\ensuremath{\mathcal{CP}}\xspace}

\newcommand{\vev}[1]{\ensuremath{\langle #1 \rangle}}

\newcommand{\nphantom}[1]{\sbox0{#1}\hspace{-\the\wd0}}

\advance \headheight by 3.0truept       
\usepackage[pdftex]{hyperref}

\hypersetup{
    pdftitle = {Eclectic flavor scheme from ten-dimensional string theory - II Detailed technical analysis},
    pdfauthor = {Nilles, Ramos-Sanchez, Vaudrevange}
}
 
\begin{document}

\begin{titlepage}

\begin{flushright}
\normalsize{TUM-HEP 1288/20}
\end{flushright}

\vspace*{1.0cm}

\begin{center}
{\Large\textbf{\boldmath Eclectic flavor scheme from ten-dimensional string theory - II\\ Detailed technical analysis\unboldmath}}

\vspace{1cm}

\textbf{%
Hans Peter Nilles$^{a}$, Sa\'ul Ramos--S\'anchez$^{b}$, Patrick K.S. Vaudrevange$^{c}$
}
\\[8mm]
\textit{$^a$\small Bethe Center for Theoretical Physics and Physikalisches Institut der Universit\"at Bonn,\\ Nussallee 12, 53115 Bonn, Germany}
\\[2mm]
\textit{$^b$\small Instituto de F\'isica, Universidad Nacional Aut\'onoma de M\'exico,\\ POB 20-364, Cd.Mx. 01000, M\'exico}
\\[2mm]
\textit{$^c$\small Physik Department T75, Technische Universit\"at M\"unchen,\\ James-Franck-Stra\ss e 1, 85748 Garching, Germany}
\end{center}

\vspace*{1cm}

\begin{abstract}
String theory leads to a flavor scheme where modular symmetries play a crucial role. Together with 
the traditional flavor symmetries they combine to an eclectic flavor group, which we determine via 
outer automorphisms of the Narain space group. Unbroken flavor symmetries are subgroups of this 
eclectic group and their size depends on the location in moduli space. This mechanism of local 
flavor unification allows a different flavor structure for different sectors of the theory (such as 
quarks and leptons) and also explains the spontaneous breakdown of flavor- and \CP-symmetries (via 
a motion in moduli space). We derive the modular groups, including \CP and $R$-symmetries, for 
different sub-sectors of six-dimensional string compactifications and determine the general 
properties of the allowed flavor groups from this top-down perspective. It leads to a very 
predictive flavor scheme that should be confronted with the variety of existing bottom-up 
constructions of flavor symmetry in order to clarify which of them could have a consistent top-down 
completion.
\end{abstract}

\end{titlepage}

\newpage

\section{Introduction}

The consideration of finite modular flavor groups has been suggested in the pioneering work of 
Feruglio~\cite{Feruglio:2017spp}. A combination of these modular symmetries and traditional flavor 
symmetries leads to the eclectic flavor scheme~\cite{Nilles:2020nnc,Nilles:2020kgo}. Here, we 
present a study of the eclectic flavor approach towards ten-dimensional string theory with $D=6$ 
compactified spatial dimensions\footnote{For further work on modular flavor symmetries in string 
theory, see e.g.\ refs.~\cite{Kobayashi:2016ovu,Kobayashi:2018rad,Kobayashi:2018bff,Kariyazono:2019ehj,Ohki:2020bpo,Kikuchi:2020frp,Kikuchi:2020nxn,Ishiguro:2020nuf}}. 
This generalizes previous work with $D=2$ compact dimensions which is shown to fail to capture all 
the relevant properties of the eclectic picture. A few of the highlights of this discussion have 
been presented in an earlier paper~\cite{Nilles:2020tdp} without explaining the full technical 
details. These highlights include
\begin{itemize}
\item a further enhancement of the eclectic group,

\item a new interpretation of discrete $R$-symmetries originating from modular transformations of 
the complex structure modulus,

\item a relation between the $R$-charges and the modular weights of matter fields, and

\item new insights in the nature of \CP-symmetry and its spontaneous breakdown.
\end{itemize}
These results have been illustrated previously~\cite{Nilles:2020kgo,Nilles:2020tdp} in an example 
of a $D=2$ sublattice based on the $\mathbbm T^2/\Z3$ orbifold. Here we shall now present the 
complete results in the generic situation and provide the full technical details of the derivation. 
In addition, we shall illustrate an interesting novel feature: the existence of accidental 
continuous (gauge) symmetries of the effective theory at special loci in moduli space, which turn 
out to be further enhancements of the traditional flavor symmetries.

In section~\ref{sec:ModularSymmetries} we shall discuss basic background material concerning the 
$D=2$ Narain lattice formulation~\cite{Narain:1985jj,Narain:1986am,Narain:1986qm,GrootNibbelink:2017usl} 
and its outer automorphisms relevant for (discrete) modular symmetries~\cite{Baur:2019kwi,Baur:2019iai}. 
These include the groups $\SL{2,\Z{}}_{T}$ and $\SL{2,\Z{}}_{U}$ for the K\"ahler modulus $T$ 
and the complex structure modulus $U$, respectively, mirror symmetry as well as a \CP-like 
transformation. Results for the generators are summarized in table~\ref{tab:OuterAutomorphisms} for 
various $\mathbbm T^2/\Z{K}$ orbifolds. We include a discussion of the modular properties of the 
superpotential and K\"ahler potential in the modular invariant field theory, including modular weights
and automorphy factors.

Section~\ref{sec:SL2ZU} concentrates on the discussion of those aspects of the modular group 
$\SL{2,\Z{}}_U$ of the complex structure modulus which will become relevant for the analysis in 
$D=6$ compact dimensions. In contrast to $\SL{2,\Z{}}_T$ (which exchanges windings and momenta), 
$\SL{2,\Z{}}_U$ can be given a fully geometrical interpretation. In previous discussions of $D=2$ 
orbifolds $\mathbbm T^2/\Z{K}$, the subtleties of this analysis have not yet been completely 
considered, because the modulus $U$ is frozen as a result of the orbifold twist (except for the 
case $\mathbbm T^2/\Z2$ which has been discussed elsewhere~\cite{Baur:2020jwc}). We show here 
that, even with a frozen value $\vev{U}$ of the complex structure $U$, there are nontrivial 
elements of $\SL{2,\Z{}}_U$ that remain unbroken. These lead to additional discrete $R$-symmetries 
according to the stabilizer subgroups given in formula~\eqref{eq:StabilizersOfSL2ZU}. These symmetries 
will be relevant for the $D=6$ case in the form of sublattice rotations.

In section~\ref{sec:Rsymmetries}, we consider the $\mathbbm T^6/P$ orbifold (with point group $P$). 
We start with the discussion in full generality (including the restrictions of $\mathcal N=1$ 
supersymmetry in four space-time dimensions) and then illustrate the results for the $\Z6$-II 
orbifold. We shall identify the sublattice rotations of $\SL{2,\Z{}}_U$, their action on matter 
fields and determine which (fractional) modular weights are relevant for the emerging $R$-symmetries. 
Section~\ref{sec:CPin6D} includes a discussion of \CP-transformations, including the full $D=6$ 
picture (summarized in table~\ref{tab:CP}).

Section~\ref{sec:RandCPinZ3} presents the complete description for a sublattice $\mathbbm T^2/\Z3$ 
of the $\Z6$-II orbifold. This includes the traditional flavor symmetry $\Delta(54)$, the finite 
modular symmetry $T^\prime$ from $\SL{2,\Z{}}_T$, a discrete $R$-symmetry $\Z9^R$ from 
$\SL{2,\Z{}}_U$ as well as a \CP-like transformation (which is separately discussed in 
subsections~\ref{sec:CPforZ3} and~\ref{sec:CP}). The maximal eclectic group is shown to have 
$3,888$ elements, extending $\Omega(2)\cong[1944,3448]$ (see table~\ref{tab:Z3FlavorGroupsExtended}) 
by the \CP-like transformation. We discuss restrictions of this symmetry on the form of the 
superpotential and K\"ahler potential of the $\mathcal N=1$ supersymmetric field theory.

Only part of the eclectic flavor symmetry is realized linearly. For generic values of the moduli 
only the traditional flavor symmetry is unbroken. This symmetry will be enhanced at special points 
(or sub-loci) of the moduli space. In section~\ref{sec:LocalFlavorUnification} we present this 
phenomenon of {\it local flavor unification} for the $\mathbbm T^2/\Z3$ orbifold and identify 
these enhanced groups (see table~\ref{tab:UnifiedFlavorSymmetries} and 
figure~\ref{fig:Z3KaehlerModuliSpace}) as well as the specific forms of the superpotential. 
Moving away from these sub-loci in moduli space will correspond to a 
spontaneous breakdown of the enhanced flavor groups (in particular also for the violation of 
\CP-symmetry). A novel observation concerns the emergence of (accidental) continuous gauge 
symmetries at special loci in moduli space, here discussed in section~\ref{sec:accidentalU1s}.

Section~\ref{sec:conclusions} will present conclusions and an outlook for future applications of 
our results.

Note to the reader: As our discussion is partially very technical we include a short specific 
summary at the end of each section~\ref{sec:ModularSymmetries}--\ref{sec:LocalFlavorUnification} 
to make the results of the discussion as transparent as possible.

\section[Modular symmetries in string theory]{Modular symmetries in string theory}
\label{sec:ModularSymmetries}

In general, the modular group $\SL{2,\Z{}}$ can be defined as the group generated by two abstract 
generators, $\mathrm{S}$ and $\mathrm{T}$, that satisfy the defining relations 
\begin{equation}
\label{eq:DefiningRelationsOfSL2Z}
  \mathrm{S}^4 ~=~ (\mathrm{S}\,\mathrm{T})^3 ~=~ \Id\qquad\mathrm{and}\qquad\mathrm{S}^2\,\mathrm{T} ~=~ \mathrm{T}\,\mathrm{S}^2\;.
\end{equation}
It turns out convenient to represent these generators by the matrices
\begin{equation}\label{eq:SL2ZGeneratorsSandT}
\mathrm{S} ~:=~ \left(\begin{array}{cc} 0 & 1\\ -1 & 0\end{array}\right) \qquad\mathrm{and}\qquad 
\mathrm{T} ~:=~ \left(\begin{array}{cc} 1 & 1\\ 0 & 1\end{array}\right)\;.
\end{equation}
In this context, the elements $\gamma\in\SL{2,\Z{}}$ can be expressed by $2\x2$
matrices with integer entries and determinant one, i.e.
\begin{equation}
\label{eq:SL2ZElements}
  \gamma~:=~ \left(\begin{array}{cc} a&b\\c&d\end{array}\right)\,, \qquad\mathrm{with}\qquad ad-bc=1,\,a,b,c,d\in\Z{}\;.
\end{equation}
Particularly relevant in string models (and in bottom-up flavor model building) are the so-called 
finite modular groups. Their generators are also denoted by $\mathrm S$ and $\mathrm T$, which
fulfill the following defining relations
\begin{subequations}
\begin{eqnarray}
\label{eq:DefiningRelationsOfGammaN}
  \Gamma'_N~:&&~\mathrm{S}^4 ~=~ (\mathrm{S}\,\mathrm{T})^3 ~=~ \mathrm{T}^N ~=~\Id\qquad\mathrm{and}\qquad
               \mathrm{S}^2\,\mathrm{T} ~=~ \mathrm{T}\,\mathrm{S}^2\;,\\
  \Gamma_N~:&&~\mathrm{S}^2 ~=~ (\mathrm{S}\,\mathrm{T})^3 ~=~ \mathrm{T}^N ~=~\Id\;,
\end{eqnarray}
\end{subequations}
where $\Gamma'_N$ is the double cover of $\Gamma_N$. The finite modular groups $\Gamma'_N$ and 
$\Gamma_N$ can be obtained from the modular groups \SL{2,\Z{}} and 
$\mathrm{PSL}(2,\Z{})\cong\SL{2,\Z{}}/\Z2$ as the quotients
$\Gamma'_N\cong \SL{2,\Z{}}/\Gamma(N)$ and $\Gamma_N\cong \mathrm{PSL}(2,\Z{})/\tilde\Gamma(N)$,
where 
$\Gamma(N)\cong\left\langle \gamma\in\SL{2,\Z{}}\, |\, \gamma = \Id\mod{N}\right\rangle$
and 
$\tilde\Gamma(N)\cong\left\langle \gamma\in\mathrm{PSL}(2,\Z{})\, |\, \gamma = \Id\mod{N}\right\rangle$
are modular congruence subgroups.

\subsection{Narain lattice of torus compactifications}
\label{sec:NarainTorus}

In string theory, target-space modular symmetries can arise from the two-dimensional worldsheet 
CFT. To see this, let us first consider string theory compactified on a $D$-dimensional torus 
$\mathbbm{T}^D$, which is most naturally described in the Narain 
formulation~\cite{Narain:1985jj,Narain:1986am,Narain:1986qm}. There, one considers right- and 
left-moving string coordinates $y_\mathrm{R}$ and $y_\mathrm{L}$, which are related to the ordinary 
internal coordinates $y$ and their dual coordinates $\tilde{y}$ via
\begin{equation}\label{eq:CoordLR}
y ~:=~ \sqrt{\frac{\alpha'}{2}}\left(y_\mathrm{L}+y_\mathrm{R}\right) \qquad\mathrm{and}\qquad \tilde{y} ~:=~ \sqrt{\frac{\alpha'}{2}}\left(y_\mathrm{L}-y_\mathrm{R}\right)\;.
\end{equation}
Then, $(y_\mathrm{R}, y_\mathrm{L})$ are compactified on a $2D$-dimensional auxiliary torus. This 
torus is defined by the so-called Narain lattice $\Gamma$ that is spanned by the so-called Narain 
vielbein $E$. In more detail, the Narain torus boundary condition in this formalism reads
\begin{equation}\label{eq:NarainLattice}
\left(\begin{array}{c} y_\mathrm{R}\\ y_\mathrm{L}\end{array}\right) ~\sim~ \left(\begin{array}{c} y_\mathrm{R}\\ y_\mathrm{L}\end{array}\right) + \lambda\;, \quad\mathrm{where}\quad \lambda ~\in~ \Gamma ~:=~ \Bigg\{ E\,\hat{N} ~\Big|~ \hat{N} = \left(\begin{array}{c}n\\m\end{array}\right) ~\in~ \Z{}^{2D}\Bigg\}\;,
\end{equation}
and $n,m\in\Z{}^D$ are known as winding and Kaluza--Klein (KK) numbers, respectively. Due to 
worldsheet modular invariance of the one-loop string partition function, the Narain lattice 
$\Gamma$ spanned by $E$ has to be an even, integer, self-dual lattice with metric of signature 
$(D,D)$. This translates into the following condition on the Narain vielbein $E$:
\begin{equation}\label{eq:NarainCondition}
E^\mathrm{T} \eta\,E ~=~ \hat\eta ~:=~ \left(\begin{array}{cc}0&\Id_D\\\Id_D&0\end{array}\right)\;, \quad\mathrm{where}\quad \eta ~:=~ \left(\begin{array}{cc}-\Id_D&0\\0&\Id_D\end{array}\right)
\end{equation}
denotes the metric of signature $(D,D)$ and $\Id_D$ is the $(D\times D)$-dimensional identity 
matrix. Then, the Narain scalar product of two Narain vectors $\lambda_i = E\,\hat{N}_i\in\Gamma$ 
with $\hat{N}_i = (n_i, m_i) \in\Z{}^{2D}$ for $i\in\{1,2\}$ reads
\begin{equation}\label{eq:NarainSP}
\lambda_1^\mathrm{T}\, \eta\, \lambda_2 ~=~ \hat{N}_1^\mathrm{T} \hat\eta\, \hat{N}_2 ~=~ n_1^\mathrm{T} m_2 + m_1^\mathrm{T} n_2 ~\in~\Z{}\;.
\end{equation}
This confirms that the lattice $\Gamma$ is integer and even, as the scalar product is integer and, 
if $\hat{N}_1 = \hat{N}_2$, even. Also, as expected from eq.~\eqref{eq:NarainCondition}, we note that
$\hat\eta$ is the Narain metric in the lattice basis.

In the absence of (discrete) Wilson lines~\cite{Ibanez:1986tp}, one can choose the $(2D \x 2D)$-dimensional 
Narain vielbein $E$ fulfilling eq.~\eqref{eq:NarainCondition} as
\begin{equation}\label{eq:VielbeinMatrixE}
E~:=~\dfrac{1}{\sqrt{2}}\, \begin{pmatrix}
\dfrac{e^{-\mathrm{T}}}{\sqrt{\alpha'}}\,(G - B) &          -\sqrt{\alpha'}\,e^{-\mathrm{T}} \\[0.4cm]
\dfrac{e^{-\mathrm{T}}}{\sqrt{\alpha'}}\,(G + B) &\phantom{-}\sqrt{\alpha'}\,e^{-\mathrm{T}} \end{pmatrix}\;,
\end{equation}
see refs.~\cite{GrootNibbelink:2017usl,Baur:2019kwi,Baur:2019iai}. Here, $\alpha'$ is the Regge 
slope that enters the Narain vielbein $E$, such that $E$ is dimensionless. Then, $E$ is 
parameterized by the $(D\times D)$-dimensional geometrical vielbein $e$, yielding the metric 
$G:=e^\mathrm{T}e$ of the geometrical torus $\mathbbm{T}^D$, and the $(D\times D)$-dimensional 
anti-symmetric $B$-field $B$. 

Using the explicit parameterization~\eqref{eq:VielbeinMatrixE}, we can translate the Narain boundary 
condition~\eqref{eq:NarainLattice} to the coordinates $y$ and their duals. We obtain
\begin{equation}\label{eq:NarainLatticeForYandtildeY}
\left(\begin{array}{c} y\\ \tilde{y}\end{array}\right) ~\sim~ \left(\begin{array}{c} y\\ \tilde{y}\end{array}\right) \,+\, \left(\begin{array}{c} e\,n\\ e^{-\mathrm{T}}\,\left(B\,n + \alpha'\,m\right)\end{array}\right)\;.
\end{equation}
Note that the boundary condition $y \sim y + e\,n$ for the geometrical coordinates $y$ of closed 
strings motivates the name of $n\in\Z{}^D$ as winding numbers of the geometrical torus $\mathbbm{T}^D$ 
spanned by $e$.

The Narain lattice $\Gamma$ is beneficial as it incorporates the winding numbers $n\in\Z{}^D$ and 
the KK numbers $m\in\Z{}^D$ on equal footing. Hence, it allows for a natural formulation of 
$T$-duality in string theory, where KK and winding numbers get interchanged. 

\subsection{Outer automorphisms of the Narain lattice}
\label{sec:OuterAutomorphismsOfTheNarainLattice}

In the Narain lattice basis, the Narain metric $\hat\eta$ can be used to define the group of outer 
automorphisms\footnote{Here, $\mathrm{O}_{\hat{\eta}}(D,D,\Z{})$ gives the discrete ``rotational'' 
outer automorphisms of the Narain lattice. Furthermore, there are continuous translations that act 
by conjugation on Narain lattice vectors and, therefore, correspond to the trivial automorphism of 
$\Gamma$.} of the Narain lattice $\Gamma$ that preserve the Narain metric $\hat\eta$ as
\begin{equation}
\label{eq:OuterGroupNarainLattice}
\mathrm{O}_{\hat{\eta}}(D,D,\Z{}) ~:=~ \big\langle ~\hat\Sigma~\big|~\hat\Sigma~\in~\mathrm{GL}(2D,\Z{})\quad\mathrm{with}\quad\hat\Sigma^\mathrm{T} \hat\eta\, \hat\Sigma~=~\hat\eta ~\big\rangle\;,
\end{equation}
such that each Narain lattice vector $\lambda = E\,\hat{N}\in\Gamma$ is transformed by an outer 
automorphism $\hat\Sigma\in\mathrm{O}_{\hat{\eta}}(D,D,\Z{})$ according to
\begin{equation}
\label{eq:TrafoOfNarainLattice}
\lambda = E\,\hat{N} ~\stackrel{\hat\Sigma}{\longrightarrow}~ E\,\hat\Sigma^{-1}\hat{N} ~\in~\Gamma\;.
\end{equation}
Here, we use $\hat\Sigma^{-1}$ instead of $\hat\Sigma$ for later convenience, see 
section~\ref{sec:ModularForOrbifolds}. Then, $\hat\Sigma^{-1}\hat{N} \in \Z{}^{2D}$ and the Narain 
scalar product $\lambda^\mathrm{T}\, \eta\, \lambda$ from eq.~\eqref{eq:NarainSP} is invariant 
under the transformation~\eqref{eq:TrafoOfNarainLattice}. Note that the transformation of the 
Narain lattice vector eq.~\eqref{eq:TrafoOfNarainLattice} can be interpreted as induced by the 
transformation of the Narain vielbein under outer automorphisms given by
\begin{equation}
\label{eq:TrafoOfNarainVielbein}
E ~\stackrel{\hat\Sigma}{\longrightarrow}~ E\,\hat\Sigma^{-1} \qquad\mathrm{for}\qquad \hat\Sigma~\in~\mathrm{O}_{\hat{\eta}}(D,D,\Z{})\;.
\end{equation}
The group of outer automorphisms of the Narain lattice $\mathrm{O}_{\hat{\eta}}(D,D,\Z{})$ with these 
properties contains the modular group, as we discuss next.

For concreteness, let us focus on a geometrical two-torus $\mathbbm{T}^2$ with $D=2$. One can easily 
verify that $\mathrm{O}_{\hat{\eta}}(2,2,\Z{})$ contains two copies of $\SL{2,\Z{}}$ as follows: 
First, we can define two $\mathrm{O}_{\hat{\eta}}(2,2,\Z{})$ elements
\begin{equation}\label{eq:KSandKT}
\hat{K}_\mathrm{S} ~:=~ \left(\begin{array}{cc}0&-\epsilon\\-\epsilon&0\end{array}\right) \quad,\quad \hat{K}_\mathrm{T} ~:=~ \left(\begin{array}{cc}\Id_2&0\\-\epsilon&\Id_2\end{array}\right)\;, \quad\mathrm{where}\quad \epsilon ~:=~ \left(\begin{array}{cc}0&1\\-1&0\end{array}\right)\;,
\end{equation}
which satisfy the defining relations~\eqref{eq:DefiningRelationsOfSL2Z} of $\SL{2,\Z{}}$,
\begin{equation}
\left(\hat{K}_\mathrm{S}\right)^4 ~=~ \left(\hat{K}_\mathrm{S}\,\hat{K}_\mathrm{T}\right)^3 ~=~ \Id_4\qquad\mathrm{and}\qquad\left(\hat{K}_\mathrm{S}\right)^2\,\hat{K}_\mathrm{T} ~=~ \hat{K}_\mathrm{T}\,\left(\hat{K}_\mathrm{S}\right)^2\;.
\end{equation}
We denote this modular group by $\SL{2,\Z{}}_T$. Moreover, we identify the 
$\mathrm{O}_{\hat{\eta}}(2,2,\Z{})$ elements
\begin{equation}\label{eq:CSandCT}
\hat{C}_\mathrm{S} ~:=~ \left(\begin{array}{cc}-\epsilon&0\\0&-\epsilon\end{array}\right) \quad,\quad \hat{C}_\mathrm{T} ~:=~ \left(\begin{array}{cc}\gamma&0\\0&\gamma^{-\mathrm{T}}\end{array}\right)\;, \quad\mathrm{where}\quad \gamma ~:=~ \left(\begin{array}{cc}1&-1\\0&1\end{array}\right)\;,
\end{equation}
which also satisfy the \SL{2,\Z{}} defining relations,
\begin{equation}
\left(\hat{C}_\mathrm{S}\right)^4 ~=~ \left(\hat{C}_\mathrm{S}\,\hat{C}_\mathrm{T}\right)^3 ~=~ \Id_4\qquad\mathrm{and}\qquad\left(\hat{C}_\mathrm{S}\right)^2\,\hat{C}_\mathrm{T} ~=~ \hat{C}_\mathrm{T}\,\left(\hat{C}_\mathrm{S}\right)^2\;.
\end{equation}
Hence, $\hat{C}_\mathrm{S}$ and $\hat{C}_\mathrm{T}$ give rise to another factor of the modular 
group, denoted by $\SL{2,\Z{}}_U$. Even though all elements of $\SL{2,\Z{}}_T$ 
commute with those of $\SL{2,\Z{}}_U$, these two factors are not fully independent since 
they share a common element, given by
\begin{equation}
\left(\hat{K}_\mathrm{S}\right)^2 ~=~ \left(\hat{C}_\mathrm{S}\right)^2 ~=~ -\Id_4\;.
\end{equation}
The group $\mathrm{O}_{\hat{\eta}}(2,2,\Z{})$ of outer automorphisms of the Narain lattice 
$\Gamma$ contains two additional generators, which we can choose for example as
\begin{equation}\label{eq:MandSigmaStar}
\hat\Sigma_*~:=~\left(\begin{array}{cccc}
-1& 0&  0& 0\\
 0& 1&  0& 0\\
 0& 0& -1& 0\\
 0& 0&  0& 1 
\end{array}\right) \qquad\mathrm{and}\qquad 
\hat{M}~:=~\left(\begin{array}{cccc}
 0& 0& 1&0\\
 0& 1& 0&0\\
 1& 0& 0&0\\
 0& 0& 0&1 
\end{array}\right)\;.
\end{equation}
They give rise to $\Z{2}\times\Z{2}$. Motivated by the relations
\begin{subequations}
\begin{eqnarray}
\hat\Sigma_*\,\hat{K}_\mathrm{S}\,\hat\Sigma_*^{-1} ~=~ \left(\hat{K}_\mathrm{S}\right)^{-1} \quad & , &\quad \hat\Sigma_*\,\hat{K}_\mathrm{T}\,\hat\Sigma_*^{-1} ~=~ \left(\hat{K}_\mathrm{T}\right)^{-1}\;,\\
\hat\Sigma_*\,\hat{C}_\mathrm{S}\,\hat\Sigma_*^{-1} ~=~ \left(\hat{C}_\mathrm{S}\right)^{-1} \quad & , &\quad \hat\Sigma_*\,\hat{C}_\mathrm{T}\,\hat\Sigma_*^{-1} ~=~ \left(\hat{C}_\mathrm{T}\right)^{-1}\;,
\end{eqnarray}
\end{subequations}
we call $\hat\Sigma_*$ a \CP-like transformation. Furthermore, we call $\hat{M}$ a mirror 
transformation as it interchanges $\SL{2,\Z{}}_T$ and $\SL{2,\Z{}}_U$, i.e.
\begin{equation}
\hat{M}\,\hat{K}_\mathrm{S}\,\hat{M}^{-1} ~=~ \hat{C}_\mathrm{S} \qquad\mathrm{and}\qquad \hat{M}\,\hat{K}_\mathrm{T}\,\hat{M}^{-1} ~=~ \hat{C}_\mathrm{T}\;.
\end{equation}

\subsection[Moduli and modular symmetries of T2]{\boldmath Moduli and modular symmetries of $\mathbbm T^2$ \unboldmath}
\label{sec:ModuliAndSymmetriesT2}

The modular subgroups $\SL{2,\Z{}}_T$ and $\SL{2,\Z{}}_U$ of $\mathrm{O}_{\hat{\eta}}(2,2,\Z{})$ 
are associated with the transformations of the moduli of a $\mathbbm T^2$ torus, as we now discuss.

The Narain vielbein $E$ of the geometrical two-torus $\mathbbm T^2$ with background $B$-field 
allows for certain deformations while satisfying the conditions eq.~\eqref{eq:NarainCondition} 
required for $E$ to span a Narain lattice. These deformations correspond to two moduli fields:
\begin{subequations}\label{eq:TorusModuli}
\begin{eqnarray}
\textrm{K{\"a}hler\ modulus}\qquad         T & := & \frac{1}{\alpha'}\left(B_{12}+\I\sqrt{\mathrm{det}G}\right)\quad\mathrm{and}\\
\textrm{complex\ structure\ modulus}\qquad U & := & \frac{1}{G_{11}}\left(G_{12}+\I\sqrt{\mathrm{det}G}\right)\;.\label{eq:DefU}
\end{eqnarray}
\end{subequations}
Comparing these definitions with eq.~\eqref{eq:VielbeinMatrixE}, we realize that the Narain 
vielbein can be expressed in terms of the $\mathbbm T^2$ moduli, $E=E(T,U)$, up to some unphysical 
transformations, see for example ref.~\cite{GrootNibbelink:2017usl}. 
Hence, the action of $\hat\Sigma \in \mathrm{O}_{\hat\eta}(2,2,\Z{})$ on the Narain vielbein given 
by eq.~\eqref{eq:TrafoOfNarainVielbein} is equivalent to a transformation of the $T$ and $U$ 
moduli. To determine explicitly the transformation properties of the moduli, it is convenient to 
define the generalized metric (valid for an arbitrary dimension $D$)
\begin{equation}
\label{eq:GeneralizedMetric}
\mathcal{H} ~:=~ E^\mathrm{T}E ~=~
\begin{pmatrix}
\frac{1}{\alpha'}\left(G-B\,G^{-1}B\right) & -B\,G^{-1}\\
G^{-1}B                                    & \alpha'\,G^{-1}
\end{pmatrix}\;.
\end{equation}
For a two-torus $\mathbbm T^2$, the generalized metric in terms of the torus moduli reads
\begin{equation}
\label{eq:ExplicitGeneralizedMetric}
\mathcal{H}(T,U) ~=~ \frac{1}{\im T\im U}
 \begin{pmatrix}
   |T|^2       & |T|^2 \re U  & \re T\re U  & -\re T \\
   |T|^2 \re U & |T\,U|^2     & |U|^2\re T  & -\re T\re U\\
   \re T\re U  & |U|^2\re T   & |U|^2       & -\re U \\
   -\re T      & -\re T\re U  & -\re U      & 1
 \end{pmatrix}\,.
\end{equation}
Due to the transformation of the vielbein eq.~\eqref{eq:TrafoOfNarainVielbein}, the action of 
$\hat\Sigma \in \mathrm{O}_{\hat\eta}(2,2,\Z{})$ on the generalized metric is given by
\begin{equation}
\label{eq:GeneralizedMetricTrafo}
\mathcal{H}(T,U) ~\stackrel{\hat\Sigma}{\longmapsto}~ \hat\Sigma^{-\mathrm{T}} \mathcal{H}(T,U)\, \hat\Sigma^{-1} ~=:~ \mathcal{H}(T',U')\;.
\end{equation}
From this equation and the expression~\eqref{eq:ExplicitGeneralizedMetric}, given a transformation $\hat\Sigma$, 
one can readily obtain the forms of the transformed moduli $T'$ and $U'$.
For example, considering as $\hat\Sigma$ the $\SL{2,\Z{}}_T$ modular generators $\hat{K}_\mathrm{S}$ 
and $\hat{K}_\mathrm{T}$ in eq.~\eqref{eq:KSandKT}, we observe that the associated moduli
transformations are
\begin{equation}
 T \xrightarrow{\hat{K}_\mathrm{S}} T' = -\frac{1}{T}\,,\quad U \xrightarrow{\hat{K}_\mathrm{S}} U' = U\,,\qquad\text{and}\qquad  
 T \xrightarrow{\hat{K}_\mathrm{T}} T' = T+1\,,\quad U \xrightarrow{\hat{K}_\mathrm{T}} U' = U\,,
\end{equation}
which explains the index chosen for the modular group $\SL{2,\Z{}}_T$. From this observation,
as shown e.g.\ in ref.~\cite{Baur:2019iai}, we know that under a modular transformation 
$\gamma_T\in\SL{2,\Z{}}_T$ the moduli $T$ and $U$ transform as
\begin{equation}\label{eq:TrafoOfTModulus}
T ~\stackrel{\gamma_T}{\longrightarrow}~ \gamma_T\,T ~:=~ \frac{a_T\,T+b_T}{c_T\,T+d_T} \quad\mathrm{and}\quad U ~\stackrel{\gamma_T}{\longrightarrow}~ U \quad\mathrm{for}\quad\gamma_T = \left(\begin{array}{cc} a_T & b_T\\ c_T & d_T\end{array}\right) \in~ \SL{2,\Z{}}_T\;,
\end{equation}
in the $2\x2$ matrix notation used in eq.~\eqref{eq:SL2ZElements}.

Analogously, letting the generalized metric~\eqref{eq:ExplicitGeneralizedMetric} transform
under the $\SL{2,\Z{}}_U$ generators defined in eq.~\eqref{eq:CSandCT}, one finds that the 
$T$ and $U$ moduli transform according to
\begin{equation}
 T \xrightarrow{\hat{C}_\mathrm{S}} T' = T\,,\quad U \xrightarrow{\hat{C}_\mathrm{S}} U' = -\frac{1}{U}\,,\qquad\text{and}\qquad  
 T \xrightarrow{\hat{C}_\mathrm{T}} T' = T\,,\quad U \xrightarrow{\hat{C}_\mathrm{T}} U' = U+1\,.
\end{equation}
Hence, a general modular element $\gamma_U\in\SL{2,\Z{}}_U$ acts on the moduli as
\begin{equation}\label{eq:TrafoOfUModulus}
T ~\stackrel{\gamma_U}{\longrightarrow}~ T \quad\mathrm{and}\quad U ~\stackrel{\gamma_U}{\longrightarrow}~ \gamma_U\,U ~:=~ \frac{a_U\,U+b_U}{c_U\,U+d_U} \quad\mathrm{for}\quad\gamma_U = \left(\begin{array}{cc} a_U & b_U\\ c_U & d_U\end{array}\right) \in~ \SL{2,\Z{}}_U\;,
\end{equation}
as we will also re-derive later in section~\ref{sec:GeomInterpretationSL2ZU} using an alternative 
approach. Moreover, repeating the previous steps, the mirror transformation $\hat{M}$ interchanges 
$T$ and $U$,
\begin{equation}
T ~\stackrel{\hat{M}}{\longrightarrow}~ T' ~=~ U \quad\mathrm{and}\quad U ~\stackrel{\hat{M}}{\longrightarrow}~ U' ~=~ T\;,
\end{equation}
while the \CP-like transformation $\hat\Sigma_*$ acts as
\begin{equation}
T ~\stackrel{\hat\Sigma_*}{\longrightarrow}~ T' ~=~ -\bar{T} \quad\mathrm{and}\quad U ~\stackrel{\hat\Sigma_*}{\longrightarrow}~ U' ~=~ -\bar{U}\;.
\end{equation}
A brief remark is in order: The modular transformation $(\hat{K}_\mathrm{S})^2$, which equals 
$(\hat{C}_\mathrm{S})^2$, acts trivially on both moduli. Hence, the modular groups restricted to 
their action only on the moduli is $\mathrm{PSL}(2,\Z{})_T$ and $\mathrm{PSL}(2,\Z{})_U$, where for 
example $\gamma_T\in\SL{2,\Z{}}_T$ and $-\gamma_T\in\SL{2,\Z{}}_T$ are identified in 
$\mathrm{PSL}(2,\Z{})_T$. However, in full string theory $(\hat{K}_\mathrm{S})^2$ acts 
nontrivially, for example, on massive winding strings.

\subsection{Orbifold compactifications in the Narain formulation}
\label{sec:NarainOrbifolds}

Defining $Y:=(y_\mathrm{R},y_\mathrm{L})^\mathrm{T}$ for the Narain coordinates, a $\mathbbm T^D/\Z{K}$ 
orbifold in the Narain formulation is obtained by extending the boundary conditions of closed 
strings eq.~\eqref{eq:NarainLattice} to
\begin{equation}
\label{eq:OrbifoldCondition}
Y ~\sim~ \Theta^k Y + E\,\hat{N}\;, \quad\mathrm{where}\quad \Theta^K ~=~ \Id_{2D} \quad\mathrm{and}\quad k=0,\ldots,K-1\;.
\end{equation}
Moreover, the so-called Narain twist $\Theta$ shall not interchange right- and left-movers. 
Thus, it is given by
\begin{equation}\label{eq:ConditionsOnNarainTwist}
\Theta ~:=~ \begin{pmatrix} \theta_\mathrm{R} & 0\\ 0 & \theta_\mathrm{L} \end{pmatrix} ~\in~ \mathrm{O}(D)\times\mathrm{O}(D)\;.
\end{equation}
Hence, $\Theta$ generates a $\Z{K}$ rotation that is modded out from the Narain torus in the 
orbifold. The Narain twist has to be a (rotational) symmetry of the Narain lattice of the torus, 
i.e.\ $\Theta\,\Gamma=\Gamma$. Thus, $\Theta$ has to be an outer automorphism of the Narain lattice 
$\Gamma$. This translates into the following condition on the so-called Narain twist in the Narain 
lattice basis
\begin{equation}\label{eq:NarainTwistLatticeBasis}
\hat\Theta ~:=~ E^{-1}\Theta\, E ~\stackrel{!}{\in}~ \mathrm{GL}(2D,\Z{})\;.
\end{equation}
Furthermore, from eq.~\eqref{eq:ConditionsOnNarainTwist} we find that the Narain 
metric~\eqref{eq:NarainCondition} and the generalized metric~\eqref{eq:GeneralizedMetric} are 
orbifold invariant,
\begin{equation}
 \hat\Theta^\mathrm{T}\hat\eta\,\hat\Theta~=~\hat\eta\qquad\mathrm{and}\qquad
 \hat\Theta^\mathrm{T}\mathcal{H}\,\hat\Theta~=~\mathcal{H}\;.
\end{equation}

In summary, the Narain twist in the Narain lattice basis $\hat\Theta:=E^{-1}\Theta\, E$ has to be 
a Narain-metric preserving outer automorphism of the Narain lattice, which, in addition, leaves the 
generalized metric (and therefore all torus moduli) invariant, i.e.
\begin{equation}\label{eq:OrbifoldConditionsOnModuli}
\hat\Theta \stackrel{!}{\in} \mathrm{O}_{\hat\eta}(D,D,\Z{})\,,\qquad 
T_a ~\stackrel{\hat\Theta}{\longrightarrow}~ T_a' ~\stackrel{!}{=}~ T_a\qquad\mathrm{and}\qquad
U_b ~\stackrel{\hat\Theta}{\longrightarrow}~ U_b' ~\stackrel{!}{=}~ U_b\;,
\end{equation}
where $T_a$ and $U_b$ denote all K\"ahler and complex structure moduli of a $\mathbbm T^D/\Z{K}$ 
orbifold, respectively. This invariance condition of the moduli results in a stabilization of 
some of them: One says that some moduli are frozen geometrically in order to satisfy 
eq.~\eqref{eq:OrbifoldConditionsOnModuli}.

\subsection{Modular symmetries of orbifold compactifications}
\label{sec:ModularForOrbifolds}

We shall restrict ourselves here to symmetric orbifolds $\mathbbm T^D/\Z{K}$, for which right- and 
left-movers are equally affected by the orbifold twist, which is granted if we additionally impose 
in eq.~\eqref{eq:ConditionsOnNarainTwist} the condition $\theta:=\theta_\mathrm{R}=\theta_\mathrm{L}\in\mathrm{O}(D)$. 
Then, the orbifold twist $\theta$ generates the so-called geometrical point group 
$P \cong \Z{K} \subset \mathrm{O}(D)$.

The group structure and the symmetries of the orbifold can be realized by expressing the boundary conditions 
eq.~\eqref{eq:OrbifoldCondition} of the Narain orbifold in terms of the so-called Narain space group 
$S_\mathrm{Narain}$ as
\begin{equation}
\label{eq:OrbifoldConditions}
Y ~\sim~ g\,Y ~:=~ \Theta^k Y + E\,\hat{N}\;,\quad\mathrm{where}\quad g~=~(\Theta^k,E\,\hat N) ~\in~ S_\mathrm{Narain}\;.
\end{equation}
The Narain space group is equipped with the product defined by 
\begin{equation}
 g' g ~=~ (\Theta^{k'},E\,\hat N')(\Theta^k,E\,\hat N) ~:=~ (\Theta^{k'+k},E\,\hat N' + \Theta^{k'}E\,\hat N)\,,\quad g,g'\in S_\mathrm{Narain}\,.
\end{equation}
An arbitrary element $g\in S_\mathrm{Narain}$ can be expressed in the Narain lattice basis
by conjugating with $h:=(E,0)$ as
\begin{equation}
\label{eq:NarainLatticeConjugation}
\hat g ~:=~ h^{-1}\,g\,h ~=~ (E^{-1},0)(\Theta^k,E\,\hat N)(E,0) ~=~ (E^{-1}\Theta^k E, \hat N)~=~(\hat\Theta^k,\hat N)\,,
\end{equation}
where we have used the Narain twist in the Narain lattice basis defined in 
eq.~\eqref{eq:NarainTwistLatticeBasis}. In our notation, the Narain space group and its elements 
are indicated by a hat in the Narain lattice basis, i.e.\ $\hat g\in \hat S_\mathrm{Narain}$. 

To identify the modular symmetries of the four-dimensional effective theory after orbifolding 
amounts to determining the outer automorphisms of the Narain space group $\hat S_\mathrm{Narain}$. 
An outer automorphism of the Narain space group is given by a transformation
\begin{equation}
 \hat h ~:=~ (\hat\Sigma,\,\hat T)~\notin~\hat S_\mathrm{Narain}\,,
\end{equation}
such that $\hat h^{-1}\hat g\,\hat h \in \hat S_\mathrm{Narain}$ for all $\hat g=(\hat\Theta^k,\hat N)\in\hat S_\mathrm{Narain}$
is satisfied. Explicitly, the condition for $\hat h$ to be an outer automorphism of 
$\hat S_\mathrm{Narain}$ reads
\begin{equation}
\label{eq:OuterAutomorphismSNarain}
 (\hat\Sigma,\,\hat T)^{-1} (\hat\Theta^k,\,\hat N)\, (\hat\Sigma,\,\hat T) ~=~ 
     (\hat\Sigma^{-1}\hat\Theta^k\hat\Sigma,\, \hat\Sigma^{-1}(\hat\Theta^k-\Id_{2D})\,\hat T + \hat\Sigma^{-1}\hat N) ~\stackrel{!}{\in}~\hat S_\mathrm{Narain}\,,
\end{equation}
see refs.~\cite{lutowski2013finite,Baur:2020jwc} for an algorithm to classify the outer 
automorphisms of a (Narain) space group. If $\hat S_\mathrm{Narain}$ is generated by pure rotations 
$(\hat\Theta,0)$ and pure translations $(\Id_{2D},\hat N)$ with $\hat N\in\Z{}^{2D}$, the 
condition~\eqref{eq:OuterAutomorphismSNarain} is equivalent to demanding that
\begin{equation}
\label{eq:conditionsOnOuter}
\hat\Sigma^{-1}\hat\Theta^k\hat\Sigma ~\stackrel{!}{=}~ \hat\Theta^{k'}\;,\;\;\;
\hat\Sigma^{-1}\hat N ~\stackrel{!}{\in}~\Z{}^{2D}  \quad\mathrm{and}\quad  
\left(\Id_{2D} - \hat\Theta^k\right)\,\hat T ~\stackrel{!}{\in}~ \Z{}^{2D}\,,
\end{equation}
for all $k=0,\ldots,K-1$ and $\hat N\in\Z{}^{2D}$. Then, also the group of outer automorphism is 
generated by pure rotations $(\hat\Sigma,0)$ and pure translations $(\Id_{2D}, \hat T)$, such that 
$\hat\Sigma\neq\hat\Theta^k$ and $\hat T\notin\Z{}^{2D}$. The translational outer automorphisms 
contribute to the so-called traditional flavor symmetry of the theory, as we will review in 
section~\ref{sec:TFSZ3orbifold} in the case of a $\mathbbm T^2/\Z{3}$ orbifold sector. Let us focus 
now on the rotational elements $(\hat\Sigma,0)$. Note that choosing $k=0$ in 
eq.~\eqref{eq:conditionsOnOuter} delivers precisely the outer automorphisms of the Narain lattice, 
studied in section~\ref{sec:OuterAutomorphismsOfTheNarainLattice}. In particular, we find the 
condition $\hat\Sigma^{-1}\hat N\in\Z{}^{2D}$, which is satisfied only if 
$\hat\Sigma\in O_{\hat\eta}(D,D,\Z{})$, cf.\ eqs.~\eqref{eq:OuterGroupNarainLattice} 
and~\eqref{eq:TrafoOfNarainLattice}. However, for an orbifold compactification, the outer 
automorphism $\hat\Sigma$ is further constrained by the first condition in 
eq.~\eqref{eq:conditionsOnOuter} for $k \neq 0$.

\begin{table}[t!]
\center
\begin{tabular}{|c|c|c|}
\hline
$K$ & $\mathbbm Z_K$ Narain twist $\hat\Theta\in\mathrm{O}_{\hat{\eta}}(2,2,\Z{})$ & outer automorphisms $\hat\Sigma$\\
\hline
\hline
$2$ & $\left(\begin{array}{cccc}-1&0&0&0\\0&-1&0&0\\0&0&-1&0\\0&0&0&-1\end{array}\right)$ & 
      $\hat C_\mathrm{S},\,\hat C_\mathrm{T},\,\hat K_\mathrm{S},\,\hat K_\mathrm{T},\,\hat M,\,\hat\Sigma_*$ \\
\hline
$3$ & $\left(\begin{array}{cccc}0&-1&0&0\\1&-1&0&0\\0&0&-1&-1\\0&0&1&0\end{array}\right)$ & 
      $\hat K_\mathrm{S},\,\hat K_\mathrm{T},\,\hat C_\mathrm{S}\hat\Sigma_*$ \\
\hline
$4$ & $\left(\begin{array}{cccc}0&-1&0&0\\1&0&0&0\\0&0&0&-1\\0&0&1&0\end{array}\right)$   & 
      $\hat K_\mathrm{S},\,\hat K_\mathrm{T},\,\hat\Sigma_*$ \\
\hline
$6$ & $\left(\begin{array}{cccc}1&-1&0&0\\1&0&0&0\\0&0&0&-1\\0&0&1&1\end{array}\right)$   & 
      $\hat K_\mathrm{S},\,\hat K_\mathrm{T},\,\hat C_\mathrm{S}\hat\Sigma_*$ \\
\hline
\end{tabular}
\caption{(Rotational) Outer automorphisms $\hat\Sigma$ of all consistent symmetric orbifolds
$\mathbbm T^2/\Z{K}$ with Narain twist $\hat\Theta$.}
\label{tab:OuterAutomorphisms}
\end{table}

Let us focus now on $D=2$. As we have seen in section~\ref{sec:OuterAutomorphismsOfTheNarainLattice},
in this case the outer automorphisms of $\mathbbm T^2$ build the group of modular symmetries of a 
torus compactification, generated by $\hat C_\mathrm{S}$, $\hat C_\mathrm{T}$, $\hat K_\mathrm{S}$, 
$\hat K_\mathrm{T}$, $\hat M$ and $\hat\Sigma_*$, given in eqs.~\eqref{eq:KSandKT},~\eqref{eq:CSandCT} 
and~\eqref{eq:MandSigmaStar}. Thus, the outer automorphisms $\hat\Sigma$ of the orbifold must be a 
modular subgroup thereof, depending on the orbifold geometry $\mathbbm T^2/\Z{K}$, with $K\in\{2,3,4,6\}$. 
Table~\ref{tab:OuterAutomorphisms} presents the Narain twists $\hat\Theta$ and a choice of 
generators of the modular symmetries arising from the outer automorphisms of all symmetric 
$\mathbbm T^2/\Z{K}$ orbifolds. As we shall discuss explicitly in 
section~\ref{sec:GeometricalRotationsFromSL2ZU}, even though the Narain twist $\hat\Theta$ is an 
inner automorphism of the orbifold $\mathbbm T^2/\Z{K}$, it becomes an outer automorphism if the 
$\mathbbm T^2/\Z{K}$ orbifold is only a subsector of a six-dimensional factorized orbifold.

\subsection{Modular invariant field theory from strings}
\label{sec:modularInvariantTheory}

In string models, matter fields are associated with the excitation modes of strings. In a two-dimensional 
compactification of heterotic strings on a $\mathbbm T^2/\Z{K}$ orbifold, matter fields arise from
strings that close on the orbifold. They can be arranged in three categories:\\
{\it (i) Untwisted strings}, which are trivially closed strings, even before compactification;\\
{\it (ii) Winding strings}, which close only after winding along the directions $e_1$ and/or $e_2$ 
of the $\mathbbm T^2$ torus. At a generic point in moduli space, winding strings have masses at or 
beyond the string scale. Thus, they do not lead to matter fields that appear in the low-energy 
effective field theory; and\\
{\it (iii) Twisted strings}, which close only due to the action of the twist.

This classification can be stated in terms of elements of the orbifold space group, which set the 
boundary conditions for the strings to close. This relation is revealed by regarding 
eq.~\eqref{eq:OrbifoldConditions} as the boundary conditions of closed strings in the orbifold, 
according to
\begin{equation}
\label{eq:OrbifoldBoundaryConditions}
Y(\tau,\sigma+1) ~=~ g\,Y(\tau,\sigma) ~:=~ \Theta^k Y(\tau,\sigma) + E\,\hat{N}\;,
   \quad\mathrm{where}\quad g~=~(\Theta^k,E\,\hat N) ~\in~ S_\mathrm{Narain}\;.
\end{equation}
$Y=Y(\tau,\sigma)$ corresponds to the (bosonic) worldsheet string field in terms of the 
worldsheet coordinates $\tau$ and $\sigma$. Hence, we note that untwisted strings are 
associated with the trivial space group element $(\Id_4,0)$, winding strings are connected 
with translational space group elements $(\Id_4,E\,\hat N)$, and twisted strings are related 
to more general space group elements $(\Theta^k,E\,\hat N)$, with $\Theta^k\neq\Id_4$.

The properties and dynamics of string matter fields are inextricably linked to the attributes of 
the compact space on which these strings live. In particular, if two of the extra dimensions are 
associated with a $\mathbbm T^2/\Z{K}$ orbifold sector, matter fields inherit target space modular 
symmetries, whose generators are listed in table~\ref{tab:OuterAutomorphisms} for each twist order 
$K$. Matter fields of heterotic orbifold compactifications are endowed with a modular weight for 
each $\SL{2,\Z{}}$ modular symmetry, and we denote the set of all modular weights by $\mathfrak{n}$. 
Let us discuss the example of two modular symmetries $\SL{2,\Z{}}_T$ and $\SL{2,\Z{}}_U$ of the K\"ahler 
modulus $T$ and the complex structure modulus $U$, where we have $\mathfrak{n}=(n_T,n_U)\in\mathbbm Q^2$ (see 
e.g.\ refs.~\cite{Ibanez:1992hc,Olguin-Trejo:2017zav}). We denote untwisted and twisted matter 
fields of the orbifold theory as $\Phi_\mathfrak{n}$ given that matter fields can be distinguished in general 
by their modular weights $\mathfrak{n}$. In the case of twisted matter fields, they build multiplets 
$\Phi_\mathfrak{n}=(\phi_1,\phi_2,\ldots)^\mathrm{T}$, where the fields $\phi_i$ are associated with different 
closed strings, whose boundary conditions are produced by inequivalent space group elements with 
the same power $k$ of $\Theta$ (i.e.\ by space group elements from the same twisted sector but 
different conjugacy classes). Consequently, in the absence of discrete Wilson 
lines~\cite{Ibanez:1986tp} in the $\mathbbm T^2/\Z{K}$ orbifold sector, the twisted fields $\phi_i$ 
in the multiplet $\Phi_\mathfrak{n}$ share identical gauge quantum numbers and modular weights.

Ignoring \CP-like and mirror modular transformations that shall be addressed in the following 
section, a general matter field $\Phi_\mathfrak{n}$ transforms under general modular transformations 
$\gamma_T \in \SL{2,\Z{}}_T$ and $\gamma_U \in \SL{2,\Z{}}_U$, as defined by 
eqs.~\eqref{eq:TrafoOfTModulus} and~\eqref{eq:TrafoOfUModulus}, according to
\begin{subequations}\label{eq:ModularTransformationOfPhi}
\begin{eqnarray}
\label{eq:ModularTransformationOfPhiGammaT}
\Phi_\mathfrak{n} & \stackrel{\gamma_T}{\longrightarrow} & \Phi'_\mathfrak{n} ~:=~ (c_T\,T+d_T)^{n_T}\, \rho_{\rep{s}}(\gamma_T)\,\Phi_\mathfrak{n}\;,\\
\label{eq:ModularTransformationOfPhiGammaU}
\Phi_\mathfrak{n} & \stackrel{\gamma_U}{\longrightarrow} & \Phi'_\mathfrak{n} ~:=~ (c_U\,U+d_U)^{n_U}\, \rho_{\rep{s}}(\gamma_U)\,\Phi_\mathfrak{n}\;,
\end{eqnarray}
\end{subequations}
where $(c_T\,T + d_T)^{n_T}$ and $(c_U\,U + d_U)^{n_U}$ are the so-called automorphy factors with 
modular weights $\mathfrak{n}=(n_T,n_U)\in\mathbbm{Q}^2$. The matrices $\rho_{\rep{s}}(\gamma_T)$ 
and $\rho_{\rep{s}}(\gamma_U)$ build a (reducible or irreducible) representation $\rep s$ of some 
finite modular group (for example $\Gamma'_{N_T}\x \Gamma'_{N_U}$, where 
in terms of the $2\x2$ matrices $\mathrm S$, $\mathrm T$ given in eq.~\eqref{eq:SL2ZGeneratorsSandT}, 
the pair of transformation matrices $\rho_{\rep{s}}(\gamma_T=\mathrm S)$ and 
$\rho_{\rep{s}}(\gamma_T=\mathrm T)$ satisfy the defining relations 
eq.~\eqref{eq:DefiningRelationsOfGammaN} for some integer $N_T$ and analogously for $\gamma_U$).

As we can infer from the modular symmetry generators listed in table~\ref{tab:OuterAutomorphisms} and 
shall be discussed in detail in section~\ref{sec:SL2ZU}, in all $\mathbbm T^2/\Z{K}$ orbifold 
compactifications but those with $K=2$, the complex structure $U$ is not dynamic as it is
fixed at a value $\vev U$. We will see that this fact together with the transformation~\eqref{eq:ModularTransformationOfPhiGammaU}
leads to new insights about the effective nature of these symmetry transformations: they give 
rise to the well-known $R$-symmetries in heterotic orbifold compactifications. If the complex 
structure modulus $U$ is fixed, the strength of a superpotential coupling is given by a modular form 
$\hat Y^{(n_Y)}_{\rep{s_Y}}(T)$ of the K\"ahler modulus $T$. Then, under a general modular 
transformation $\gamma_T\in\SL{2,\Z{}}_T$ such a modular form transforms as
\begin{equation}\label{eq:YModularTrafoGeneral}
\hat Y^{(n_Y)}_{\rep{s_Y}}(T) ~\stackrel{\gamma_T}{\longrightarrow}~ \hat Y^{(n_Y)}_{\rep{s_Y}}\!\left(\tfrac{a_T\,T+b_T}{c_T\,T+d_T}\right) ~=~ (c_T\,T+d_T)^{n_Y} \rho_{\rep{s_Y}}(\gamma_T) \,\hat Y^{(n_Y)}_{\rep{s_Y}}(T)\,.
\end{equation}
Here, $\rep{s_Y}$ denotes an $s_Y$-dimensional representation of the finite modular group (e.g.\ 
$\Gamma_{N_T}$ or $\Gamma'_{N_T}$) and $(c_T\,T+d_T)^{n_Y}$ is the automorphy factor of $\gamma_T$ 
with modular weight $n_Y$. If the finite modular group is $\Gamma_N$, $n_Y$ must be even. However, 
for the double cover groups $\Gamma'_N$ that appear in string orbifold compactifications also odd 
$n_Y$ are allowed~\cite{Liu:2019khw}.

String models based on orbifold compactifications yield typically an $\mathcal N=1$ effective field
theory. Here, we are mainly interested in its K\"ahler potential $K$ and superpotential 
$\mathcal W$. The superpotential is a holomorphic function of moduli and matter fields 
$\Phi_\mathfrak{n}$. Under $\gamma_T \in\SL{2,\Z{}}_T$ and $\gamma_U \in\SL{2,\Z{}}_U$ 
modular transformations, the superpotential becomes
\begin{subequations}\label{eq:WgammaTrafo}
\begin{eqnarray}
\label{eq:WgammaTrafoT}
\!\!\!\!\!\!\mathcal{W}(T, U, \Phi_\mathfrak{n}) \!\!\!\!\! & \stackrel{\gamma_T}{\longrightarrow} &\!\!\!\!\! \mathcal{W}\!\left(\frac{a_T\,T+b_T}{c_T\,T+d_T},U, \Phi'_\mathfrak{n}\right) \!=\! (c_T\,T+d_T)^{-1}\, \mathcal{W}(T,U, \Phi_\mathfrak{n}),\\
\!\!\!\!\!\!\mathcal{W}(T, U, \Phi_\mathfrak{n}) \!\!\!\!\! & \stackrel{\gamma_U}{\longrightarrow} &\!\!\!\!\! \mathcal{W}\!\left(T,\frac{a_U\,U+b_U}{c_U\,U+d_U}, \Phi'_\mathfrak{n}\right) \!=\! (c_U\,U+d_U)^{-1}\, \mathcal{W}(T,U, \Phi_\mathfrak{n}),
\end{eqnarray}
\end{subequations}
i.e.\ $\mathcal W$ behaves similar to a matter field with modular weights $n_U=n_T=-1$ and 
which is invariant under the finite modular group of the theory. Consequently, each allowed 
superpotential coupling of matter fields $\Phi_\mathfrak{n}$ has to satisfy the conditions
\begin{equation}\label{eq:SumOfModularWeights}
n_Y\, + \sum_{\mathrm{matter\ fields}\ \Phi_\mathfrak{n}} n_T ~=~ -1 \qquad\mathrm{and}\qquad 
\rho_{\rep{s_Y}}(\gamma_T) \,\otimes \bigotimes_{\mathrm{matter\ fields}\ \Phi_\mathfrak{n}} \rho_{\rep{s}}(\gamma_T) ~\supset~ \Id\;,
\end{equation}
for $\gamma_T \in\SL{2,\Z{}}_T$ with $\mathfrak{n}=(n_T,n_U)$, $\rho_{\rep{s}}$ is the representation 
$\rep{s}$ associated with $\Phi_\mathfrak{n}$, and the coupling $\hat{Y}_{\rep{s_Y}}^{(n_Y)}$ of weight 
$n_Y$ transforms in the representation $\rep{s_Y}$. An analogous result holds for 
$\gamma_U \in\SL{2,\Z{}}_U$.

On the other hand, the general $\Phi$-independent contribution to the K\"ahler potential is given 
by~\cite{Dixon:1989fj}
\begin{equation}\label{eq:KahlerOfT}
K ~\supset~ K_0(T,U) ~:=~ -\ln\left(-\I\,T+\I\,\bar{T}\right) -\ln\left(-\I\,U+\I\,\bar{U}\right)\;.
\end{equation}
This universal K\"ahler contribution transforms under $\gamma_T \in\SL{2,\Z{}}_T$ and 
$\gamma_U \in\SL{2,\Z{}}_U$ as
\begin{subequations}\label{eq:KgammaTrafo}
\begin{eqnarray}
K_0(T,U) & \stackrel{\gamma_T}{\longrightarrow} &K_0\!\left(\frac{a_T\,T+b_T}{c_T\,T+d_T},U\right) ~=~ K_0(T,U) + f_T(T) + \overline{f_T(T)}\;,\\
K_0(T,U) & \stackrel{\gamma_U}{\longrightarrow} &K_0\!\left(T,\frac{a_U\,U+b_U}{c_U\,U+d_U}\right) ~=~ K_0(T,U) + f_U(U) + \overline{f_U(U)}\;,
\end{eqnarray}
\end{subequations}
where $f_T(T)=\ln(c_T T+d_T)$ and $f_U(U)=\ln(c_U U+d_U)$. Note that the terms $f_T(T)+\overline{f_T(T)}$ 
and $f_U(U)+\overline{f_U(U)}$ can be removed by performing a K\"ahler transformation after each 
modular transformation, rendering the K\"ahler potential modular invariant. 
A general K\"ahler transformation is defined as~\cite[ch.~23]{Wess:1992cp}
\begin{equation}
 K ~\to~ K + \mathcal F + \overline{\mathcal F}\,,\qquad \mathcal W ~\to~ e^{-\mathcal F}\,\mathcal W\,,
\end{equation}
where $\mathcal F$ is a holomorphic function of chiral superfields.
Therefore, any additional $\Phi$-dependent contribution to $K$ must be invariant under modular transformations, 
cf.\ ref.~\cite{Chen:2019ewa}. On the other hand, it turns out that the superpotential is also 
modular invariant since e.g.\ the modular transformation eq.~\eqref{eq:WgammaTrafoT} followed by 
the K\"ahler transformation applied for achieving invariance of $K$ yields
\begin{equation}
 \mathcal{W}~\stackrel{\gamma_T}{\longrightarrow}~(c_T\,T+d_T)^{-1}\, \mathcal{W}~\xyrightarrow{\text{K\"ahler}}
            ~(c_T\,T+d_T)^{-1}\,e^{f_T(T)} \mathcal{W} ~=~ \mathcal W\;,
\end{equation}
with $f_T(T) = \ln(c_T T +d_T)$.

\subsection{Summary}

The modular symmetry groups \SL{2,\Z{}} and $\mathrm{PSL}(2,\Z{})$ as well as their respective 
finite modular groups $\Gamma'_N$ and $\Gamma_N$ are natural to string compactifications. For 
example, as explained in section~\ref{sec:ModuliAndSymmetriesT2} strings on an internal two-torus 
yield two moduli: a complex structure modulus $U$ as well as a K\"ahler modulus $T$. This is in 
contrast to bottom-up models of flavor that typically consider only one modulus. Using the Narain 
description of a torus compactification introduced in section~\ref{sec:NarainTorus}, one can find 
the modular transformations acting on both moduli by computing the outer automorphisms 
of the Narain lattice associated with the compact space (see 
section~\ref{sec:OuterAutomorphismsOfTheNarainLattice}). These modular transformations build in 
general a large group that includes $\SL{2,\Z{}}_T$ and $\SL{2,\Z{}}_U$ for the standard modular 
transformations of $T$ and $U$, as well as two additional special transformations: a \Z2 mirror 
duality that exchanges $\SL{2,\Z{}}_T$ and $\SL{2,\Z{}}_U$, and a \Z2 \CP-like transformation. From 
this result, one can explore how this changes for all admissible $\mathbbm T^2/\Z{K}$ toroidal 
orbifolds, whose Narain formulation is introduced in section~\ref{sec:NarainOrbifolds}. As 
displayed in table~\ref{tab:OuterAutomorphisms}, we find that only $\mathbbm T^2/\Z2$ orbifolds 
preserve all the modular symmetries of the torus, while for $K\in\{3,4,6\}$ the modular groups are 
subgroups of the group of modular symmetries of the torus. Interestingly, under modular 
transformations string matter fields $\Phi_\mathfrak{n}$ and the couplings among them transform as 
representations of finite modular groups $\Gamma'_N$. In these terms, as detailed in 
section~\ref{sec:modularInvariantTheory}, we review the modular properties of the superpotential 
and K\"ahler potential that yield a modular invariant effective field theory.

\section[SL(2,Z)U of the complex structure]{\boldmath $\SL{2,\Z{}}_U$ of the complex structure\unboldmath}
\label{sec:SL2ZU}

$\SL{2,\Z{}}_T$ of the K\"ahler modulus $T$ is of stringy nature as it relates, for 
example, compactifications on compact spaces with small and large volumes. In contrast, in this 
section we will show explicitly using the Narain formulation of $D=2$ extra dimensions that the 
$\SL{2,\Z{}}_U$ factor of the complex structure modulus $U$ allows for a pure geometrical 
interpretation: On the one hand, $\SL{2,\Z{}}_U$ modular transformations are defined in 
terms of a special class of outer automorphisms of the Narain lattice. On the other hand, 
$\SL{2,\Z{}}_U$ only affects the geometrical vielbein $e$. Hence, we will see in a second 
step that rotational $\Z{N}$ symmetries of the two-torus are described by those $\SL{2,\Z{}}_U$ 
modular transformations that leave the complex structure modulus $U$ invariant. Moreover, the 
automorphy factors of $\SL{2,\Z{}}_U$ modular transformations turn out to be related to 
discrete $\Z{N}$ charges. Furthermore, they promote the two-dimensional rotational $\Z{N}$ symmetry 
to a discrete $R$-symmetry.

\subsection[Geometrical interpretation of SL(2,Z)U of the complex structure]{\boldmath Geometrical interpretation of $\SL{2,\Z{}}_U$ of the complex structure\unboldmath}
\label{sec:GeomInterpretationSL2ZU}

In the Narain lattice basis, the generators $\hat{C}_\mathrm{S}$ and $\hat{C}_\mathrm{T}$ of 
$\SL{2,\Z{}}_U$ (and also the \CP-like transformation $\hat\Sigma_*$) as defined in 
section~\ref{sec:OuterAutomorphismsOfTheNarainLattice} yield elements of the form
\begin{equation}\label{eq:hatSigma}
\hat\Sigma ~=~ \left(\begin{array}{cc}\hat\sigma&0\\0&\hat\sigma^{-\mathrm{T}}\end{array}\right) ~\in~ \langle~ \hat{C}_\mathrm{S}\;,\;\hat{C}_\mathrm{T}\;,\;\hat\Sigma_*~\rangle ~\subset~ \mathrm{O}_{\hat{\eta}}(2,2,\Z{})\;, 
\end{equation}
where $\hat\sigma\in\mathrm{GL}(2,\Z{})$. Note that, in contrast to  $\SL{2,\Z{}}_T$, 
modular transformations of this kind do not interchange winding and KK numbers. Next, we can change 
the basis to right- and left-moving coordinates (in analogy to eq.~\eqref{eq:NarainLatticeConjugation})
\begin{equation}\label{eq:SL2ZUSigmaFromHat}
\Sigma ~:=~ E\,\hat\Sigma\,E^{-1}\;.
\end{equation}
Hence, $\Sigma$ acts on right- and left-moving string coordinates as
\begin{equation}
\left(\begin{array}{c}y_\mathrm{R}\\y_\mathrm{L}\end{array}\right) ~\stackrel{\hat\Sigma}{\longrightarrow}~ \Sigma\,\left(\begin{array}{c}y_\mathrm{R}\\y_\mathrm{L}\end{array}\right)\;,
\end{equation}
see section~\ref{sec:ModularForOrbifolds}. Consequently, the coordinates $y$ and their dual 
coordinates $\tilde{y}$, as defined in eq.~\eqref{eq:CoordLR}, transform as
\begin{equation}\label{eq:SigmaOnCoordLR}
\begin{pmatrix}y\\\tilde{y}\end{pmatrix} ~\stackrel{\hat\Sigma}{\longrightarrow}~ \begin{pmatrix}\sigma & 0 \\ e^{-\mathrm{T}}\hat\sigma^{-\mathrm{T}} \left(\hat\sigma^\mathrm{T} B\,\hat\sigma-B \right)\,e^{-1} & \sigma^{-\mathrm{T}}\end{pmatrix}\,\begin{pmatrix}y\\\tilde{y}\end{pmatrix}\;,
\end{equation}
where $\sigma~:=~e\,\hat\sigma\,e^{-1}$. It is easy to see that
\begin{equation}
\hat\sigma^\mathrm{T} B\, \hat\sigma ~=~ \mathrm{det}(\hat\sigma)\,B \qquad\mathrm{for}\qquad B ~=~ B_{12}\,\left(\begin{array}{cc}0&1\\-1&0\end{array}\right)
\end{equation}
and $\hat\sigma\in\mathrm{GL}(2,\Z{})$. 

Now let us discuss the case $\mathrm{det}(\hat\sigma) =+1$, i.e.\ when 
$\hat\sigma\in\SL{2,\Z{}}$ corresponds to a modular transformation without \CP. Then, 
$\hat\sigma\in\SL{2,\Z{}}$ is compatible with the $B$-field in the two-torus 
$\mathbbm T^2$, i.e.\ $\hat\sigma^\mathrm{T}\, B\, \hat\sigma = B$, and we obtain
\begin{equation}\label{eq:SL2ZUSigma}
\Sigma ~=~ E\,\hat\Sigma\,E^{-1} ~=~ \frac{1}{2}\left(\begin{array}{cc}\sigma+\sigma^{-\mathrm{T}}&\sigma-\sigma^{-\mathrm{T}}\\\sigma-\sigma^{-\mathrm{T}}&\sigma+\sigma^{-\mathrm{T}}\end{array}\right)\;.
\end{equation}
Furthermore, eq.~\eqref{eq:SigmaOnCoordLR} yields in the case $\mathrm{det}(\hat\sigma) =+1$ the 
simple transformations
\begin{equation}\label{eq:SL2ZUTrafoOfYandTildeY}
y ~\stackrel{\hat\Sigma}{\longrightarrow}~ \sigma\,y \qquad\mathrm{and}\qquad \tilde{y} ~\stackrel{\hat\Sigma}{\longrightarrow}~ \sigma^{-\mathrm{T}}\,\tilde{y}\;.
\end{equation}
These transformations can be absorbed completely into a redefinition of the geometrical torus 
vielbein $e$. Explicitly, the torus boundary condition
\begin{equation}
y ~\sim~ y + e\,n \qquad\mathrm{with}\quad n\in\Z{}^2
\end{equation}
is mapped under $y \rightarrow \sigma\,y$ to 
\begin{equation}
\sigma\,y ~\sim~ \sigma\,y + e\,n  \qquad\Leftrightarrow\qquad y ~\sim~ y + \sigma^{-1}e\,n ~=~ y + \left(e\,\hat\sigma^{-1}\right)\,n\;.
\end{equation}
Hence, an $\SL{2,\Z{}}_U$ transformation $\hat\Sigma$ can be performed alternatively by a 
transformation of only the vielbein $e$, which then induces a change of the torus metric 
$G=e^{\mathrm{T}}e$, i.e.
\begin{equation}\label{eq:Sl2ZUonLattice}
e ~\stackrel{\hat\Sigma}{\longrightarrow}~ \sigma^{-1}\,e ~=~ e\,\hat\sigma^{-1} \qquad\Rightarrow\qquad G ~\stackrel{\hat\Sigma}{\longrightarrow}~ \hat\sigma^{-\mathrm{T}}\,G\,\hat\sigma^{-1}\;,
\end{equation}
while the $B$-field is invariant. Note that the boundary condition 
\begin{equation}
\tilde{y} ~\sim~ \tilde{y} + e^{-\mathrm{T}}\,\left(B\,n + \alpha'\,m\right) \qquad\mathrm{with}\quad n,m\in\Z{}^2
\end{equation}
of the dual coordinates $\tilde{y}$ (as given in eq.~\eqref{eq:NarainLatticeForYandtildeY}) 
transforms analogously under $\tilde{y} \rightarrow \sigma^{-\mathrm{T}}\,\tilde{y}$ from 
eq.~\eqref{eq:SL2ZUTrafoOfYandTildeY}. Since $\hat\sigma\in\SL{2,\Z{}}$, the 
two-dimensional lattice spanned by $e$ is mapped to itself under $\hat\sigma^{-1}$ in 
eq.~\eqref{eq:Sl2ZUonLattice}: in other words, $\hat\sigma\in\SL{2,\Z{}}$ is an outer 
automorphism of the two-dimensional lattice spanned by the geometrical vielbein $e$. For example, 
under modular $\mathrm{S}$ and $\mathrm{T}$ transformations of the complex structure modulus $U$, 
the geometrical lattice transforms as
\begin{subequations}\label{eq:TrafoOfeUnderSL2ZU}
\begin{eqnarray}
e_1 ~\stackrel{\hat{C}_\mathrm{S}}{\longrightarrow} \hspace{3pt}          -e_2 \qquad \mathrm{and} \qquad e_2 & \stackrel{\hat{C}_\mathrm{S}}{\longrightarrow} & e_1    \hspace{1.82cm} \mathrm{for\ S\ transformation}\;,\\
e_1 ~\stackrel{\hat{C}_\mathrm{T}}{\longrightarrow} \hspace{3pt}\phantom{-}e_1 \qquad \mathrm{and} \qquad e_2 & \stackrel{\hat{C}_\mathrm{T}}{\longrightarrow} & e_1+e_2\hspace{1cm}    \mathrm{for\ T\ transformation}\;,
\end{eqnarray}
\end{subequations}
using eq.~\eqref{eq:Sl2ZUonLattice} with $\hat\sigma=-\epsilon$ for $\hat{C}_\mathrm{S}$ and 
$\hat\sigma=\gamma$ for $\hat{C}_\mathrm{T}$ as given in eq.~\eqref{eq:CSandCT}. Moreover, using 
eq.~\eqref{eq:TrafoOfeUnderSL2ZU} the complex structure modulus $U$ defined in eq.~\eqref{eq:DefU} 
transforms as
\begin{subequations}\label{eq:SL2ZUTrafoSandT}
\begin{eqnarray}
U & \stackrel{\hat{C}_\mathrm{S}}{\longrightarrow} & -\frac{1}{U} \hspace{1.82cm} \mathrm{for\ S\ transformation}\;,\\
U & \stackrel{\hat{C}_\mathrm{T}}{\longrightarrow} & U + 1        \hspace{1.54cm} \mathrm{for\ T\ transformation}\;.
\end{eqnarray}
\end{subequations}
Note that these transformations~\eqref{eq:SL2ZUTrafoSandT} can also be obtained directly following 
ref.~\cite{Kikuchi:2020frp} if one takes the torus lattice vectors $e_1$ and $e_2$ to be complex 
numbers, i.e.\ $e_1,e_2\in\mathbb{C}$, and rewrites the complex structure modulus as $U=e_2/e_1$. 
Then, the transformations~\eqref{eq:TrafoOfeUnderSL2ZU} of $e_1$ and $e_2$ imply 
eqs.~\eqref{eq:SL2ZUTrafoSandT}. 

In addition, we take the transformation eq.~\eqref{eq:Sl2ZUonLattice} of the metric $G$ under $\hat\Sigma$ 
given in eq.~\eqref{eq:hatSigma} in order to obtain the general transformation property of the 
complex structure modulus $U$, i.e.\ (see also ref.~\cite{Love:1994ms})
\begin{equation}
U ~\stackrel{\hat\Sigma}{\longrightarrow}~ \frac{a_U\,U+b_U}{c_U\,U+d_U} \quad\mathrm{for}\quad \hat\Sigma = \left(\begin{array}{cc} \hat\sigma & 0\\ 0 &\hat\sigma^{-\mathrm{T}}\end{array}\right) \in\mathrm{O}_{\hat{\eta}}(2,2,\Z{})\;, \quad\mathrm{where}\quad \hat\sigma =: \left(\begin{array}{cc} a_U & -b_U\\-c_U & d_U\end{array}\right)\;,
\end{equation}
for $a_U,b_U,c_U,d_U \in \Z{}$ satisfying $a_U\,d_U-b_U\,c_U=+1$, while the K\"ahler modulus $T$ is invariant.

To summarize and to be specific, on the one hand we have identified a special class of outer 
automorphisms $\hat\Sigma$ of the Narain lattice $\Gamma$ for $D=2$, given by
\begin{equation}
\hat\Sigma ~=~ \left(\begin{array}{cc} \hat\sigma & 0\\ 0 &\hat\sigma^{-\mathrm{T}}\end{array}\right) ~\in~ \mathrm{O}_{\hat{\eta}}(2,2,\Z{}) \qquad\mathrm{with}\qquad \hat\sigma ~=:~  \left(\begin{array}{cc} a_U & -b_U\\-c_U & d_U\end{array}\right) ~\in~ \SL{2,\Z{}}\;.
\end{equation}
On the other hand, we have shown that an outer automorphism of this class acts geometrically on the 
vielbein $e$ that defines the two-torus as
\begin{equation}
e ~\stackrel{\hat\Sigma}{\longrightarrow}~ e\,\hat\sigma^{-1} \quad\Leftrightarrow\quad \begin{pmatrix}e_2\\e_1\end{pmatrix} ~\stackrel{\hat\Sigma}{\longrightarrow}~ \begin{pmatrix}a_U&b_U\\c_U&d_U\end{pmatrix}\,\begin{pmatrix}e_2\\e_1\end{pmatrix}\;.
\end{equation}
Hence, it gives rise to a modular transformation of the complex structure modulus
\begin{equation}
U ~\stackrel{\gamma_U}{\longrightarrow}~ \frac{a_U\,U+b_U}{c_U\,U+d_U}\;, \qquad\mathrm{where}\qquad \gamma_U~:=~\left(\begin{array}{cc} a_U & b_U\\c_U & d_U\end{array}\right) ~\in~ \SL{2,\Z{}}_U\;,
\end{equation}
while $T$ is invariant. In other words, there exist outer automorphisms of the Narain lattice 
$\Gamma$ for $D=2$ that are specified by their geometrical action $\hat\sigma\in\SL{2,\Z{}}$ 
on the torus vielbein $e$ and translate to modular transformations $\gamma_U\in\SL{2,\Z{}}_U$ 
of the complex structure modulus $U$, where the dictionary between these two reads explicitly
\begin{equation}\label{eq:sigma2gamma}
\gamma_U ~:=~ \left(\begin{array}{cc} 0 & 1\\1 & 0\end{array}\right)\, \hat\sigma^{-\mathrm{T}} \left(\begin{array}{cc} 0 & 1\\1 & 0\end{array}\right) ~\in~ \SL{2,\Z{}}_U\;.
\end{equation}

\subsection[Geometrical rotations from SL(2,Z)U]{\boldmath Geometrical rotations from $\SL{2,\Z{}}_U$\unboldmath}
\label{sec:GeometricalRotationsFromSL2ZU}

Based on the discussion from the last section, we now analyze a special class of outer automorphisms 
of the Narain lattice that corresponds to geometrical rotations in the extra-dimensional space, see 
also ref.~\cite{Kikuchi:2020frp} for a related discussion. We focus on two extra dimensions ($D=2$), 
but the generalization to six extra dimensions is straightforward, since a rotation in six dimensions 
can be decomposed into three rotations in three orthogonal two-dimensional planes.
Thus, the results of this section will be important for both: i) to define an orbifold twist and 
ii) to perform a two-dimensional ``sublattice rotation'' of a six-dimensional orbifold.

We begin with a discrete rotation $\theta_\subind{K} \in\Z{K} \subset \SO{2}$ that acts as
\begin{equation}\label{eq:SublatticeRotationGeom}
y ~\stackrel{\theta_\subind{K}}{\longrightarrow}~ \theta_\subind{K}\,y\;, \qquad\mathrm{where}\qquad \theta_\subind{K} ~:=~ \left(\begin{array}{cc} \cos(2\pi v_\subind{K}) & -\sin(2\pi v_\subind{K}) \\ \sin(2\pi v_\subind{K}) & \cos(2\pi v_\subind{K}) \end{array}\right) ~\in~ \SO{2}\;,
\end{equation}
while it leaves all orthogonal coordinates inert. The geometrical rotation angle $v_\subind{N}$ is 
defined as
\begin{equation}\label{eq:RotationAngle}
v_\subind{K} ~:=~ \frac{1}{K}\,, \qquad\mathrm{such\ that}\qquad \left(\theta_\subind{K}\right)^K ~=~ \Id_2\;.
\end{equation}
In order to be a rotational symmetry of the geometrical torus, $\theta_\subind{K}$ has to map the 
lattice spanned by the torus vielbein $e$ to itself. Hence, we have to impose the condition
\begin{equation}\label{eq:sigma}
\hat\sigma_\subind{K} ~:=~ e^{-1}\theta_\subind{K}\,e ~\stackrel{!}{\in}~ \SL{2,\Z{}} \qquad\mathrm{for}\qquad \theta_\subind{K} ~\in~ \SO{2}\;.
\end{equation}
This constrains the order $K$ to the allowed orders of the $\Z{K}$ wallpaper groups, being 
$K\in\{2,3,4,6\}$, see e.g.\ ref.~\cite{Fischer:2012qj}.

Now, we translate this discrete geometrical rotation into the Narain formulation of string theory. 
To do so, we split $y$ into right- and left-moving string coordinates using eq.~\eqref{eq:CoordLR} 
and define the action on these coordinates as
\begin{equation}\label{eq:LRRotation}
\left(\begin{array}{c}y_\mathrm{R}\\y_\mathrm{L}\end{array}\right) ~\stackrel{\Theta_\subind{K}}{\longrightarrow}~ \Theta_\subind{K}\,\left(\begin{array}{c}y_\mathrm{R}\\y_\mathrm{L}\end{array}\right)\;, \qquad\mathrm{where}\qquad \Theta_\subind{K} ~:=~ \left(\begin{array}{cc}\theta_{\subind{K},\,\mathrm{R}}&0\\0&\theta_{\subind{K},\,\mathrm{L}}\end{array}\right)\;,
\end{equation}
and $\theta_{\subind{K},\,\mathrm{R}}$, $\theta_{\subind{K},\,\mathrm{L}}\in\Z{K}\subset\SO{2}$ are 
of order $K$. Using eq.~\eqref{eq:SublatticeRotationGeom} we obtain
\begin{equation}
y ~:=~ \sqrt{\frac{\alpha'}{2}}\left(y_\mathrm{R} + y_\mathrm{L}\right) ~\stackrel{\Theta_\subind{K}}{\longrightarrow}~ \sqrt{\frac{\alpha'}{2}}\left(\theta_{\subind{K},\,\mathrm{R}}\,y_\mathrm{R} + \theta_{\subind{K},\,\mathrm{L}}\,y_\mathrm{L}\right) ~\stackrel{!}{=}~ \theta_\subind{K}\,y\;,
\end{equation}
such that the rotation has to be left-right symmetric
\begin{equation}\label{eq:LRSymmetricRotation}
\theta_\subind{K} ~:=~ \theta_{\subind{K},\,\mathrm{R}} ~=~ \theta_{\subind{K},\,\mathrm{L}}\;.
\end{equation}
Thus, the left-right-symmetric rotation $\Theta_\subind{K}$ defined in eq.~\eqref{eq:LRRotation} 
together with eq.~\eqref{eq:LRSymmetricRotation} corresponds to the transformation $\Sigma$ from 
eq.~\eqref{eq:SL2ZUSigma} with $\sigma = \theta_\subind{K}\in\SO{2}$, i.e.\ 
$\sigma^{-\mathrm{T}} = \sigma$. Consequently, we can follow the discussion around 
eq.~\eqref{eq:SL2ZUSigma} and write the rotation in terms of an outer automorphism of 
the Narain lattice as 
\begin{equation}
\hat\Theta_\subind{K} ~:=~ E^{-1}\Theta_\subind{K}\,E ~=~ \left(\begin{array}{cc}\hat\sigma_\subind{K}&0\\0&\hat\sigma_\subind{K}^{-\mathrm{T}}\end{array}\right) ~\in~ \mathrm{O}_{\hat{\eta}}(2,2,\Z{})\;.
\end{equation}
Due to this block-structure together with $\hat\sigma_\subind{K}:=e^{-1}\theta_\subind{K}\,e\in\SL{2,\Z{}}$, 
the Narain twist $\hat\Theta_\subind{K}$ can be decomposed in terms of the generators 
$\hat{C}_\mathrm{S}$ and $\hat{C}_\mathrm{T}$, given in eq.~\eqref{eq:CSandCT}. Then, using 
eq.~\eqref{eq:sigma2gamma} we can translate the action of the outer automorphism 
$\hat\Theta_\subind{K}$ to a corresponding modular transformation from $\SL{2,\Z{}}_U$ as
\begin{equation}\label{eq:SublatticeRotationToSL2Z}
e^{-1}\theta_\subind{K}\,e ~=~ \hat\sigma_\subind{K} ~=:~ \left(\begin{array}{cc}a_U&-b_U\\-c_U&d_U\end{array}\right) ~\in~ \SL{2,\Z{}} \quad\Leftrightarrow\quad \gamma_\subind{K} ~:=~ \left(\begin{array}{cc}a_U&b_U\\c_U&d_U\end{array}\right) ~\in~ \SL{2,\Z{}}_U\;,
\end{equation}
where the modular transformation $\gamma_\subind{K}$ is of order $K$, i.e.\ 
$\left(\gamma_\subind{K}\right)^K = \Id_2$.

\subsection[Stabilizing the complex structure modulus by geometrical rotations]{\boldmath Stabilizing the complex structure modulus by geometrical rotations\unboldmath}
\label{sec:StabilizingU}

Now, we can use the transformation property of the torus metric $G=e^\mathrm{T}e$ given in 
eq.~\eqref{eq:Sl2ZUonLattice} for a rotational symmetry eq.~\eqref{eq:sigma} in order to find
\begin{equation}
G ~\stackrel{\hat\Theta_\subind{K}}{\longrightarrow}~ \hat\sigma_\subind{K}^{-\mathrm{T}}\,G\,\hat\sigma_\subind{K}^{-1} ~=~ G\;.
\end{equation}
Hence, the metric $G$ and, therefore, the complex structure modulus $U$ (as defined in 
eq.~\eqref{eq:DefU}) must be invariant under a Narain twist $\hat\Theta_\subind{K}$. 
In other words, the vev of the complex structure modulus $\vev U$ is a 
fixed point of those modular transformations $\gamma_\subind{K}\in\SL{2,\Z{}}_U$ that 
correspond to rotational symmetries of the geometrical torus,
\begin{equation}\label{eq:FixedPointInU}
U ~\stackrel{\gamma_\subind{K}}{\longrightarrow}~ \gamma_\subind{K} \;U ~=~ \frac{a_U\,U+b_U}{c_U\,U+d_U} ~=~ U \qquad\mathrm{at}\qquad U ~=~ \vev U\;,
\end{equation}
where the vacuum expectation value $\vev U$ of the complex structure modulus 
parameterizes the torus vielbein $e$, except for the overall size of its two-torus. 

In our case of two-dimensional rotational symmetries of order $K$, the K\"ahler modulus $T$ is 
invariant but the complex structure modulus $U$ is frozen geometrically to, for example,
\begin{equation}\label{eq:ComplexStructureForZK}
\begin{array}{cclcl}
U                & = & \mathrm{unstabilized} \qquad &\mathrm{if}& K~=~2\;,\\
\langle U\rangle & = & \omega                \qquad &\mathrm{if}& K~=~3\mathrm{\ or\ } 6\;,\\
\langle U\rangle & = & \I                    \qquad &\mathrm{if}& K~=~4\;.
\end{array}
\end{equation}
The detailed results for $\Z{K}$ rotational symmetries of the Narain lattice in $D=2$ for 
$K\in\{2,3,4,6\}$ are listed in table~\ref{tab:ZNRotations}. Stabilizing the complex structure 
modulus by geometrical rotations has two important effects:
\begin{itemize}
\item[i)] Even if the complex structure modulus is stabilized, some modular transformations from 
$\SL{2,\Z{}}_U$ will remain unbroken, i.e.\ there are elements in $\SL{2,\Z{}}_U$ 
that leave $\langle U\rangle$ invariant. In order to analyze this, we define the so-called 
stabilizer subgroup $H_{\langle U\rangle}$ of $\SL{2,\Z{}}_U$ modular transformations 
that leave $\vev U$ invariant,
\begin{equation}\label{eq:SubgroupOfSL2ZU}
H_{\vev U} ~=~ \bigg\{ \begin{pmatrix} a_U & b_U \\ c_U & d_U \end{pmatrix}\in \SL{2,\Z{}}_U ~\bigg|~ \frac{a_U\,\vev U+b_U}{c_U\,\vev U+d_U} ~=~ \vev U\bigg\}\;,
\end{equation}
i.e.\ $\vev U$ is a fixed point of the elements of $H_{\vev U}$. Hence, the 
vev $\vev U$ of the stabilized complex structure modulus spontaneously breaks 
$\SL{2,\Z{}}_U$ to an unbroken symmetry group, being 
\begin{equation}\label{eq:StabilizersOfSL2ZU}
\begin{array}{cclcl}
H_{U}             & \cong & \Z{2}                      \qquad &\mathrm{if}& K~=~2\;,\\
H_{\vev U=\omega} & \cong & \Z{6}\cong\Z{2}\times\Z{3} \qquad &\mathrm{if}& K~=~3\mathrm{\ or\ } 6\;,\\
H_{\vev U=\I}     & \cong & \Z{4}                      \qquad &\mathrm{if}& K~=~4\;,
\end{array}
\end{equation}
where the $\Z{2}$ factors are generated by $\mathrm{S}^2 \in \SL{2,\Z{}}_U$ and $\Z{3}$ is 
generated by $\mathrm{S}\,\mathrm{T} \in \SL{2,\Z{}}_U$. These unbroken modular 
transformations will be of importance for the symmetries after orbifolding.
\item[ii)] As we will analyze next, stabilizing the complex structure modulus $U$ promotes the 
automorphy factor $(c_U\, U+d_U)$ to a phase.
\end{itemize}

\begin{table}[t!]
\center
\begin{tabular}{|c||ll|c|c|c|}
\hline
order             & \multicolumn{2}{c|}{\multirow{2}{*}{rotational symmetry $\hat\Theta_\subind{K}\in\mathrm{O}_{\hat{\eta}}(2,2,\Z{})$}} & \multirow{2}{*}{$\gamma_\subind{K}\in \SL{2,\Z{}}_U$} & \multirow{2}{*}{$\vev U$} & \multirow{2}{*}{$v_\subind{K}$}\\
$K$               &  & & & & \\
\hline
\hline
$2$ & $\hat\Theta_\subind{2} := \left(\hat{C}_\mathrm{S}\right)^2$                     & $\!\!\!\!\!\!=\left(\begin{array}{cccc}-1&0&0&0\\0&-1&0&0\\0&0&-1&0\\0&0&0&-1\end{array}\right)$\!\!&
$\mathrm{S}^2=\left(\begin{array}{cc}-1&0\\0&-1\end{array}\right)$ & arb. & $\nicefrac{1}{2}$\\
\hline
$3$ & $\hat\Theta_\subind{3} := \hat{C}_\mathrm{S}\,\hat{C}_\mathrm{T}$                & $\!\!\!\!\!\!=\left(\begin{array}{cccc}0&-1&0&0\\1&-1&0&0\\0&0&-1&-1\\0&0&1&0\end{array}\right)$&
$\mathrm{S}\,\mathrm{T}=\left(\begin{array}{cc}0&1\\-1&-1\end{array}\right)$ & $\omega$ & $\nicefrac{1}{3}$\\
\hline
$4$ & $\hat\Theta_\subind{4} := \hat{C}_\mathrm{S}$                                    & $\!\!\!\!\!\!=\left(\begin{array}{cccc}0&-1&0&0\\1&0&0&0\\0&0&0&-1\\0&0&1&0\end{array}\right)$&
$\mathrm{S}=\left(\begin{array}{cc}0&1\\-1&0\end{array}\right)$ & $\I$ & $\nicefrac{1}{4}$\\
\hline
$6$ & $\hat\Theta_\subind{6} := \left(\hat{C}_\mathrm{S}^3\hat{C}_\mathrm{T}\right)^5$ & $\!\!\!\!\!\!=\left(\begin{array}{cccc}1&-1&0&0\\1&0&0&0\\0&0&0&-1\\0&0&1&1\end{array}\right)$&
\!\!$(\mathrm{S}^3\mathrm{T})^5=\left(\begin{array}{cc}1&1\\-1&0\end{array}\right)$\!\! & $\omega$ & $\nicefrac{1}{6}$\\
\hline
\end{tabular}
\caption{Geometrical $\Z{K}$ symmetries $\hat\Theta_\subind{K}\in\mathrm{O}_{\hat{\eta}}(2,2,\Z{})$ 
of the Narain lattice, translated to modular transformations $\gamma_\subind{K}\in \SL{2,\Z{}}_U$ 
that leave the vev $\vev U$ of the associated complex structure modulus invariant. As a 
consequence, $\hat\Theta_\subind{K}$ acts left-right symmetrically and rotates the geometrical 
coordinate $y$ by an angle $v_\subind{K}$. At the same time, the superpotential picks up a phase 
$\exp\left(2\pi\I\,v_\subind{K}\right)$ such that the geometrical rotation $\hat\Theta_\subind{K}$ 
generates an $R$-symmetry, as shown in section~\ref{sec:U-Rsymmetries}}.
\label{tab:ZNRotations}
\end{table}

\subsection[R-symmetries from SL(2,Z)U]{\boldmath $R$-symmetries from $\SL{2,\Z{}}_U$\unboldmath}
\label{sec:U-Rsymmetries}

In addition, we can compute the automorphy factor $(c_U\,\vev U+d_U)^{-1}$ of the 
superpotential $\mathcal{W}$, see eq.~\eqref{eq:WgammaTrafo}, for a rotational $\Z{K}$ symmetry 
that is described by a modular transformation $\gamma_\subind{K}\in\SL{2,\Z{}}_U$. It has 
to be a phase of order $K$ since $(\gamma_\subind{K})^K = \Id_2$. Indeed, we obtain
\begin{equation}\label{eq:AutomorphyFactorAtU}
c_U\,\vev U+d_U ~=~ \exp\left(-2\pi\I\,v_\subind{K}\right)\;,
\end{equation}
where $c_U$ and $d_U$ are given by $\gamma_\subind{K}\in\SL{2,\Z{}}_U$ and 
$v_\subind{K}=\nicefrac{1}{K}$ turns out to coincide with the geometrical rotation angle 
corresponding to $\gamma_\subind{K}$ as defined in section~\ref{sec:GeometricalRotationsFromSL2ZU}
and listed in table~\ref{tab:ZNRotations}. This yields
\begin{equation}\label{eq:SublatticeRotationOfW}
\mathcal{W} ~\stackrel{\gamma_\subind{K}}{\longrightarrow}~ (c_U\,\vev U+d_U)^{-1}\, \mathcal{W} ~=~ \exp\left(2\pi\I\,v_\subind{K}\right)\,\mathcal{W}\;.
\end{equation}
Since the modular transformation $\gamma_\subind{K}\in\SL{2,\Z{}}_U$ leaves the (stabilized) 
complex structure modulus $\vev U$ invariant (cf.\ eq.~\eqref{eq:FixedPointInU}), the 
K\"ahler potential eq.~\eqref{eq:KahlerOfT} is invariant under $\gamma_\subind{K}$. Hence, the 
phase in eq.~\eqref{eq:SublatticeRotationOfW} has to be compensated by a transformation 
\begin{equation}\label{eq:SublatticeRotationOfVartheta}
\vartheta ~\stackrel{\gamma_\subind{K}}{\longrightarrow}~ \exp\left(2\pi\I\,R_\vartheta\right)\,\vartheta\;, 
   \qquad\mathrm{where}\qquad R_\vartheta ~:=~ \frac{1}{2K}
\end{equation}
of the Grassmann number $\vartheta$ of $\mathcal{N}=1$ superspace. Thus, the rotational modular 
transformation $\gamma_\subind{K}\in\SL{2,\Z{}}_U$ of the complex structure 
modulus is an $R$-transformation. In more detail, using $\int\!\mathrm{d}\vartheta\,\vartheta = 1$ we 
know that the $R$-charge of $\int\!\mathrm{d}\vartheta$ is $-R_\vartheta$, such that 
$\mathcal{L}\supset\int\!\mathrm{d}^2\vartheta\, \mathcal{W}$ is invariant.

\subsection{Summary}

In summary, a two-dimensional rotational $\Z{K}$ symmetry of the extra dimensional space is an 
outer automorphism of the Narain lattice that leaves the torus metric $G$ and the $B$-field 
invariant. Because of its geometric nature, it corresponds to a modular transformation 
$\gamma_\subind{K}$ from $\SL{2,\Z{}}_U$ of the complex structure modulus $U$ that leaves $U$ 
evaluated at its vev and $T$ invariant, as might have been expected from geometrical intuition. 
Moreover, a geometric rotation in two dimensions acts as an $R$-transformation, where the 
$R$-charge of the superpotential originates from the automorphy factor $(c_U\,\vev U+d_U)^{-1}$ 
of the modular transformation $\gamma_\subind{K}\in\SL{2,\Z{}}_U$.

\section[R-symmetries and CP in six-dimensional orbifolds]{\boldmath $R$-symmetries and $\CP$ in six-dimensional orbifolds\unboldmath}
\label{sec:Rsymmetries}

It is known that toroidal orbifold compactifications of string theory lead to discrete
$R$-symmetries of the four-dimensional effective theory. In the traditional approach, 
$R$-symmetries are explained to originate from rotational isometries of the six-dimensional 
orbifold geometry~\cite{Kobayashi:2004ya,Bizet:2013gf,Nilles:2013lda,Bizet:2013wha,Nilles:2017heg}. 
Here, we give a novel interpretation for the origin of $R$-symmetries as a special class of outer 
automorphisms of the Narain space group: $R$-symmetries correspond to elements from the modular 
group $\SL{2,\Z{}}_U$ that leave the vev of the (stabilized) complex structure modulus $U$ 
invariant.

\subsection[Six-dimensional orbifolds T6/P]{\boldmath Six-dimensional orbifolds $\mathbbm T^6/P$\unboldmath}
\label{sec:FactorizedOrbifold}

In order to specify a geometrical point group $P\cong \Z{K}$ of a six-dimensional orbifold, we 
define a six-dimensional orbifold twist $\theta\in\SO{6}$ by combining three two-dimensional 
$\Z{K_i}$ rotations $\theta_\subind{K_i}\in\SO{2}$ given in eq.~\eqref{eq:SublatticeRotationGeom} 
for $i\in\{1,2,3\}$. In detail, the orbifold twist is chosen as
\begin{equation}
\theta ~:=~ \left(\theta_\subind{K_1}\right)^{k_1} \oplus \left(\theta_\subind{K_2}\right)^{k_2} \oplus \left(\theta_\subind{K_3}\right)^{k_3} \quad\mathrm{such\ that}\quad y ~\stackrel{\theta}{\longrightarrow}~ \theta\,y\;,
\end{equation}
for $k_i \in\{0,1,\ldots,K_i-1\}$ and $y\in\mathbbm{R}^6$. One can choose complex coordinates 
$z^i:=y^{2i-1}+\I\,y^{2i}$ for $i\in\{1,2,3\}$. Then, each two-dimensional rotation $\theta_\subind{K_i}$ 
acts only in the $i$-th complex plane $z^i$. The orbifold twist $\theta$ generates a $\Z{K}$ 
rotation group, whose order $K$ is given by the least common multiple of $K_1$, $K_2$ and $K_3$. By 
combining the three rotation angles $k_i\,v_\subind{K_i}=\nicefrac{k_i}{K_i}$ of 
$(\theta_\subind{K_i})^{k_i}$ we obtain the $\Z{K}$ orbifold twist vector
\begin{equation}\label{eq:ZKTwist}
v ~:=~ \left( 0, \frac{k_1}{K_1}, \frac{k_2}{K_2}, \frac{k_3}{K_3}\right)\;.
\end{equation}
We have extended $v$ by an additional null entry for its action in string light-cone coordinates, 
including those of the uncompactified space. Using the results from table~\ref{tab:ZNRotations}, 
the orbifold twist vector $v$ corresponds to a modular transformation 
\begin{equation}\label{eq:CombinedGamma}
\gamma ~:=~ \left(\gamma_\subind{K_1}\right)^{k_1} \oplus \left(\gamma_\subind{K_2}\right)^{k_2} \oplus \left(\gamma_\subind{K_3}\right)^{k_3} ~\in~ \SL{2,\Z{}}_{U_1} \times \SL{2,\Z{}}_{U_2} \times \SL{2,\Z{}}_{U_3}\;.
\end{equation}
This transformation leaves the three complex structure moduli $U_i$ invariant, where $i\in\{1,2,3\}$ 
corresponds to the three complex planes, i.e.\
\begin{equation}
U_i ~\stackrel{\gamma}{\longrightarrow}~ \left(\gamma_\subind{K_i}\right)^{k_i}U_i ~\stackrel{!}{=}~ U_i \qquad\mathrm{for}\qquad U_i ~=~ \langle U_i\rangle
\end{equation}
and $\langle U_i\rangle$ is given in eq.~\eqref{eq:ComplexStructureForZK} for the cases 
$K_i\in\{2,3,4,6\}$. Hence, a complex structure modulus $U_i$ is either stabilized if $K_i \neq 2$ 
or unstabilized if $K_i = 2$.

In addition, using the three automorphy factors arising from the modular transformations 
$\gamma_\subind{K_i}\in\SL{2,\Z{}}_{U_i}$ for $i\in\{1,2,3\}$ in combination with 
eqs.~\eqref{eq:SublatticeRotationOfW} and~\eqref{eq:SublatticeRotationOfVartheta}, we confirm that 
the superpotential $\mathcal{W}$ and the Grassmann number $\vartheta$ are invariant under the 
combined modular transformation $\gamma$ from eq.~\eqref{eq:CombinedGamma},
\begin{subequations}
\begin{eqnarray}
\mathcal{W} & \stackrel{\gamma}{\longrightarrow} & \exp\left(2\pi\I\,\left(k_1\,v_\subind{K_1}+k_2\,v_\subind{K_2}+k_3\,v_\subind{K_3}\right)\right)\,\mathcal{W} ~=~ \mathcal{W}\;,\\
\vartheta   & \stackrel{\gamma}{\longrightarrow} & \exp\left(\frac{2\pi\I}{2}\!\left(k_1\,v_\subind{K_1}+k_2\,v_\subind{K_2}+k_3\,v_\subind{K_3}\right)\right)\,\vartheta   ~=~ \vartheta\;,\label{eq:ZNTwistOnVartheta}
\end{eqnarray}
\end{subequations}
if the associated orbifold twist vector $v$ preserves $\mathcal{N}=1$ supersymmetry, i.e.\ if
\begin{equation}\label{eq:N1SUSY}
\frac{k_1}{K_1} + \frac{k_2}{K_2} + \frac{k_3}{K_3} ~=~ 0\ \mathrm{mod}\ 2\;.
\end{equation}
On the other hand, the K\"ahler potential is invariant under the modular transformation 
$\gamma$ given in eq.~\eqref{eq:CombinedGamma}: For $K_i\in\{3,4,6\}$, the complex structure modulus 
$U_i$ is fixed to the values $\vev{U_i}$ provided in eq.~\eqref{eq:ComplexStructureForZK}. Hence, 
the K\"ahler potential of the theory does not exhibit any field-dependence on $U_i$. For $K_i=2$, 
the K\"ahler potential does depend on the (unstabilized) modulus $U_i$ explicitly, but is left 
invariant by the rotational modular transformation $\gamma$ because neither the complex 
structure modulus itself nor the $\Phi$-dependent terms of the form $|\Phi|^2$ are altered 
under these transformations.

\paragraph{Factorized orbifolds.} In the following we concentrate on factorized $\mathbbm T^6/\Z{K}$ 
orbifold geometries, where the six-torus factorizes as $\mathbbm T^6 = \mathbbm T^2\times\mathbbm T^2\times\mathbbm T^2$ 
and each two-dimensional $\Z{K_i}$ rotation $\theta_\subind{K_i}$ is a symmetry of the respective 
two-torus $\mathbbm T^2$. In this case, the orbifold twist $\hat\Theta$ in the Narain lattice basis 
reads
\begin{equation}\label{eq:FactorizedOrbifoldTwist}
\hat\Theta ~:=~ \hat\Theta_\subind{K_1} \oplus \hat\Theta_\subind{K_2} \oplus \hat\Theta_\subind{K_3} ~\in~ \mathrm{O}_{\hat{\eta}}(2,2,\Z{}) \times \mathrm{O}_{\hat{\eta}}(2,2,\Z{}) \times \mathrm{O}_{\hat{\eta}}(2,2,\Z{}) ~\subset~ \mathrm{O}_{\hat{\eta}}(6,6,\Z{})\;,
\end{equation}
where the $\Z{K_i}$ Narain twists $\hat\Theta_\subind{K_i}\in\mathrm{O}_{\hat{\eta}}(2,2,\Z{})$ of 
order $K_i\in\{2,3,4,6\}$ are given in table~\ref{tab:ZNRotations}.

For example, for the phenomenological promising $\Z{6}$-II $(1,1)$
orbifold geometry~\cite{Kobayashi:2004ya,Buchmuller:2005jr,Lebedev:2006kn}, we have an orbifold 
twist vector
\begin{equation}\label{eq:Z6IITwist}
v ~=~ \left( 0, \nicefrac{1}{6}, \nicefrac{1}{3}, \nicefrac{-1}{2}\right)\;,
\end{equation}
which preserves $\mathcal{N}=1$ supersymmetry since $\nicefrac{1}{6}+\nicefrac{1}{3}-\nicefrac{1}{2}=0$. 
For the $\Z{6}$-II $(1,1)$ orbifold geometry~\cite{Fischer:2012qj} this twist vector corresponds to 
an orbifold twist $\hat\Theta$ in the Narain lattice basis
\begin{equation}\label{eq:NarainTwistFactorizedOrbifold}
\hat\Theta ~:=~ \hat\Theta_\subind{6} \oplus \hat\Theta_\subind{3} \oplus \hat\Theta_\subind{2} ~\in~ \mathrm{O}_{\hat{\eta}}(2,2,\Z{}) \times \mathrm{O}_{\hat{\eta}}(2,2,\Z{}) \times \mathrm{O}_{\hat{\eta}}(2,2,\Z{}) ~\subset~ \mathrm{O}_{\hat{\eta}}(6,6,\Z{})\;,
\end{equation}
Note that we have chosen the rotation angle of $\hat\Theta_\subind{2}$ as $\nicefrac{-1}{2}$ in 
order to fix the action on spacetime spinors, see for example eq.~\eqref{eq:ZNTwistOnVartheta}.
Using the results from table~\ref{tab:ZNRotations}, the $\Z{6}$-II orbifold twist vector $v$ 
corresponds to a modular transformation 
\begin{equation}
\gamma ~:=~ \gamma_\subind{6}\oplus\gamma_\subind{3}\oplus\gamma_\subind{2} ~\in~ \SL{2,\Z{}}_{U_1} \times \SL{2,\Z{}}_{U_2} \times \SL{2,\Z{}}_{U_3}\;.
\end{equation}
This transformation leaves the three complex structure moduli $U_i$ of the three complex planes 
$i\in\{1,2,3\}$ invariant,
\begin{equation}
\left(U_1, U_2, U_3\right) ~\stackrel{\gamma}{\longrightarrow}~ \left(\gamma_\subind{6}U_1, \gamma_\subind{3}U_2, \gamma_\subind{2}U_3\right) ~=~ \left(U_1, U_2, U_3\right)\;,
\end{equation}
if the moduli are evaluated at 
\begin{equation}
\left(U_1, U_2, U_3\right) ~=~ \left(\omega,\omega,U_3\right)\;,
\end{equation}
see table~\ref{tab:ZNRotations}. Hence, the complex structure moduli $U_1$ and $U_2$ of the 
$\Z{6}$-II orbifold geometry are stabilized (for example) at $\langle U_1\rangle=\langle U_2\rangle=\omega$, 
while $U_3$ in the $\Z{2}$ orbifold plane is unstabilized.

\subsection[Sublattice rotations from SL(2,Z)U of the complex structure modulus]{\boldmath Sublattice rotations from $\SL{2,\Z{}}_U$ of the complex structure modulus\unboldmath}
\label{sec:SublatticeRotations}

For simplicity, we have chosen the six-torus $\mathbbm T^6$ that underlies the $\mathbbm T^6/\Z{K}$ 
orbifold geometry to be factorized as $\mathbbm T^2\times\mathbbm T^2\times\mathbbm T^2$. Then, the 
six-dimensional orbifold has three independent $\Z{K_i}$ rotational isometries, whose orders 
$K_i\in\{2,3,4,6\}$ for $i\in\{1,2,3\}$ are given in general by the components of the $\Z{K}$ 
orbifold twist vector defined in eq.~\eqref{eq:ZKTwist}: there is one rotational isometry per 
two-torus $\mathbbm T^2$. These $\Z{K_i}$ rotations are also called sublattice rotations.

In detail, as the $\mathbbm T^6/\Z{K}$ orbifold geometry (with $\Z{K}$ orbifold twist vector given 
in eq.~\eqref{eq:ZKTwist}) is assumed to be factorized, it contains three two-dimensional 
subsectors, which we call $\mathbbm T^2/\Z{K_i}$ orbifold {\it sectors} for $i\in\{1,2,3\}$. Then, a 
sublattice rotation $\theta_\subind{K_i}$ acts on the geometrical coordinate $y$ of the 
$\mathbbm T^2/\Z{K_i}$ orbifold sector as
\begin{equation}
y ~\stackrel{\theta_\subind{K_i}}{\longrightarrow}~ \theta_\subind{K_i}\,y\;, \qquad\mathrm{where}\qquad \theta_\subind{K_i} ~\in~ \Z{K_i} ~\subset~\SO{2}\;,
\end{equation}
while it leaves the orthogonal coordinates invariant. Following 
section~\ref{sec:GeometricalRotationsFromSL2ZU}, these three geometrical transformations for 
$i\in\{1,2,3\}$ correspond to three outer automorphisms $\hat{\mathrm{R}}_i$ of the full 
six-dimensional Narain space group, given by
\begin{subequations}\label{eq:T6OrbifoldOuterAuts}
\begin{eqnarray}
\hat{\mathrm{R}}_1 & := & \hat\Theta_\subind{K_1} \oplus \Id_4 \oplus \Id_4\;,\\
\hat{\mathrm{R}}_2 & := & \Id_4 \oplus \hat\Theta_\subind{K_2} \oplus \Id_4\;,\\
\hat{\mathrm{R}}_3 & := & \Id_4 \oplus \Id_4 \oplus \hat\Theta_\subind{K_3}\;,
\end{eqnarray}
\end{subequations}
where $\hat{\mathrm{R}}_i \in \mathrm{O}_{\hat{\eta}}(2,2,\Z{}) \times \mathrm{O}_{\hat{\eta}}(2,2,\Z{}) \times \mathrm{O}_{\hat{\eta}}(2,2,\Z{}) \subset \mathrm{O}_{\hat{\eta}}(6,6,\Z{})$ and
\begin{equation}
\hat\Theta_\subind{K_i} ~=~ \left(\begin{array}{cc}\hat\sigma_\subind{K_i}&0\\0&\hat\sigma_\subind{K_i}^{-\mathrm{T}}\end{array}\right) ~\in~ \mathrm{O}_{\hat{\eta}}(2,2,\Z{}) \quad\mathrm{and}\quad \hat\sigma_\subind{K_i} ~=:~ \left(\begin{array}{cc}a_i&-b_i\\-c_i&d_i\end{array}\right) ~\in~ \SL{2,\Z{}}\;,
\end{equation}
for $i\in\{1,2,3\}$. Each of them is associated with a modular transformation 
\begin{equation}
\gamma_\subind{K_i} ~:=~ \left(\begin{array}{cc}a_i&b_i\\c_i&d_i\end{array}\right) ~\in~ \SL{2,\Z{}}_{U_i}
\end{equation}
corresponding to the complex structure modulus $U_i$, which leaves \emph{all} moduli invariant. 
Note that the product $\hat{\mathrm{R}}_1\,\hat{\mathrm{R}}_2\,\hat{\mathrm{R}}_3$ from 
eq.~\eqref{eq:T6OrbifoldOuterAuts} equals the Narain twist $\hat\Theta$ given in 
eq.~\eqref{eq:NarainTwistFactorizedOrbifold}. Hence, $\hat{\mathrm{R}}_1\,\hat{\mathrm{R}}_2\,\hat{\mathrm{R}}_3$ 
is an inner automorphism of the Narain space group.

Next, we analyze the action of the modular transformation $\gamma_\subind{K_i}\in\SL{2,\Z{}}_{U_i}$ 
on matter fields $\Phi_\mathfrak{n}$, where $\mathfrak{n}$ collectively denotes the set of all 
modular weights, containing the modular weights $n_{U_i}$ of $\SL{2,\Z{}}_{U_i}$ for 
$i\in\{1,2,3\}$, see e.g.\ ref.~\cite{Ibanez:1992hc,Olguin-Trejo:2017zav} for the definition of the 
(fractional) modular weights for various orbifold geometries. In this case, the modular 
transformation~\eqref{eq:ModularTransformationOfPhi} of matter fields reads
\begin{equation}\label{eq:ModularTransformationOfPhiUnderRotation}
\Phi_\mathfrak{n} ~\stackrel{\gamma_\subind{K_i}}{\longrightarrow}~ (c_i\,\langle U_i \rangle+d_i)^{n_{U_i}}\,\rho_{\rep{s}}(\gamma_\subind{K_i})\,\Phi_\mathfrak{n} ~=~ \exp\left(-2\pi\I\,v_\subind{K_i}\,n_{U_i}\right)\,\rho_{\rep{s}}(\gamma_\subind{K_i})\,\Phi_\mathfrak{n}\;,
\end{equation}
such that the superpotential picks up a phase
\begin{equation}\label{eq:SublatticeRotationOfW2}
\mathcal{W} ~\stackrel{\gamma_\subind{K_i}}{\longrightarrow}~ (c_i\,\langle U_i \rangle+d_i)^{-1}\, \mathcal{W} ~=~ \exp\left(2\pi\I\,v_\subind{K_i}\right)\,\mathcal{W}\;.
\end{equation}
Here, we used that the automorphy factor evaluated at $\langle U_i \rangle$ becomes a 
modulus-independent phase, see eq.~\eqref{eq:AutomorphyFactorAtU}. In order to ensure that 
$(\gamma_\subind{K_i})^{K_i}=\Id_2$, the matrix $\rho_{\rep{s}}(\gamma_\subind{K_i})$ of the modular 
transformation has to be of order $K_i$. Furthermore, note that $\rho_{\rep{s}}(\gamma_\subind{K_i})$ can be 
diagonal or non-diagonal: On the one hand, it is diagonal for example for $K_i\in\{2,3\}$ since the 
corresponding $\mathrm{O}_{\hat{\eta}}(6,6,\Z{})$ sublattice rotation maps each string (i.e.\ each 
conjugacy class of the constructing element) to itself, possibly multiplied by a phase. On the 
other hand, $\rho_{\rep{s}}(\gamma_\subind{K_i})$ is non-diagonal for example for a $K_i=4$ sublattice 
rotation in a $\mathbbm T^6/\Z{4}$ orbifold geometry, see ref.~\cite{Bizet:2013wha}.

Eq.~\eqref{eq:ModularTransformationOfPhiUnderRotation} defines the discrete $R$-charges of matter 
fields $\Phi_\mathfrak{n}$ for modular transformations $\gamma_\subind{K_i}\in\SL{2,\Z{}}_{U_i}$ 
that correspond to geometrical sublattice rotations. On the other hand, it is 
known~\cite{Kobayashi:2004ya,Bizet:2013gf,Nilles:2013lda,Bizet:2013wha,Nilles:2017heg} that a 
matter field $\Phi_\mathfrak{n}$ transforms under a geometrical sublattice rotation 
$\theta_\subind{K_i}$ as
\begin{equation}
  \Phi_\mathfrak{n} ~\stackrel{\theta_\subind{K_i}}{\longrightarrow}~ \exp\left(\frac{2\pi\I\,R_i}{K_i^2}\right)\,\Phi_\mathfrak{n}\;.
\end{equation}
Using $K_i^2$ as denominator, the $R$-charges $R_i$ of superfields $\Phi_\mathfrak{n}$ become 
integer, as defined in ref.~\cite{Nilles:2013lda} for $K_i\in\{2,3,6\}$. Consequently, $R$-charges 
$R_i$ and modular weights $n_{U_i}$ are related. For example, if the representation matrix 
$\rho_{\rep{s}}(\gamma_\subind{K_i})$ from eq.~\eqref{eq:ModularTransformationOfPhiUnderRotation} 
is diagonal, we find
\begin{equation}
\label{eq:RchargesVsModularWeights}
R_i ~=~ \left(-n_{U_i} + \alpha_i\right)K_i\;,
\end{equation}
where $\alpha_i$ gives the order $K_i$ phase of the $i$-th entry on the diagonal of 
$\rho_{\rep{s}}(\gamma_\subind{K_i})$, i.e.\ $\rho_{\rep{s}}(\gamma_\subind{K_i}) 
 = \mathrm{diag}(\exp(\nicefrac{2\pi\I\alpha_i(\phi_1)}{K_i}),\ldots,\exp(\nicefrac{2\pi\I\alpha_i(\phi_s)}{K_i}))$.

Couplings $\hat{Y}_{\rep{s_Y}}^{(n_Y)}$ do not depend on the complex structure modulus $U_i$ 
if $K_i\in\{3,4,6\}$. In contrast, for $K_i=2$ they do depend on $U_i$ but are invariant under the 
``rotational'' modular transformation $\gamma_\subind{2}$. Hence, we can set $n_Y=0$ and 
$\rho_{\rep{s_Y}}(\gamma_\subind{2})=1$. Then, we obtain from eq.~\eqref{eq:SumOfModularWeights} the 
invariance conditions
\begin{equation}\label{eq:RInvarianceFromModularInvariance}
\sum_{\mathrm{matter\ fields\ } \Phi_\mathfrak{n}} n_{U_i} ~=~ -1
\qquad\mathrm{and}\qquad \bigotimes_{\mathrm{matter\ fields}\ \Phi_\mathfrak{n}} \rho_{\rep{s}}(\gamma_\subind{K_i}) ~\supset~ \Id\;.
\end{equation}
The later condition can be rewritten (for diagonalized representation matrices 
$\rho_{\rep{s}}(\gamma_\subind{K_i})$) as
\begin{equation}
\sum_{\mathrm{matter\ fields\ }\Phi_\mathfrak{n}} \alpha_i ~=~ 0\ \mathrm{mod}\ K_i\;.
\end{equation}
Consequently, we see that using eqs.~\eqref{eq:RInvarianceFromModularInvariance} we have to impose 
the constraint
\begin{equation}\label{eq:RChargeConservation}
\sum_{\mathrm{matter\ fields\ } \Phi_\mathfrak{n}} R_i ~=~ K_i\ \mathrm{mod}\ (K_i)^2
\end{equation}
on the $R$-charges $R_i$ for each superpotential coupling of matter fields $\Phi_\mathfrak{n}$ to 
be allowed, cf.\ refs.~\cite{Kobayashi:2004ya,Bizet:2013gf,Nilles:2013lda,Bizet:2013wha,Nilles:2017heg}. 
Hence, the order $K_i$ sublattice rotation yields an $R$-symmetry that is in general $\Z{(K_i)^2}^R$, 
where the nontrivial right-hand side of eq.~\eqref{eq:RChargeConservation} follows from eq.~\eqref{eq:SublatticeRotationOfW2}. 
As we have shown, this $R$-symmetry originates from unbroken modular transformations $\SL{2,\Z{}}_{U_i}$ 
(that leave the complex structure modulus $U_i$ invariant), even in the case when $U_i$ is frozen 
geometrically by the orbifolding.

Finally, let us note that for special points in moduli space (e.g.\ for special values of a complex 
structure modulus or for vanishing discrete Wilson lines) there can be additional $R$-symmetries 
originating from the modular group $\SL{2,\Z{}}_{U}$ of a complex structure modulus \emph{and} from 
$\SL{2,\Z{}}_{T}$ of the K\"ahler modulus. A detailed example of sublattice rotations in a 
$\mathbbm T^2/\Z3$ orbifold sector will be discussed later in section~\ref{sec:Z3SublatticeRotation}.

\subsection[CP for six-dimensional orbifolds]{\boldmath$\CP$ for six-dimensional orbifolds\unboldmath}
\label{sec:CPin6D}

A \CP-like transformation has to map a field to a $\CP$-conjugate partner with inverse (discrete 
and gauge) charges. For heterotic orbifolds, a string state with constructing element 
$\hat{g}\in \hat S_\mathrm{Narain}$ finds its $\CP$-partner in a string state with constructing 
element $\hat{g}^{-1}\in \hat S_\mathrm{Narain}$. Hence, a $\CP$-like transformation is an outer 
automorphism of the Narain space group $\hat S_\mathrm{Narain}$, such that 
\begin{equation}
\hat{g} ~\stackrel{\CP}{\longrightarrow}~ \left(\hat{\CP},0\right)^{-1}\hat{g}\,\left(\hat{\CP},0\right) ~\stackrel{!}{=}~ \hat{g}^{-1}\;,
\end{equation}
for all constructing elements $\hat{g} \in \hat S_\mathrm{Narain}$. This means in particular that 
the Narain twist is mapped to its inverse under a $\CP$-like transformation. In detail, in the 
Narain lattice basis we have to impose the conditions
\begin{equation}\label{eq:CPForTheta}
\hat\Theta ~\stackrel{\CP}{\longrightarrow}~ \left(\hat{\CP}\right)^{-1}\hat\Theta\,\left(\hat{\CP}\right) ~\stackrel{!}{=}~ \hat\Theta^{-1} \quad\mathrm{and}\quad \hat{\CP} ~\in~ \mathrm{O}_{\hat{\eta}}(6,6,\Z{})\;.
\end{equation}

As in section~\ref{sec:FactorizedOrbifold}, we now assume our six-dimensional orbifold geometry 
$\mathbbm T^6/\Z{K}$ under consideration to be factorized, i.e.\ 
$\mathbbm T^6=\mathbbm T^2\times\mathbbm T^2\times\mathbbm T^2$. In more detail, we take the $\Z{K}$ 
Narain twist $\hat\Theta$ to be built out of three two-dimensional orbifold sectors
\begin{equation}\label{eq:FactorizedTwist}
\hat\Theta ~:=~ \hat\Theta_\subind{K_1} \oplus \hat\Theta_\subind{K_2} \oplus \hat\Theta_\subind{K_3} ~\in~ \mathrm{O}_{\hat{\eta}}(2,2,\Z{}) \times \mathrm{O}_{\hat{\eta}}(2,2,\Z{}) \times \mathrm{O}_{\hat{\eta}}(2,2,\Z{}) ~\subset~ \mathrm{O}_{\hat{\eta}}(6,6,\Z{})\;,
\end{equation}
see eq.~\eqref{eq:FactorizedOrbifoldTwist}. Then, eq.~\eqref{eq:CPForTheta} is solved by
\begin{equation}\label{eq:CPin2D}
\hat\Theta_\subind{K_i} ~\stackrel{\CP}{\longrightarrow}~ \left(\hat{\CP}_\subind{K_i}\right)^{-1}\hat\Theta_\subind{K_i}\,\left(\hat{\CP}_\subind{K_i}\right) ~\stackrel{!}{=}~ \left(\hat\Theta_\subind{K_i}\right)^{-1}\;,
\end{equation}
where
\begin{equation}
\label{eq:6DCP}
\hat{\CP} ~:=~ \hat{\CP}_\subind{K_1} \oplus \hat{\CP}_\subind{K_2} \oplus \hat{\CP}_\subind{K_3} \quad\mathrm{and}\quad \hat{\CP}_\subind{K_i} ~\in~ \mathrm{O}_{\hat{\eta}}(2,2,\Z{})\;.
\end{equation}
This means that for each complex plane $i\in\{1,2,3\}$ (or each $\mathbbm T^2/\Z{K_i}$ orbifold 
sector) on which a Narain twist $\hat\Theta_\subind{K_i}$ of order $K_i\in\{2,3,4,6\}$ acts, we 
look for a corresponding $\CP$-like transformation that we call $\hat{\CP}_\subind{K_i}$. In 
addition to eq.~\eqref{eq:CPin2D}, a two-dimensional $\CP$-like transformation has to act on the 
K\"ahler modulus $T_i$ of the $\mathbbm T^2/\Z{K_i}$ orbifold sector as
\begin{equation}
T_i ~\stackrel{\CP}{\longrightarrow}~ -\bar{T}_i\;,
\end{equation}
and in addition, if $K_i=2$, on the complex structure modulus as $U_i \rightarrow -\bar{U}_i$, see 
ref.~\cite{Dent:2001cc,Baur:2019kwi,Novichkov:2019sqv}. The results are given in table~\ref{tab:CP}. 
Hence, the six-dimensional $\CP$-like transformation given by eq.~\eqref{eq:6DCP} acts 
simultaneously on all six dimensions of the compactified space, see also 
ref.~\cite{Strominger:1985it}.

At first sight, there seems to be a special case if $K_i=2$, i.e.\ if the six-dimensional orbifold 
geometry contains a $\mathbbm T^2/\Z{2}$ orbifold sector. Let us discuss an example, where the 
first complex plane contains a $\mathbbm T^2/\Z{2}$ orbifold sector ($K_1=2$), i.e.
\begin{equation}\label{eq:FactorizedTwistK1=2}
\hat\Theta ~:=~ \hat\Theta_\subind{2} \oplus \hat\Theta_\subind{K_2} \oplus \hat\Theta_\subind{K_3}\;,
\end{equation}
and $K_i \neq 2$ for $i\in\{2,3\}$. Then, we can define two independent transformations 
\begin{equation}
\hat\CP ~:=~ \hat\CP_\subind{2} \oplus \hat\CP_\subind{K_2} \oplus \hat\CP_\subind{K_3}\quad\mathrm{and}\quad \hat\CP' ~:=~ \hat\CP_\subind{2} \oplus \Id_4 \oplus \Id_4\;,
\end{equation}
with $\hat\CP_\subind{2}=\hat\Sigma_*$ given in table~\ref{tab:CP}. Both transformations, 
$\hat\CP$ and $\hat\CP'$, satisfy eq.~\eqref{eq:CPForTheta} since 
$(\hat\Theta_\subind{2})^{-1}=\hat\Theta_\subind{2}$ for $\hat\Theta_\subind{2} := -\Id_4$. 
However, only $\hat\CP$ is a consistent transformation but not $\hat\CP'$: $\hat\CP'$ acts 
nontrivially on the K\"ahler modulus $T_1$ and the complex structure modulus $U_1$ associated with 
the first complex plane of the six-dimensional orbifold, but leaves the moduli associated with the 
four orthogonal compact dimensions invariant. Nonetheless, the superpotential $\mathcal{W}$ has 
to be a holomorphic function of all fields (matter fields {\it and} moduli) and after 
$\CP$-conjugation it needs to become anti-holomorphic. Therefore, assuming that there are some 
superpotential couplings that depend on moduli associated with all six compact dimensions, $\CP$ has 
to act nontrivially in all three complex planes of the six-dimensional orbifold simultaneously. In 
this case, $\CP$ is broken spontaneously if any modulus departs from one of its $\CP$-conserving 
points in moduli space.

\begin{table}[t!]
\center
\begin{tabular}{|c||ll|c|c|}
\hline
order             & \multicolumn{2}{c|}{\multirow{2}{*}{orbifold twist $\hat\Theta_\subind{K_i}\in\mathrm{O}_{\hat{\eta}}(2,2,\Z{})$}} & \multirow{2}{*}{$\hat{\CP}_\subind{K_i}\in\mathrm{O}_{\hat{\eta}}(2,2,\Z{})$} & action of $\hat{\CP}_\subind{K_i}$\\
$K_i$             &  & & & on moduli\\
\hline
\hline
$2$ & $\hat\Theta_\subind{2}$ & $\!\!\!\!\!\!=\left(\begin{array}{cccc}-1&0&0&0\\0&-1&0&0\\0&0&-1&0\\0&0&0&-1\end{array}\right)$\!\!& 
$\hat{\Sigma}_*=\left(\begin{array}{cccc}-1&0&0&0\\0&1&0&0\\0&0&-1&0\\0&0&0&1\end{array}\right)$ & $\begin{array}{c}T_i \rightarrow -\bar{T}_i\\U_i \rightarrow -\bar{U}_i\end{array}$\\
\hline
$3$ & $\hat\Theta_\subind{3}$ & $\!\!\!\!\!\!=\left(\begin{array}{cccc}0&-1&0&0\\1&-1&0&0\\0&0&-1&-1\\0&0&1&0\end{array}\right)$&
$\hat{K}_*=\left(\begin{array}{cccc}1&0&0&0\\1&-1&0&0\\0&0&1&1\\0&0&0&-1\end{array}\right)$ & $\begin{array}{c}T_i \rightarrow -\bar{T}_i\\\langle U_i\rangle = \omega \mathrm{\ invariant}\end{array}$\\
\hline
$4$ & $\hat\Theta_\subind{4}$ & $\!\!\!\!\!\!=\left(\begin{array}{cccc}0&-1&0&0\\1&0&0&0\\0&0&0&-1\\0&0&1&0\end{array}\right)$&
$\hat{\Sigma}_*=\left(\begin{array}{cccc}-1&0&0&0\\0&1&0&0\\0&0&-1&0\\0&0&0&1\end{array}\right)$ & $\begin{array}{c}T_i \rightarrow -\bar{T}_i\\\langle U_i\rangle = \I \mathrm{\ invariant}\end{array}$\\
\hline
$6$ & $\hat\Theta_\subind{6}$ & $\!\!\!\!\!\!=\left(\begin{array}{cccc}1&-1&0&0\\1&0&0&0\\0&0&0&-1\\0&0&1&1\end{array}\right)$&
$\hat{K}_*=\left(\begin{array}{cccc}1&0&0&0\\1&-1&0&0\\0&0&1&1\\0&0&0&-1\end{array}\right)$ & $\begin{array}{c}T_i \rightarrow -\bar{T}_i\\\langle U_i\rangle = \omega \mathrm{\ invariant}\end{array}$\\
\hline
\end{tabular}
\caption{$\CP$-like transformations in the Narain formulation of two-dimensional 
$\mathbbm T^2/\Z{K_i}$ orbifold sectors for $K_i\in\{2,3,4,6\}$.}
\label{tab:CP}
\end{table}

\subsection{Summary}

Many properties of factorized six-dimensional orbifolds can be understood in terms of three 
two-dimensional $\mathbbm T^2/\Z{K_i}$ orbifold sectors, $i\in\{1,2,3\}$, as we show in 
section~\ref{sec:FactorizedOrbifold}. Nevertheless, these six-dimensional orbifolds display a 
richer structure compared to two-dimensional ones. In particular, we observe two important new 
features:
\begin{itemize}
\item[i)] Inner automorphisms of a Narain space group of a two-dimensional orbifold sector can 
become {\it outer} automorphisms of a Narain space group of a suitable six-dimensional orbifold. 
This applies especially to two-dimensional sublattice rotations of six-dimensional orbifolds, which 
are given by (unbroken) $\SL{2,\Z{}}_{U_i}$ modular transformations. Interestingly, this uncovers 
an unexpected relation, eq.~\eqref{eq:RchargesVsModularWeights}, between the $R$-charge $R_i$ of a 
matter field associated with a sublattice rotation in the $i$-th two-torus and its 
$\SL{2,\Z{}}_{U_i}$ modular weight.
\item[ii)] \CP-like transformations in six-dimensional factorized orbifolds must act simultaneously 
in all two-dimensional orbifold sectors, in order to not spoil the holomorphicity of the 
superpotential. This is based on the assumption that at least one superpotential coupling depends 
on the moduli from all three two-dimensional orbifold sectors. We list in table~\ref{tab:CP} the 
\CP-like transformations for all $\mathbbm T^2/\Z{K_i}$ orbifold sectors ($K_i\in\{2,3,4,6\}$).
\end{itemize}

\section[R-symmetries and CP in a T2/Z3 orbifold sector]{\boldmath $R$-symmetries and $\CP$ in a $\mathbbm T^2/\Z{3}$ orbifold sector\unboldmath}
\label{sec:RandCPinZ3}

The four-dimensional effective field theory (obtained from a heterotic compactification on an 
orbifold geometry that contains a $\mathbbm T^2/\Z{3}$ orbifold sector without discrete Wilson 
lines) is equipped with various types of symmetries:
\begin{itemize}
\item a $\Delta(54)$ traditional flavor symmetry, 
\item an $\SL{2,\Z{}}_T$ modular symmetry that acts as a $T'$ finite modular group on 
matter fields and couplings,
\item a $\Z{9}^R$ discrete $R$-symmetry,
\item a $\CP$-like transformation,
\end{itemize}
see also refs.~\cite{Kobayashi:2006wq,Beye:2014nxa,Lauer:1989ax,Lerche:1989cs,Lauer:1990tm,Kobayashi:2004ya,Bizet:2013gf,Nilles:2013lda,Bizet:2013wha,Nilles:2017heg}. 
These discrete symmetries combine to the eclectic flavor symmetry $\Omega(2)\cong[1944, 3448]$. 
Furthermore, as we have seen in the previous sections, all of them have a common origin from string theory: they can 
be defined as outer automorphisms of the Narain space group~\cite{Baur:2019kwi,Baur:2019iai,Nilles:2020tdp}. 
This novel approach has revealed several interconnections between discrete symmetries of the 
various types, with the ultimate aim to connect bottom-up flavor model building with a consistent 
top-down approach from strings.

In this section, we first recall the main results from the analysis of the effective field theory 
obtained from a heterotic string compactification on a $\mathbbm T^6/\Z{K}$ orbifold geometry that 
contains a $\mathbbm T^2/\Z{3}$ orbifold sector, for example in the first complex plane. The moduli 
in this plane are denoted for simplicity by $T$ and $U$ (without subindex). In addition, we focus 
on three copies (labeled by an index $i\in\{1,2,3\}$) of twisted matter fields $(X_i,Y_i,Z_i)$ 
without string oscillator excitations. These matter fields are localized in the extra dimensions at 
the three fixed points of the $\mathbbm T^2/\Z{3}$ orbifold sector. We call them collectively 
$\Phi_{\nicefrac{-2}{3}}^i=(X_i,Y_i,Z_i)^\mathrm{T}$ for $i\in\{1,2,3\}$, where 
$n_T=\nicefrac{-2}{3}$ is the modular weight with respect to $\SL{2,\Z{}}_T$. In this section, we 
will obtain new results concerning the origin and nature of discrete $R$-symmetries and $\CP$, 
related to the modular symmetries of the theory.

\subsection[Modular symmetry in a T2/Z3 orbifold sector]{\boldmath Modular symmetry in a $\mathbbm T^2/\Z{3}$ orbifold sector\unboldmath}

One defines a so-called finite modular group $\Gamma'_N$ using the 
relations~\eqref{eq:DefiningRelationsOfSL2Z} combined with the additional constraint
\begin{equation}\label{eq:DefiningRelationsOfGammaNPrime}
\mathrm{T}^N ~=~ \Id\;, 
\end{equation}
which renders the group finite. For the $\mathbbm T^2/\Z{3}$ orbifold sector, we are interested in 
the case $N=3$ that corresponds to the finite modular group
\begin{equation}
T' ~\cong~ \Gamma_3' ~\cong~ \mathrm{SL}(2,3)\;.
\end{equation}
Then, a modular form of $T'$ with modular weight $n_Y \in \mathbbm{N}$ is defined as a (vector of) 
holomorphic function(s) of the K\"ahler modulus $T$ that transforms under general modular 
transformations $\gamma_T\in\SL{2,\Z{}}_T$ as
\begin{equation}\label{eq:YModularTrafo}
\hat Y^{(n_Y)}_{\rep{s_Y}}(T) ~\stackrel{\gamma_T}{\longrightarrow}~ \hat Y^{(n_Y)}_{\rep{s_Y}}\!\left(\tfrac{a_T\,T+b_T}{c_T\,T+d_T}\right) ~=~ (c_T\,T+d_T)^{n_Y} \rho_{\rep{s_Y}}(\gamma_T) \,\hat Y^{(n_Y)}_{\rep{s_Y}}(T)\,.
\end{equation}
Here, $\rep{s_Y}$ denotes an $s$-dimensional representation of $T'$ and $(c_T\,T+d_T)^{n_Y}$ is the 
automorphy factor of $\gamma_T$ with modular weight $n_Y$.

It turns out that there exists only one (two-component) modular form of weight $n_Y=1$
\begin{equation}\label{eq:YTWeight1}
\hat Y^{(1)}_{\rep{2''}}(T) ~:=~ \begin{pmatrix}\hat{Y}_1(T)\\\hat{Y}_2(T)\end{pmatrix} ~:=~ \begin{pmatrix}-3\sqrt{2}\frac{\eta^3(3\,T)}{\eta(T)}\\ 3\frac{\eta^3(3\,T)}{\eta(T)} + \frac{\eta^3(T/3)}{\eta(T)}\end{pmatrix}\;.
\end{equation}
Using eq.~\eqref{eq:YModularTrafo} it can be checked explicitly that this modular form transforms 
in a two-dimensional (unitary) representation $\rep{2''}$ of $T'$, where
\begin{equation}
\label{eq:SL2ZTOn2}
\rho_{\rep{2''}}(\hat K_\mathrm{S}) ~:=~ -\frac{\I}{\sqrt{3}}\begin{pmatrix}1 & \sqrt{2} \\ \sqrt{2} & -1\end{pmatrix}\quad\mathrm{and}\quad 
\rho_{\rep{2''}}(\hat K_\mathrm{T}) ~:=~ \begin{pmatrix}\omega & 0\\ 0 & 1\end{pmatrix}\;.
\end{equation}
Note that we use the $T'$ conventions from ref.~\cite{Ishimori:2010au} (with $p=\I$).

In addition, twisted matter fields $\Phi_{\nicefrac{-2}{3}}^i=(X_i,Y_i,Z_i)^\mathrm{T}$ carry a 
modular weight $n_T = \nicefrac{-2}{3}$ and build a representation $\rep{s}=\rep{2'}\oplus\rep{1}$ of $T'$ 
for $i\in\{1,2,3\}$, as shown in ref.~\cite{Baur:2019iai}, i.e.
\begin{equation}
\label{eq:ModularTrafosOfPhi_n}
\Phi_{\nicefrac{-2}{3}}^i ~\stackrel{\gamma_T}{\longrightarrow}~ {{\Phi^i}'}_{\nicefrac{-2}{3}} ~:=~ (c_T\,T+d_T)^{\nicefrac{-2}{3}}\, \rho_{\rep{s}}(\gamma_T)\,\Phi_{\nicefrac{-2}{3}}^i\;,\\
\end{equation}
for $\gamma_T\in\SL{2,\Z{}}_T$, where the representation matrices $\rho_{\rep{s}}(\gamma_T)$ of 
$\SL{2,\Z{}}_T$ elements $\gamma_T$ are generated by~\cite{Baur:2019iai,Nilles:2020kgo}
\begin{equation}
\label{eq:ModularKTrafoOfTwistedStringsRep}
\rho(\hat K_\mathrm{S}) ~:=~ \frac{\I}{\sqrt{3}}\begin{pmatrix} 1 & 1 & 1 \\ 1 & \omega^2 &\omega \\ 1 & \omega & \omega^2 \end{pmatrix} \qquad\mathrm{and}\qquad
\rho(\hat K_\mathrm{T}) ~:=~ \begin{pmatrix} \omega^2 & 0 & 0 \\ 0 & 1 &0 \\ 0 & 0 & 1 \end{pmatrix}\,.
\end{equation}

\begin{figure*}[t]
\begin{center}
\subfloat[]{\label{fig:Z3OrbifoldA}
\includegraphics[width=0.4\linewidth]{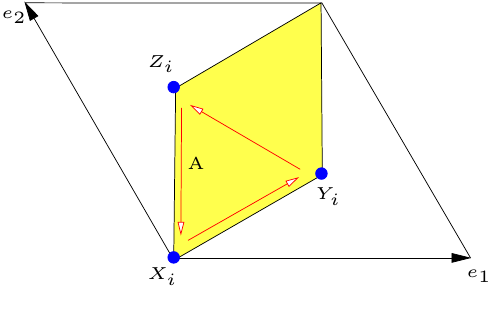}}
\subfloat[]{\label{fig:Z3OrbifoldC}
\includegraphics[width=0.4\linewidth]{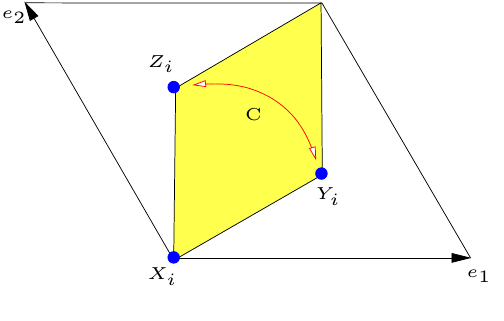}}
\end{center}
\vspace{-0.6cm}
\caption{\label{fig:Z3Orbifold}The $\mathbbm{T}^2/\Z{3}$ orbifold sector and the geometrical 
actions of the Narain outer automorphisms: (a) the $\Z{3}$ translation $\mathrm{A}$ rotates the 
three twisted states $(X_i,Y_i,Z_i)$ and (b) the 180$^\circ$ rotation $\mathrm{C}$ maps $X_i$ to 
itself and interchanges $Y_i$ and $Z_i$.}
\end{figure*}

\subsection[Traditional flavor symmetry in a T2/Z3 orbifold sector]{\boldmath Traditional flavor symmetry in a $\mathbbm T^2/\Z{3}$ orbifold sector\unboldmath}
\label{sec:TFSZ3orbifold}

It is well-known that a $\mathbbm T^2/\Z{3}$ orbifold sector yields a $\Delta(54)$ traditional 
flavor symmetry, where twisted matter fields build triplet-representations~\cite{Kobayashi:2006wq,Carballo-Perez:2016ooy}. 
New insight was gained using the Narain space group and its outer automorphisms~\cite{Baur:2019kwi,Baur:2019iai}: 
In the absence of discrete Wilson lines, there exist two translational outer automorphisms, 
$\mathrm{A}$ and $\mathrm{B}$, of the $\mathbbm T^2/\Z{3}$ Narain space group,
\begin{equation}
\mathrm{A} ~:=~ (\Id, \hat{T}_1)\;,\;  \mathrm{B} ~:=~ (\Id, \hat{T}_2)\;,  \quad\mathrm{where}\quad \hat{T}_1 ~:=~ \begin{pmatrix}\frac{1}{3}\\\frac{2}{3}\\0\\0\end{pmatrix} \;,\; \hat{T}_2 ~:=~ \begin{pmatrix}0\\0\\\frac{1}{3}\\\frac{1}{3}\\\end{pmatrix}\;,
\end{equation}
see the discussion around eq.~\eqref{eq:conditionsOnOuter} in section~\ref{sec:ModularForOrbifolds}. 
Similar to eq.~\eqref{eq:NarainLatticeForYandtildeY}, one can see that $\mathrm{A}$ and $\mathrm{B}$ 
induce the following geometrical translations
\begin{equation}
\label{eq:GeomAandB}
y ~\stackrel{\mathrm{A}}{\longrightarrow}~ y + \frac{1}{3}\left(e_1 + 2e_2\right) \qquad\mathrm{and}\qquad y ~\stackrel{\mathrm{B}}{\longrightarrow}~ y\;,
\end{equation}
as illustrated in figure~\ref{fig:Z3OrbifoldA}. Note that, in addition to eq.~\eqref{eq:GeomAandB}, 
there is a nontrivial action on $\tilde{y}$. These Narain-translations leave the K\"ahler modulus 
inert. Hence, they belong to the traditional flavor symmetry. Acting on twisted matter fields 
$\Phi_{\nicefrac{-2}{3}}^i$, they are represented as 
\begin{equation}
\Phi_{\nicefrac{-2}{3}}^i ~\stackrel{\mathrm{A}}{\longrightarrow}~ \rho(\mathrm{A})\, \Phi_{\nicefrac{-2}{3}}^i \qquad\mathrm{and}\qquad 
\Phi_{\nicefrac{-2}{3}}^i ~\stackrel{\mathrm{B}}{\longrightarrow}~ \rho(\mathrm{B})\, \Phi_{\nicefrac{-2}{3}}^i\;,
\end{equation}
where 
\begin{equation}\label{eq:Delta54rhoAandB}
\rho(\mathrm{A}) ~=~ \begin{pmatrix}0&1&0\\0&0&1\\1&0&0\end{pmatrix}\qquad\mathrm{and}\qquad \rho(\mathrm{B}) ~=~ \begin{pmatrix}1&0&0\\0&\omega&0\\0&0&\omega^2\end{pmatrix}\;,
\end{equation}
and $\omega:=\exp\left(\nicefrac{2\pi\I}{3}\right)$. These matrices $\rho(\mathrm{A})$ and 
$\rho(\mathrm{B})$ generate a $\Delta(27)$ traditional flavor symmetry, where the twisted matter 
fields $\Phi_{\nicefrac{-2}{3}}^i$ form a triplet. This non-Abelian flavor symmetry includes the 
well-known $\Z{3}^{(\mathrm{PG})}\times\Z{3}$ point and space group selection rules of strings that
split and join while propagating on the surface of an 
orbifold~\cite{Hamidi:1986vh,Dixon:1986qv,Ramos-Sanchez:2018edc}: In our case of a 
$\mathbbm T^2/\Z{3}$ orbifold sector, it turns out that the symmetry transformations
\begin{equation}
\mathrm{A}^2 \mathrm{B}^2 \mathrm{A} \mathrm{B} \qquad\mathrm{and}\qquad \mathrm{B}
\end{equation}
correspond to the point and space group selection rules, respectively. In detail, for twisted 
matter fields $\Phi_{\nicefrac{-2}{3}}^i=(X_i,Y_i,Z_i)^\mathrm{T}$ (from the first twisted sector), 
we obtain the transformations
\begin{subequations}
\begin{eqnarray}
\Z{3}^{(\mathrm{PG})}           & : & \rho(\mathrm{A})^2\rho(\mathrm{B})^2\rho(\mathrm{A})\rho(\mathrm{B}) ~=~ \begin{pmatrix}\omega&0&0\\0&\omega&0\\0&0&\omega\end{pmatrix}\;,\label{eq:pointgroupselectionrule}\\
\Z{3}^{\phantom{(\mathrm{PG})}} & : & \rho(\mathrm{B}) ~=~ \begin{pmatrix}1&0&0\\0&\omega&0\\0&0&\omega^2\end{pmatrix}\;.
\end{eqnarray}
\end{subequations}
In addition, there is a rotational outer automorphism $\mathrm{C}:=(-\Id_4,0)$ of the 
$\mathbbm T^2/\Z{3}$ Narain space group, which corresponds to a geometrical rotation by $180^\circ$ 
in the $\mathbbm T^2/\Z{3}$ orbifold sector. Hence, $\mathrm{C}$ generates an $R$-symmetry. As we 
have seen in eq.~\eqref{eq:StabilizersOfSL2ZU}, $\mathrm{C}$ corresponds to a modular 
transformation $\mathrm{S}^2\in \SL{2,\Z{}}_U$ from the stabilizer subgroup $H_{\vev U=\omega}$.
For the twisted matter fields, $\mathrm{C}$ is represented as
\begin{equation}\label{eq:Delta54rhoC}
\rho(\mathrm{C}) ~=~ \begin{pmatrix}-1&0&0\\0&0&-1\\0&-1&0\end{pmatrix}\;.
\end{equation}
By comparing with eq.~\eqref{eq:ModularKTrafoOfTwistedStringsRep}, we note that 
$\rho(\mathrm{C})=\rho(\hat K_\mathrm{S})^2$, which is indeed a consequence of the more fundamental 
identity $\mathrm{C}=\hat K_\mathrm{S}^2$, see section 2.2. of ref.~\cite{Nilles:2020kgo} for more 
details. This additional element enhances $\Delta(27)$ to a non-Abelian 
$R$-symmetry~\cite{Chen:2013dpa} $\Delta(54)$, where the twisted matter fields 
$\Phi_{\nicefrac{-2}{3}}^i$ transform in the representation $\rep{3}_2$. Note that the non-diagonal 
structure of $\rho(\mathrm{C})$ is an example of a non-diagonal representation matrix 
$\rho_{\rep{s}}(\gamma_\subind{K_i})$ in eq.~\eqref{eq:ModularTransformationOfPhiUnderRotation}. 
Furthermore, one can understand geometrically the fact that $Y_i$ and $Z_i$ get interchanged under 
the action of $\mathrm{C}$, while $X_i$ is mapped to itself, as illustrated in 
figure~\ref{fig:Z3OrbifoldC}.

\subsection[Superpotential in a T2/Z3 orbifold sector]{\boldmath Superpotential in a $\mathbbm T^2/\Z{3}$ orbifold sector\unboldmath}
\label{sec:WZ3orbifold}

Taking only twisted matter fields from one twisted sector into account, the modular group 
$\SL{2,\Z{}}_T$, its associated finite modular group $T'$ and the traditional flavor group 
$\Delta(54)$ constrain the tri-linear superpotential 
$\mathcal W \supset \Phi_{\nicefrac{-2}{3}}^1 \Phi_{\nicefrac{-2}{3}}^2 \Phi_{\nicefrac{-2}{3}}^3$
of twisted matter fields $\Phi_{\nicefrac{-2}{3}}^i$ for $i\in\{1,2,3\}$, such that there is only 
one free parameter, denoted by $c^{(1)}$. The precise result reads~\cite{Nilles:2020kgo}
\begin{subequations}
\label{eq:Z3superpotential}
\begin{eqnarray}
\mathcal W(T,X_i,Y_i,Z_i) & \supset & c^{(1)}\, \Big[\hat{Y}_2(T) \big( X_1\,X_2\,X_3 + Y_1\,Y_2\,Y_3 + Z_1\,Z_2\,Z_3\big)  \label{eq:unsuppressedtwisted}\\
                          &         & \hspace{8mm} - \frac{\hat{Y}_1(T)}{\sqrt{2}} \big( X_1\,Y_2\,Z_3 + X_1\,Y_3\,Z_2 + X_2\,Y_1\,Z_3 \label{eq:suppressedtwisted}\\
                          &         & \hspace{22mm} +\, X_3\,Y_1\,Z_2 + X_2\,Y_3\,Z_1 + X_3\,Y_2\,Z_1\big)\Big]\;, \nonumber
\end{eqnarray}
\end{subequations}
As a remark, assuming that $\Phi_{\nicefrac{-2}{3}}^1 = (X_1,Y_1,Z_1)^\mathrm{T}$ develops a 
non-trivial vev, this superpotential results in a (symmetric) mass matrix
\begin{equation}
\mathcal W(T,X_i,Y_i,Z_i) \supset \begin{pmatrix}X_2&Y_2&Z_2\end{pmatrix}\begin{pmatrix}
  \hat{Y}_2(T) \langle X_1\rangle                 &- \frac{\hat{Y}_1(T)}{\sqrt{2}} \langle Z_1\rangle&- \frac{\hat{Y}_1(T)}{\sqrt{2}} \langle Y_1\rangle\\
- \frac{\hat{Y}_1(T)}{\sqrt{2}} \langle Z_1\rangle&\hat{Y}_2(T) \langle Y_1\rangle                   &- \frac{\hat{Y}_1(T)}{\sqrt{2}} \langle X_1\rangle\\
- \frac{\hat{Y}_1(T)}{\sqrt{2}} \langle Y_1\rangle&- \frac{\hat{Y}_1(T)}{\sqrt{2}} \langle X_1\rangle&\hat{Y}_2(T) \langle Z_1\rangle
\end{pmatrix}\begin{pmatrix}X_3\\Y_3\\Z_3\end{pmatrix}\;.
\end{equation}
This observation can be useful in semi-realistic models e.g.\ to determine the mass textures of 
quarks and leptons, considering that $\Phi_{\nicefrac{-2}{3}}^1$ corresponds to three Higgs fields 
coupled to the three generations of quark and lepton superfields.

\subsection[Z9R R-symmetry in a T2/Z3 orbifold sector]{\boldmath $\Z{9}^R$ $R$-symmetry in a $\mathbbm T^2/\Z3$ orbifold sector\unboldmath}
\label{sec:Z3SublatticeRotation}

We consider a Narain space group $S_{\mathrm{Narain}}$ of a $\mathbbm T^6/\Z{K}$ orbifold, which 
contains a $\mathbbm T^2/\Z{3}$ orbifold sector (for example in the first complex plane). Then, 
there exists an outer automorphism 
\begin{equation}\label{eq:Z3NarainSublatticeRotation}
\hat{\mathrm{R}}_1 ~:=~ \hat\Theta_\subind{3} \oplus \Id_4 \oplus \Id_4 ~\in~ \mathrm{O}_{\hat{\eta}}(2,2,\Z{}) \times \mathrm{O}_{\hat{\eta}}(2,2,\Z{}) \times \mathrm{O}_{\hat{\eta}}(2,2,\Z{}) ~\subset~ \mathrm{O}_{\hat{\eta}}(6,6,\Z{})
\end{equation}
of the Narain space group $S_{\mathrm{Narain}}$, where
\begin{equation}
\hat\Theta_\subind{3} ~:=~ \hat{C}_\mathrm{S}\,\hat{C}_\mathrm{T} ~=~ \left(\begin{array}{cc}\hat\sigma_\subind{3}&0\\0&\hat\sigma_\subind{3}^{-\mathrm{T}}\end{array}\right) ~=~ \left(\begin{array}{cccc}0&-1&0&0\\1&-1&0&0\\0&0&-1&-1\\0&0&1&0\end{array}\right) ~\in~ \mathrm{O}_{\hat{\eta}}(2,2,\Z{})\;,
\end{equation}
acts on the $D=2$ extra dimensions of the $\mathbbm T^2/\Z{3}$ orbifold sector. The columns of the 
upper $2\times 2$ block $\hat\sigma_\subind{3}=e^{-1}\theta_\subind{3}e$ of $\hat\Theta_\subind{3}$ 
can be interpreted as 
\begin{equation}
e_1 ~\stackrel{\theta_\subind{3}}{\longrightarrow}~ \theta_\subind{3}\, e_1 ~=~ e_2 \qquad\mathrm{and}\qquad e_2 ~\stackrel{\theta_\subind{3}}{\longrightarrow}~ \theta_\subind{3}\, e_2 ~=~ -e_1-e_2\;,
\end{equation}
see eqs.~\eqref{eq:SublatticeRotationGeom} with $y=e_i$ and~\eqref{eq:sigma}. In other words, the 
two-dimensional sublattice spanned by $e_1$ and $e_2$, illustrated for example in 
figure~\ref{fig:Z3Orbifold}, is left invariant by a $120^\circ$ rotation.

As a remark, let us assume that the order three discrete Wilson line in the $\mathbbm T^2/\Z{3}$ 
orbifold sector is chosen to be non-zero. Then, this $120^\circ$ rotational symmetry remains 
unbroken. In contrast, the $\Z{2}$ outer automorphism $\mathrm{C}:=(-\Id_4,0)$, discussed in 
section~\ref{sec:TFSZ3orbifold}, is broken in this case.

As we have seen in section~\ref{sec:GeometricalRotationsFromSL2ZU}, the outer automorphism 
eq.~\eqref{eq:Z3NarainSublatticeRotation} translates to a modular transformation
\begin{equation}
\gamma_\subind{3} ~:=~ \mathrm{S}\,\mathrm{T} ~=~ \begin{pmatrix}0&1\\-1&-1\end{pmatrix}  ~\in~ H_{\vev{U}=\omega} ~\subset~ \SL{2,\Z{}}_{U}\;,
\end{equation}
which leaves (by definition of the stabilizer subgroup $H_{\vev{U}=\omega}$) the complex 
structure modulus $U$ of the $\mathbbm T^2/\Z{3}$ orbifold sector invariant,
\begin{equation}
U ~\stackrel{\gamma_\subind{3}}{\longrightarrow}~ \frac{1}{-U-1} ~\stackrel{!}{=}~ U \quad\mathrm{at}\quad \langle U \rangle ~=~ \omega\;.
\end{equation}
Finally, the superpotential transforms under the modular transformation 
$\gamma_\subind{3}\in\SL{2,\Z{}}_{U}$ as
\begin{equation}\label{eq:WwithZ3R}
\mathcal{W} ~\stackrel{\gamma_\subind{3}}{\longrightarrow}~ (-\langle U \rangle-1)^{-1} \mathcal{W} ~=~ \omega\,\mathcal{W} ~=~ \exp\left(2\pi\I\,v_\subind{3}\right)\,\mathcal{W}\;, \qquad\mathrm{where}\qquad v_\subind{3} ~:=~ \nicefrac{1}{3}\;,
\end{equation}
using that the complex structure modulus of the $\mathbbm T^2/\Z{3}$ orbifold sector is frozen at 
$\langle U \rangle = \omega$.

\begin{table}[t!]
\center
\begin{tabular}{|c||c|c|c|c|c|c||c|c|c|}
\hline
        & $\Phi_0$ & $\Phi_{-1}$ & $\Phi_{\nicefrac{-2}{3}}$ & $\Phi_{\nicefrac{-5}{3}}$ & $\Phi_{\nicefrac{-1}{3}}$ & $\Phi_{\nicefrac{+2}{3}}$ & $\vartheta$ & $\mathcal W$ & $K$\\
\hline
 $R$-charges $R_1$ & $0$ & $3$ & $1$ & $-2$ & $2$ &  $5$ & $\nicefrac{3}{2}$ & $3$ & $0$\\ 
\hline
\end{tabular}
\caption{Discrete $\Z{9}^R$ $R$-charges $R_1$ for a $\mathbbm T^2/\Z3$ orbifold sector with trivial 
Wilson line of: matter fields $\Phi_{n_T}$, the Grassmann number $\vartheta$, the superpotential 
$\mathcal W$, and the K\"ahler potential $K$ in conventions, similar to those of 
ref.~\cite{Nilles:2013lda}. The $\Z{9}^R$ $R$-symmetry originates from a \Z3 sublattice rotation of 
the $\mathbbm T^2/\Z3$ orbifold sector in the normalization where all $R$-charges of bosons (but 
not of fermions) are integer. Note that the $\Z{3}^{\mathrm{(PG)}}$ point group symmetry 
eq.~\eqref{eq:pointgroupselectionrule} is a subgroup of $\Z{9}^R$ (using that 
$\vartheta \to -\vartheta$ and $\mathrm{fermions}\to -\mathrm{fermions}$ follows from 
four-dimensional Lorentz invariance).}
\label{tab:Z18Rcharges}
\end{table}

This sublattice rotation $\hat{\mathrm{R}}_1$ acts on matter fields as
\begin{equation}
\label{eq:R1sublatticeRotOnPhi}
\Phi_{n_T} ~\stackrel{\hat{\mathrm{R}}_1}{\longrightarrow}~ \exp(\nicefrac{2\pi\I\,R_1}{9})\,\Phi_{n_T}\;, 
\end{equation}
where the $R$-charges $R_1$ are normalized to be integer for superfields, see 
table~\ref{tab:Z18Rcharges}. Hence, the \Z3 sublattice rotation $\hat{\mathrm{R}}_1$ leads to a 
$\Z{9}^R$ $R$-symmetry of the theory, where the superpotential eq.~\eqref{eq:Z3superpotential} 
transforms according to eq.~\eqref{eq:WwithZ3R} with a $\Z{9}^R$ $R$-charge $3$. Note that fermions 
have half-integer $R$-charges. Hence, this symmetry acts as a $\Z{18}^R$ symmetry of the fermionic 
sector~\cite{Nilles:2013lda}. In more detail, one can show that the Grassmann number $\vartheta$ of 
$\mathcal{N}=1$ superspace has a $\Z{9}^R$ $R$-charge $R(\vartheta)=\nicefrac{3}{2}$. Furthermore, 
using $\int\!\mathrm{d}\vartheta\,\vartheta=1$ we know that the $R$-charge of 
$\int\!\mathrm{d}\vartheta$ is $-\nicefrac{3}{2}$, such that the Lagrangian 
$\mathcal{L}\supset\int\!\mathrm{d}^2\vartheta\, \mathcal{W}$ is invariant.

\begin{table}[b!]
\center
\begin{tabular}{|c|c||c|c|c|c|c|c|}
\hline
\multicolumn{2}{|c||}{nature}        & outer automorphism       & \multicolumn{5}{c|}{\multirow{2}{*}{flavor groups}} \\
\multicolumn{2}{|c||}{of symmetry}   & of Narain space group    & \multicolumn{5}{c|}{}\\
\hline
\hline
\parbox[t]{3mm}{\multirow{6}{*}{\rotatebox[origin=c]{90}{eclectic}}} &\multirow{2}{*}{modular}            & rotation $\mathrm{S}~\in~\SL{2,\Z{}}_T$ & $\Z{4}$      & \multicolumn{3}{c|}{\multirow{2}{*}{$T'$}} &\multirow{6}{*}{$\Omega(2)$}\\
&                                    & rotation $\mathrm{T}~\in~\SL{2,\Z{}}_T$ & $\Z{3}$      & \multicolumn{3}{c|}{}                      & \\
\cline{2-7}
&                                    & translation $\mathrm{A}$                & $\Z{3}$      & \multirow{2}{*}{$\Delta(27)$} & \multirow{3}{*}{$\Delta(54)$} & \multirow{4}{*}{$\Delta'(54,2,1)$} & \\
& traditional                        & translation $\mathrm{B}$                & $\Z{3}$      &                               & & & \\
\cline{3-5}
& flavor                             & rotation $\mathrm{C}=\mathrm{S}^2\in\SL{2,\Z{}}_T$      & \multicolumn{2}{c|}{$\Z{2}^R$} & & & \\
\cline{3-6}
&                                    & rotation $\hat{\mathrm{R}}_1\Leftrightarrow\gamma_\subind{3}\in\SL{2,\Z{}}_U$ & \multicolumn{3}{c|}{$\Z{9}^R$}   & & \\
\hline
\end{tabular}
\caption{Table from ref.~\cite{Nilles:2020tdp}: eclectic flavor group $\Omega(2)$ for orbifolds $\mathbbm T^6/P$ that contain a 
$\mathbbm T^2/\Z{3}$ orbifold sector. In this case, $\SL{2,\Z{}}_U$ of the stabilized complex 
structure modulus $U=\exp\left(\nicefrac{2\pi\I}{3}\right)$ is broken, resulting in a remnant 
$\Z{9}^R$ $R$-symmetry. Including $\Z{9}^R$ enhances the traditional flavor group $\Delta(54)$ to 
$\Delta'(54,2,1)\cong [162,44]$ and, thereby, the eclectic group to 
$\Omega(2) \cong [1944, 3448]$. Note that $\Omega(1)\subset \Omega(2)$.}
\label{tab:Z3FlavorGroupsExtended}
\end{table}

\subsection[The full eclectic flavor group without CP in a T2/Z3 orbifold sector]{\boldmath The full eclectic flavor group without \CP in a $\mathbbm T^2/\Z3$ orbifold sector\unboldmath}

Combining the discrete $\Z{9}^R$ $R$-symmetry with the traditional flavor symmetry $\Delta(54)$ 
yields the {\it full} traditional flavor symmetry
\begin{equation}
\Delta'(54,2,1) ~\cong~ \Delta(54) ~\cup~ \Z{9}^R ~\cong~ [162,44]\;.
\end{equation}
Hence, the full eclectic flavor group of the $\mathbbm T^2/\Z3$ orbifold sector without \CP is 
given by
\begin{equation}
\Omega(2) ~\cong~ \Omega(1) ~\cup~ \Z{9}^R ~\cong~ [1944, 3448]\;,
\end{equation}
where $\Omega(1) \cong \Delta(54) ~\cup~ T'$, see ref.~\cite{Jurciukonis:2017mjp} for the 
nomenclature. The various groups and their origins in string theory are summarized in 
table~\ref{tab:Z3FlavorGroupsExtended}. Note that $\Omega(2)$ contains the $\Delta(54)$ traditional 
flavor group and all of its enhancements (without \CP) to the various unified flavor groups at 
special points in $T$ moduli space, as we shall discuss later in 
section~\ref{sec:LocalFlavorUnification}.

\subsection[CP in a T2/Z3 orbifold sector]{\boldmath \CP in a $\mathbbm{T}^2/\Z{3}$ orbifold sector\unboldmath}
\label{sec:CPforZ3}

\begin{figure*}[t]
\centering{\includegraphics[width=0.4\linewidth]{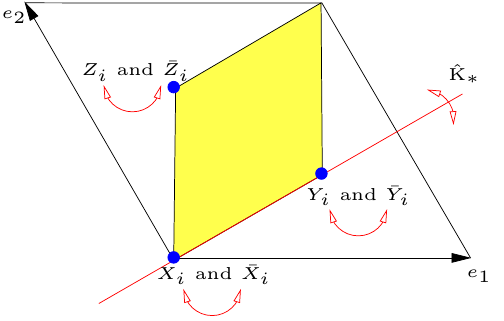}}
\caption{The fundamental domain of the $\mathbbm{T}^2/\Z{3}$ orbifold is depicted in yellow and the 
three inequivalent fixed-points are blue. $(X,Y,Z)$ denote three (left-chiral) twisted strings from the 
first twisted sector $\hat\Theta$, while $(\bar{X},\bar{Y},\bar{Z})$ are three (right-chiral) 
twisted strings from the second twisted sector $\hat\Theta^2$. The \CP-like transformation 
$\hat{K}_*$ acts as $(\hat{K}_*)^{-1} \hat\Theta\, \hat{K}_*=\hat{\Theta}^2$ 
in agreement with eq.~\eqref{eq:CPForTheta}.}
\label{fig:Z3OrbifoldCP}
\end{figure*}

For heterotic orbifold compactifications, the modular group $\SL{2,\Z{}}_T$ of the K\"ahler modulus 
$T$ originates from outer automorphisms of the Narain space group, as we reviewed in 
section~\ref{sec:ModularSymmetries}. For a $\mathbbm T^2/\Z{3}$ orbifold sector, it can be enhanced 
naturally by an additional outer automorphism
\begin{equation}
\hat{K}_* ~:=~ \hat{C}_\mathrm{S}\,\hat{C}_\mathrm{T}\,\hat{C}_\mathrm{S}\,\hat\Sigma_* ~=~ 
\begin{pmatrix}
 1& 0&  0& 0\\
 1&-1&  0& 0\\
 0& 0&  1& 1\\
 0& 0&  0&-1 
\end{pmatrix}
\end{equation}
that acts $\CP$-like~\cite{Baur:2019kwi,Baur:2019iai}, i.e.
\begin{equation}
T ~\stackrel{\hat{K}_*}{\longrightarrow}~ -\bar{T} \qquad\mathrm{and}\qquad U ~\stackrel{\hat{K}_*}{\longrightarrow}~ -\frac{\bar{U}}{1+\bar{U}} ~=~ U \qquad\mathrm{at}\qquad U ~=~ \langle U\rangle ~=~ \omega\;,
\end{equation}
see also refs.~\cite{Strominger:1985it,Dent:2001cc,Kobayashi:2020uaj}. Following the discussion in 
section~\ref{sec:GeomInterpretationSL2ZU} we see that $\hat{K}_*$ acts geometrically as 
\begin{equation}
y ~\stackrel{\hat{K}_*}{\longrightarrow}~ \sigma\,y \quad\mathrm{using}\quad
\sigma ~=~ e\,\hat\sigma\,e^{-1} \quad\mathrm{and}\quad 
\hat\sigma ~=~ 
\begin{pmatrix}
 1& 0\\
 1&-1\\
\end{pmatrix}\;.
\end{equation}
This gives rise to a reflection at the line perpendicular to $e_2$, as illustrated in 
figure~\ref{fig:Z3OrbifoldCP}. Finally, note that \CP has to act in all extra dimensions 
simultaneously, following our discussion in section~\ref{sec:CPin6D}.

\subsection[CP as a modular symmetry]{\boldmath \CP as a modular symmetry \unboldmath}
\label{sec:CP}

On the level of $2\times 2$ matrices, the $\CP$-like transformation with $\hat{K}_*$ can be 
represented by
\begin{equation}\label{eq:Kstar}
\mathrm{K}_* ~:=~ \left(\begin{array}{cc} -1 & 0 \\ 0 & 1 \end{array}\right)\;.
\end{equation}
Since $\mathrm{det}(\mathrm{K}_*)=-1$, it enhances the modular group $\SL{2,\Z{}}_T$ to 
$\mathrm{GL}(2,\Z{})_T$. Then, the modular transformations in eq.~\eqref{eq:TrafoOfTModulus} of the 
K\"ahler modulus $T$ are complemented by
\begin{equation}\label{eq:TrafoWithGamma2}
T ~\stackrel{\gamma}{\longrightarrow}~ \frac{a\,\bar{T}+b}{c\,\bar{T}+d} \quad\mathrm{if}\quad 
\gamma ~=~ \left(\begin{array}{cc} a & b 
                                \\ c & d 
                 \end{array}\right) ~\in~ \mathrm{GL}(2,\Z{})_T 
\quad\mathrm{with}\quad \mathrm{det}(\gamma)=-1\;,
\end{equation}
see also ref.~\cite{Novichkov:2019sqv} for a bottom-up derivation. 

Moreover, the transformation of matter fields eq.~\eqref{eq:ModularTransformationOfPhi} can be 
generalized to \CP-like transformations (with $\mathrm{det}(\gamma)=-1$) as
\begin{equation}\label{eq:CPLikeTrafoOfMatterField}
\Phi_{n_T} ~\stackrel{\gamma}{\longrightarrow}~ {\bar{\Phi}_{n_T}}\hspace{0.5pt}' ~:=~ (c\,\bar{T}+d)^{n_T}\, \rho_{\rep{s}}(\gamma)\,\bar{\Phi}_{n_T}\;,
\end{equation}
where the field $\bar{\Phi}_{n_T}$ on the right-hand side is evaluated at the parity-transformed 
point in spacetime. In addition, as derived in ref.~\cite{Baur:2019iai} for our concrete case of a 
$\mathbbm T^2/\Z3$ orbifold sector, twisted matter fields transform under the \CP-like generator 
$\hat{K}_*$ as
\begin{equation}\label{eq:CPTrafoOnTwistedStringsRep}
\begin{pmatrix} \Phi_{n_T} \\ \bar{\Phi}_{n_T}\end{pmatrix}
  ~\stackrel{\hat{K}_*}{\longrightarrow}~
  \rho_{\rep{s}}(\hat{K}_*) \begin{pmatrix} \Phi_{n_T} \\ \bar{\Phi}_{n_T}\end{pmatrix} 
            ~:=~ \begin{pmatrix} 0 & \Id_3 \\ \Id_3 & 0 \end{pmatrix} 
                 \begin{pmatrix} \Phi_{n_T} \\ \bar{\Phi}_{n_T}\end{pmatrix}\,,
\end{equation}
where we used that the automorphy factor $(c\,\bar{T}+d)^{n_T}$ in 
eq.~\eqref{eq:CPLikeTrafoOfMatterField} equals $1$ for $\mathrm{K}_*$ defined in 
eq.~\eqref{eq:Kstar}.

Consequently, one can show that the finite modular group $T'\cong\mathrm{SL}(2,3)$ gets enhanced by 
\CP to $\mathrm{GL}(2,3)$. Simultaneously, the eclectic flavor group $\Omega(2)$ is enhanced by \CP 
to a group of order 3,888 (while the $\Omega(1)$ subgroup of $\Omega(2)$ is enhanced to 
$[1296,2891]$ by \CP).

Modular forms are also affected by the inclusion of \CP-like modular transformations. Under a 
$\CP$-like transformation $T \rightarrow -\bar{T}$, we find for the Dedekind $\eta$-function 
$\eta(r\,T)\rightarrow \eta(-r\,\bar{T}) = (\eta(r\,T))^*$ if $r\in\mathbbm{R}$. Hence, the weight 
$1$ modular forms of $T'$ defined in eq.~\eqref{eq:YTWeight1} transform as
\begin{subequations}\label{eq:CPTrafoOfYukawas}
\begin{eqnarray}
\hat{Y}_1(T) ~\stackrel{\hat{K}_*}{\longrightarrow}~ \hat{Y}_1(-\bar{T}) & = & \left(\hat{Y}_1(T)\right)^*\;, \\
\hat{Y}_2(T) ~\stackrel{\hat{K}_*}{\longrightarrow}~ \hat{Y}_2(-\bar{T}) & = & \left(\hat{Y}_2(T)\right)^*\;,
\end{eqnarray}
\end{subequations}
cf.\ ref.~\cite{Novichkov:2018ovf}. We have chosen here a basis of $T'$ where all Clebsch--Gordan 
coefficients in $T'$ tensor products are real (see refs.~\cite{Ishimori:2010au,Chen:2014tpa} with 
$p=\I$, $p_1=1$, and $p_2=-1$). Therefore, all irreducible $T'$ modular forms of modular weight 
$n_Y\geq1$ are complex conjugated by $\hat{K}_*$, analogously to eqs.~\eqref{eq:CPTrafoOfYukawas}, 
i.e.
\begin{equation}
\hat Y^{(n_Y)}_{\rep{s}}(T) ~\stackrel{\hat{K}_*}{\longrightarrow}~ \hat Y^{(n_Y)}_{\rep{s}}(-\bar{T}) ~=~ \left(\hat Y^{(n_Y)}_{\rep{s}}(T)\right)^*\;.
\end{equation}
This can be used to generalize eq.~\eqref{eq:YModularTrafo} to all elements 
$\gamma\in\mathrm{GL}(2,\Z{})_T$ with $\mathrm{det}(\gamma)=-1$ as
\begin{equation}\label{eq:CPTrafoOfModularForms}
\hat Y^{(n_Y)}_{\rep{s}}(T) ~\stackrel{\gamma}{\longrightarrow}~ \hat Y^{(n_Y)}_{\rep{s}}\!\left(\frac{a\,\bar{T}+b}{c\,\bar{T}+d}\right) ~=~ (c\,\bar{T}+d)^{n_Y} \rho_{\rep{s}}(\gamma) \,\left(\hat Y^{(n_Y)}_{\rep{s}}(T)\right)^*\,,
\end{equation}
where we have $\rho_{\rep{s}}(\hat{K}_*)=\Id$ in our basis of $\mathrm{GL}(2,3)$ modular forms. 
Consequently, $T'\cong\mathrm{SL}(2,3)$ modular forms naturally become modular forms of the 
$\CP$-extended finite modular group $\mathrm{GL}(2,3)$.

Finally, we discuss the transformation of the superpotential $\mathcal{W}$ and the K\"ahler 
potential $K$ under general \CP-like transformations for a \CP invariant theory. Again, we take 
$\gamma\in\mathrm{GL}(2,\Z{})_T$ with $\mathrm{det}(\gamma)=-1$. Then, if the superpotential and 
the K\"ahler potential transform as
\begin{subequations}\label{eq:CPTrafoOfWAndK}
\begin{eqnarray}
\mathcal{W}\left(T,\Phi_{n_T}\right) & \stackrel{\gamma}{\longrightarrow} & \mathcal{W}\left(\frac{a\,\bar{T}+b}{c\,\bar{T}+d},{\bar{\Phi}_{n_T}}\hspace{0.5pt}'\right) ~=~ (c\,\bar{T}+d)^{-1} \left(\mathcal{W}(T,\Phi_{n_T})\right)^*\;,\label{eq:CPTrafoOfW}\\
K\left(T,\bar{T},\Phi_{n_T},\bar{\Phi}_{n_T}\right) & \stackrel{\gamma}{\longrightarrow} & K\left(\frac{a\,\bar{T}+b}{c\,\bar{T}+d},\frac{a\,T+b}{c\,T+d},{\bar{\Phi}_{n_T}}\hspace{0.5pt}',{\Phi_{n_T}}'\right)\\ 
 & = & K\left(T,\bar{T},\Phi_{n_T},\bar{\Phi}_{n_T}\right) + f(T) + f(\bar{T})\;,
\end{eqnarray}
\end{subequations}
the theory is \CP invariant, using the transformed matter fields ${\Phi_{n_T}}'$ given in 
eq.~\eqref{eq:CPLikeTrafoOfMatterField} and a K\"ahler transformation to remove the holomorphic 
function $f(T)=\mathrm{ln}(c\,T+d)$, see section~\ref{sec:modularInvariantTheory}. Let us remark 
that our convention for $\mathrm{K}_*$ given in eq.~\eqref{eq:Kstar} is chosen such that the 
automorphy factors in eqs.~\eqref{eq:CPLikeTrafoOfMatterField},~\eqref{eq:CPTrafoOfModularForms} 
and~\eqref{eq:CPTrafoOfWAndK} have a standard form.

Applied to our explicit trilinear superpotential $\mathcal{W}$ eq.~\eqref{eq:Z3superpotential}, 
we note that $\mathcal{W}$ satisfies the transformation property of a $\CP$ invariant theory. For 
example, under $\hat{K}_*$
\begin{equation}
\mathcal{W}(T,X_i,Y_i,Z_i) ~\stackrel{\hat{K}_*}{\longrightarrow}~ \mathcal{W}(-\bar{T},\bar{X}_i,\bar{Y}_i,\bar{Z}_i) ~=~ \big(\mathcal{W}(T,X_i,Y_i,Z_i)\big)^*\;,
\end{equation}
provided that the coefficient $c^{(1)}$ is real as can be chosen in our superpotential by 
appropriate field redefinitions. However, at a generic point in moduli space, the \CP-like 
transformation $\hat{K}_*$ does not leave the vacuum $\vev T$ invariant, i.e.
$\vev T \stackrel{\hat{K}_*}{\longrightarrow} -\langle \bar{T} \rangle ~\neq~ \vev T$,
which can indicate spontaneous \CP breaking. Moreover, even though the Clebsch--Gordan coefficients 
in $T'$ tensor products are real, this is not true for $\Delta(54)$ tensor products: tensor 
products of $\Delta(54)$ triplets can also yield $\Delta(54)$ doublets with complex Clebsch--Gordan 
coefficients~\cite{Ishimori:2010au}, indicating that \CP can be broken~\cite{Chen:2014tpa}. 
However, in order to see this effect in our superpotential one has to consider higher order 
couplings that, from a stringy point of view, originate from the exchange of heavy $\Delta(54)$ 
doublets, cf.\ ref.~\cite{Nilles:2018wex}.

\subsection{Summary}

In this section, we illustrate our previous findings on $R$-symmetries and \CP in the 
$\mathbbm T^2/\Z3$ orbifold sector. It is known that this case is endowed with a $T'$ finite 
modular symmetry and a $\Delta(54)$ traditional flavor symmetry. However, when embedded into a 
six-dimensional orbifold, sublattice rotations (from $\SL{2,\Z{}}_U$) yield a $\Z9^R$ $R$-symmetry, 
which acts linearly on matter fields, enhancing the traditional flavor symmetry to 
$\Delta'(54,2,1)\cong[162,44]$. This leads to the largest eclectic symmetry without \CP, being 
$\Omega(2)\cong\Delta'(54,2,1)\cup T'\cong[1944,3448]$. On the other hand, the modular group in 
$\mathbbm{T}^2/\Z{3}$ orbifolds is naturally enhanced by a \CP-like generator $\mathrm K_*$ that 
promotes $\SL{2,\Z{}}_T$ to $\mathrm{GL}(2,\Z{})_T$. We confirm that a general modular invariant 
theory is also invariant under \CP-like transformations from $\mathrm{GL}(2,\Z{})_T$ (cf.\ 
eq.~\eqref{eq:CPTrafoOfWAndK}), provided that matter fields and modular forms transform according 
to eqs.~\eqref{eq:CPLikeTrafoOfMatterField} and~\eqref{eq:CPTrafoOfModularForms}, respectively. 
Note that these conditions are expected to be valid also in general bottom-up models involving 
modular symmetries with \CP. Including \CP, the full eclectic symmetry of a $\mathbbm{T}^2/\Z{3}$ 
orbifold sector embedded in a higher-dimensional orbifold is $\Omega(2)\cup\Z{2}^{\CP}$, which 
is a symmetry of order $3,888$.

\enlargethispage{0.4cm}

\section[Local flavor unification]{Local flavor unification}
\label{sec:LocalFlavorUnification}

In this section, we discuss first in general the mechanism of local flavor unification~\cite{Baur:2019iai}
for $\mathbbm T^2/\Z{K_i}$ orbifold sectors, taking into account the effect of their embedding
in six-dimensional orbifolds. We will then focus on the $\mathbbm T^2/\Z3$ orbifold sector as 
our working example, where we additionally study the implications of this scheme on the superpotential. 

Consider a $\mathbbm T^2/\Z{K_i}$ orbifold sector, whose complex structure and K\"ahler moduli 
shall be denoted here as $U$ and $T$ (instead of $U_i$ and $T_i$), in order to simplify the 
notation as in section~\ref{sec:RandCPinZ3}. Let us focus on the case $K_i\neq2$, for which the 
$U$ modulus is geometrically fixed at a value $\vev U$ according to table~\ref{tab:ZNRotations}. 
(Some of the differences appearing in the $\mathbbm T^2/\Z2$ orbifold sector, which has been 
studied in ref.~\cite{Baur:2020jwc}, shall be briefly mentioned below.) We focus first on modular 
transformations that are not \CP-like and assume that the corresponding K\"ahler modulus $T$ has 
been stabilized~\cite{Kobayashi:2019xvz,Kobayashi:2019uyt} at a self-dual point in moduli space. 
In other words, we suppose that $T$ is stabilized at a vacuum expectation value (vev) $\vev T$, 
at which there exists a nontrivial finite~\cite{diamond2006first} subgroup 
$H_{\vev T} \subset \SL{2,\Z{}}_T$ that leaves $\vev T$ invariant. Just as 
$H_{\vev U}\subset\SL{2,\Z{}}_U$ defined in eq.~\eqref{eq:SubgroupOfSL2ZU},
$H_{\vev T}$ is also a modular stabilizer subgroup, defined as
\begin{equation}\label{eq:SubgroupOfSL2ZT}
H_{\vev T} ~=~ \bigg\{ \begin{pmatrix} a_T & b_T \\ c_T & d_T \end{pmatrix}\in \SL{2,\Z{}}_T ~\bigg|~ 
                       \frac{a_T\,\vev T+b_T}{c_T\,\vev T+d_T} ~=~ \vev T\bigg\}\;,
\end{equation}
i.e.\ $\vev T$ is a fixed point of the stabilizer subgroup $H_{\vev T}$ of $\SL{2,\Z{}}_T$. Hence, 
the vev $\vev T$ of the stabilized K\"ahler modulus spontaneously breaks $\SL{2,\Z{}}_T$ to an 
unbroken finite subgroup $H_{\vev T}$. If this subgroup is nontrivial, there are three main 
consequences:
\begin{itemize}
\item[i)] On the one hand, the vev $\vev T$ of the K\"ahler modulus and, hence, all couplings 
$\hat Y^{(n_Y)}_{\rep{s}}(\vev T)$ are invariant under $\gamma_T \in H_{\vev T}$. Thus, for all 
$\gamma_T\in H_{\vev T}$ we have 
\begin{equation}\label{eq:HInvariantYukawas}
\hat Y^{(n_Y)}_{\rep{s}}(\vev T) ~\stackrel{\gamma_T}{\longrightarrow}~ \hat Y^{(n_Y)}_{\rep{s}}\!\left(\tfrac{a_T\vev T+b_T}{c_T\vev T+d_T}\right) 
        ~=~ (c_T\,\vev T+d_T)^{n_Y} \rho_{\rep{s}}(\gamma_T) \,\hat Y^{(n_Y)}_{\rep{s}}(\vev T) ~=~ \hat Y^{(n_Y)}_{\rep{s}}(\vev T)\,,
\end{equation}
using eqs.~\eqref{eq:YModularTrafo} and~\eqref{eq:SubgroupOfSL2ZT}. This implies that the couplings 
$\hat Y^{(n_Y)}_{\rep{s}}(\vev T)$ become eigenvectors of $\rho_{\rep{s}}(\gamma_T)$ with 
eigenvalues $(c_T\,\vev T+d_T)^{-n_Y}$. Since the representation matrices $\rho_{\rep{s}}(\gamma_T)$ 
have finite order, the eigenvalues of $\rho_{\rep{s}}(\gamma_T)$ are pure phases, especially
\begin{equation}\label{eq:AutomorphyFactorIsPhaseForH}
\big|(c_T\,\vev T + d_T)^{-n_Y}\big|^2 ~=~ 1 \qquad\Rightarrow\qquad|c_T\,\vev T + d_T|^2 ~=~ 1\;.
\end{equation}
Compared to traditional flavon models, eq.~\eqref{eq:HInvariantYukawas} corresponds to the 
mechanism of flavon vev alignment, see also refs.~\cite{Novichkov:2018ovf,Novichkov:2018yse,Gui-JunDing:2019wap}.
In this scenario, since the vev $\vev T$ of the K\"ahler modulus breaks some modular symmetry
but leaves the traditional flavor symmetry unbroken, an additional flavon will be needed 
for its breaking.

\item[ii)] On the other hand, matter fields $\Phi_\mathfrak{n}$  with $\SL{2,Z}_T$ modular weight 
$n_T\in\mathfrak{n}$ (where $\mathfrak{n}$ denotes the set of all modular weights of the matter 
field $\Phi_\mathfrak{n}$) transform in general nontrivially under $\gamma_T \in H_{\vev T}$, 
according to
\begin{equation}
\label{eq:UnbrokenModularTrafoOnPhi}
\Phi_\mathfrak{n} ~\stackrel{\gamma_T}{\longrightarrow}~ {\Phi_\mathfrak{n}}' ~=~ \rho_{\rep{t},\vev T}(\gamma_T)\,\Phi_\mathfrak{n}
       ~:=~ (c_T\,\vev T+d_T)^{n_T}\, \rho_{\rep{s}}(\gamma_T)\,\Phi_\mathfrak{n}\;,
\end{equation}
see eq.~\eqref{eq:ModularTransformationOfPhi}. Since the moduli are invariant under a modular 
transformation $\gamma_T\in H_{\vev T}$, we realize that $\gamma_T$ is linearly realized. Thus, the 
transformation matrix $\rho_{\rep{t},\vev T}(\gamma_T)$ enhances the traditional flavor symmetry 
(now, also containing the discrete $\Z{(K_i)^2}^R$ $R$-symmetry that originates from 
$\SL{2,\Z{}}_U$ as discussed in section~\ref{sec:SublatticeRotations}) to the so-called {\it 
unified flavor symmetry}~\cite{Baur:2019iai} at the point $\vev T$ in moduli space. Moreover, the 
transformation matrix $\rho_{\rep{t},\vev T}(\gamma_T)$ builds a $t$-dimensional representation 
$\rep{t}$ of the unified flavor group. It is important to stress the presence of the extra phase in 
$\rho_{\rep{t},\vev T}(\gamma_T)$ that originates from the automorphy factor 
$(c_T\vev T+d_T)^{n_T}$ in eq.~\eqref{eq:UnbrokenModularTrafoOnPhi}.

\item[iii)] In addition, the superpotential transforms under $\gamma_T\in H_{\vev T}$, according to 
eq.~\eqref{eq:WgammaTrafo}, as
\begin{equation}
  \mathcal{W}(\vev T,\vev U,\Phi_\mathfrak{n})  ~\stackrel{\gamma_T}{\longrightarrow}~  \mathcal{W}\left(\vev T,\vev U,\Phi_\mathfrak{n}'\right) 
                       ~=~ (c_T\,\vev T+d_T)^{-1}\, \mathcal{W}(\vev T,\vev U,\Phi_\mathfrak{n})\;,
\end{equation}
where $(c_T\,\vev T+d_T)^{-1}$ is a complex phase, see eq.~\eqref{eq:AutomorphyFactorIsPhaseForH},
and $\vev U$ is the fixed value of the complex structure $U$ of the $\mathbbm T^2/\Z{K_i}$ 
orbifold sector, see table~\ref{tab:ZNRotations}. By definition of $H_{\vev T}$, the vev 
$\vev T$ is invariant under a transformation $\gamma_T\in H_{\vev T}$. Thus, also the K\"ahler 
potential is invariant under $\gamma_T\in H_{\vev T}$. Since the Lagrangian 
$\mathcal{L}\supset\int\!\mathrm{d}^2\vartheta\, \mathcal{W}$ must be invariant under these 
transformations as well, the Grassmann number $\vartheta$ must transform with a phase
\begin{equation}
\label{eq:thetaRtrafo}
\vartheta  ~\stackrel{\gamma_T}{\longrightarrow}~  (c_T\,\vev T+d_T)^{\nicefrac{-1}{2}}\, \vartheta\;,
\end{equation}
cf.\ section~\ref{sec:Rsymmetries}. As a consequence, unbroken modular transformations 
$\gamma_T\in H_{\vev T}$ generate $R$-symmetries at self-dual points in moduli space if 
$c_T\,\vev T+d_T \neq 1$. These $R$-symmetries appear in addition to those arising from 
$\SL{2,\Z{}}_U$ discussed in section~\ref{sec:U-Rsymmetries}.
\end{itemize} 

Taking into account the discussion of section~\ref{sec:CP}, one can extend these observations to
further include \CP. In particular, it is possible to define $H_{\vev T}$ as the subgroup of 
$\mathrm{GL}(2,\mathbbm Z)_T$ (instead of $\SL{2,\Z{}}_T$) that leaves the stabilized 
K\"ahler modulus $\vev T$ invariant, see eq.~\eqref{eq:TrafoWithGamma2}. In this case, the unified 
flavor symmetry can include $\CP$-like transformations generated by the representations
of the generators shown in table~\ref{tab:CP} combined with some $\SL{2,\Z{}}_T$ elements,
endowing the theory with a \CP symmetry. Note that in these terms, one could envisage
a scenario in which \CP violation occurs as a transition from a point \vev{T} in moduli space
where $H_{\vev T}$ contains an unbroken \CP-like symmetry to another point where 
no \CP-like transformation stabilizes \vev{T}.

In summary, taking into account in the $\mathbbm T^2/\Z{K_i}$ orbifold sector the traditional 
flavor symmetry $G_\mathrm{fl}$, the discrete $\Z{(K_i)^2}^R$ $R$-symmetry from $\SL{2,\Z{}}_U$, 
and the stabilizer subgroup $H_{\vev{T}}$ which may also include \CP-like transformations, the 
unified flavor symmetry is given by the multiplicative closure of these groups. Thus, the unified 
flavor symmetry can be expressed as
\begin{equation}
\label{eq:unifiedFlavorSymmetry}
G_{\mathrm{uf},\vev{T}} ~\cong~ G_\mathrm{fl} ~\cup~ \Z{(K_i)^2}^R ~\cup~ H_{\vev{T}} ~\subset~ G_\mathrm{eclectic}\;,
\end{equation}
where also the transformations from $\Z{(K_i)^2}^R \cup H_{\vev{T}}$ are linearly realized 
in spite of their modular origin.

\begin{figure*}[t!]
\centering{\includegraphics[width=0.42\linewidth]{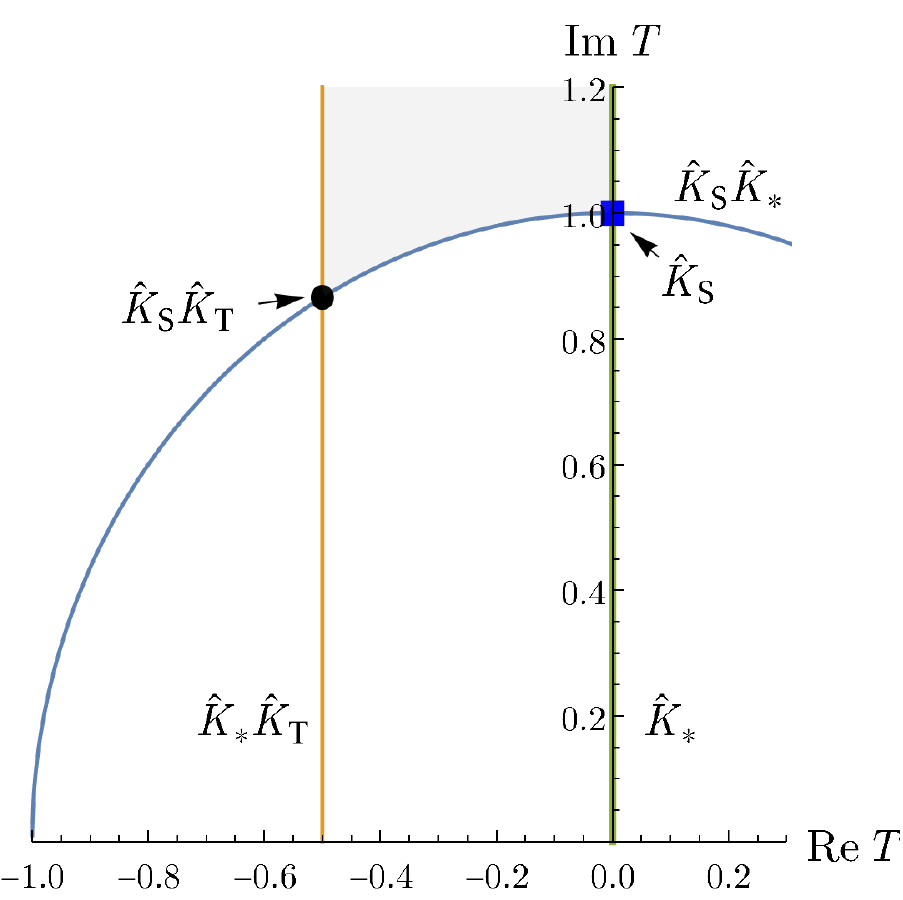}}
\caption{Fixed points and curves in the moduli space of the K\"ahler modulus 
$T=\mathrm{Re} T + \I\, \mathrm{Im} T$. The point $\vev T=\I$ (blue square) is left fixed 
by the modular $\hat K_\mathrm{S}$ transformation, whereas $\hat K_\mathrm{S}\hat K_\mathrm{T}$ leaves 
$\vev T=\omega$ (black bullet) invariant. These points lie at the intersections of the curves left 
invariant by the $\mathrm{GL}(2,\mathbbm Z)$ elements indicated next to them. These curves build the 
boundary of the fundamental domain (shaded region) of the moduli space in the $\mathbbm T^2/\Z3$ 
orbifold sector. We omit the $\hat K_{\mathrm S}^2$ modular symmetry, which acts trivially 
on the whole moduli space and is included in the traditional flavor group.}
\label{fig:Z3KaehlerModuliSpace}
\end{figure*}

Let us now spend a few words on the case of the $\mathbbm T^2/\Z2$ orbifold sector. As detailed in
ref.~\cite{Baur:2020jwc}, in this case both $T$ and $U$ moduli are not fixed. However, one can 
choose special vevs $(\vev T, \vev U)$ that are left invariant by a modular stabilizer subgroup 
$H_{(\vev T, \vev U)}$, which, in contrast to $H_{\vev{T}}$ in eq.~\eqref{eq:SubgroupOfSL2ZT}, 
includes elements not only from $\SL{2,\Z{}}_T$, but from the whole modular group of the orbifold. 
Hence, the unified flavor symmetry in this case can be expressed analogously to 
eq.~\eqref{eq:unifiedFlavorSymmetry}, replacing $H_{\vev T}$ by $H_{(\vev T, \vev U)}$.

In the following, we illustrate the scheme of local flavor unification by assuming that the K\"ahler modulus
is stabilized at various different points in moduli space of the $\mathbbm{T}^2/\Z{3}$ orbifold sector.
This moduli space is illustrated in figure~\ref{fig:Z3KaehlerModuliSpace}. We explore first the 
resulting unified flavor symmetries at the boundaries of the moduli space, depicted by the curves
in the figure, where we realize that a unique purely \CP enhancement of the flavor symmetry occurs. 
Then, we focus on two self-dual points $\vev T=\I$ and $\vev T=\omega$, displayed in
the figure~\ref{fig:Z3KaehlerModuliSpace} as a (blue) square and a (black) bullet, respectively. 
Since we suppose the $\mathbbm T^2/\Z3$ orbifold sector to be embedded in a six-dimensional orbifold, our 
starting point is the traditional flavor symmetry $\Delta'(54,2,1)\cong[162,44]\cong\Delta(54)\cup\Z9^R$, 
cf.\ table~\ref{tab:Z3FlavorGroupsExtended}. Depending on the value of $\vev T$ in moduli space, different
unbroken modular transformations contribute to enhance this traditional flavor group to various unified flavor 
symmetries contained in the full eclectic flavor group, $\Omega(2)\cup\Z2^{\CP}$, which is of order 3,888.
A brief summary of these findings is shown in table~\ref{tab:UnifiedFlavorSymmetries}.
Furthermore, we study the details of the twisted matter fields 
$\Phi_{\nicefrac{-2}{3}}^i = (X_i, Y_i, Z_i)^\mathrm{T}$ and their superpotential 
eq.~\eqref{eq:Z3superpotential} compatible with the unbroken symmetries at the different points 
in moduli space.

\subsection[Flavor enhancement at the boundary of the moduli space of T2/Z3]{\boldmath Flavor enhancement at the boundary of the moduli space of $\mathbbm T^2/\Z3$\unboldmath}
\label{sec:vevT@boundary}

\begin{table}[t!]
\center
\begin{tabular}{|c|ll|c|c|}
\hline
\multirow{2}{*}{$\vev T$} & \multicolumn{2}{c|}{stabilizer generators} & \multicolumn{2}{c|}{unified flavor symmetries}\\
                          & non \CP-like & \CP-like                    & without \CP & with \CP \\
\hline
\hline
$\I$                           & $\hat{K}_\mathrm{S}$                    & $\hat{K}_*$                    & $\Xi(2,2)\cong[324,111]$ & $[648,548]$ \\
\hline
$\omega$                       & $\hat{K}_\mathrm{S}\hat{K}_\mathrm{T}$  & $\hat{K}_* \hat{K}_\mathrm{T}$ & $H(3,2,1)\cong[486,125]$ & $[972,469]$ \\
\hline
$\re\vev T = 0$                &                                         & $\hat{K}_*$                    & $\Delta'(54,2,1)\cong[162,44]$ & $[324,125]$ \\
$\re\vev T = \nicefrac{-1}{2}$ &                                         & $\hat{K}_* \hat{K}_\mathrm{T}$ & $\Delta'(54,2,1)\cong[162,44]$ & $[324,125]$ \\
$|\vev T|=1$                   &                                         & $\hat{K}_\mathrm{S}\hat{K}_*$  & $\Delta'(54,2,1)\cong[162,44]$ & $[324,125]$ \\
\hline
\end{tabular}
\caption{Unified flavor symmetries at different K\"ahler modulus vevs \vev{T} of 
the $\mathbbm T^2/\Z3$ orbifold sector embedded in a six-dimensional orbifold. 
The $\mathrm{GL}(2,\Z{})_T$ modular transformations that leave \vev{T} invariant 
(generating the so-called stabilizer modular subgroup) enhance the traditional 
flavor symmetry $\Delta'(54,2,1)\cong\Delta(54)\cup\Z9^R$ of the orbifold sector 
to various large {\it unified flavor symmetries}, which generically include \CP-like transformations.
Note that the unified flavor symmetry is unique at all curves depicted in figure~\ref{fig:Z3KaehlerModuliSpace} 
that build the boundary of the fundamental domain of the moduli space of $T$. Furthermore, since 
$(\hat{K}_\mathrm{S})^2$ already belongs to the traditional flavor symmetry, we do not list it here.}
\label{tab:UnifiedFlavorSymmetries}
\end{table}

Let us begin the discussion of local flavor unification in the $\mathbbm T^2/\Z3$ orbifold sector
by studying the flavor symmetry enhancement at the various curves in figure~\ref{fig:Z3KaehlerModuliSpace},
which describe the boundary of the fundamental domain in moduli space for the K\"ahler modulus $T$.
The three curves are defined by the constraints $\re\vev{T}=0$, $\re\vev{T}=\nicefrac{-1}{2}$ and 
$|\vev{T}|=1$, respectively. We find that the stabilizer elements $H_{\vev T}\subset\mathrm{GL}(2,\Z{})_T$ 
compatible with the previous conditions are generated in each case by
\begin{subequations}
\label{eq:CPtrafosOnBoundary}
\begin{eqnarray}
  \text{at } \re\vev{T}=0 \hspace{5mm}: && \vev{T} ~\xrightarrow{\hat{K}_*} -\bar{\vev{T}} ~=~ \vev{T}\,, \label{eq:CPOnReT=0}\\
  \text{at } \re\vev T = \nicefrac{-1}{2}\,: && \vev{T} ~\xrightarrow{\hat{K}_*}~ -\bar{\vev{T}}
                                                        ~\xrightarrow{\hat{K}_\mathrm{T}}~ -\bar{\vev{T}}-1 ~=~ \vev{T}\,, \label{eq:CPOnReT=-1/2}\\
  \text{at } |\vev T|=1 \hspace{8mm}:&& \vev{T} ~\xrightarrow{\hat{K}_\mathrm{S}}~ -\frac{1}{\vev T}
                                                ~\xrightarrow{\hat{K}_*}~ \frac{1}{\bar{\vev{T}}} ~=~ \vev{T}\,.
\end{eqnarray}
\end{subequations}
Besides, we find the unbroken symmetry $\hat{K}_\mathrm{S}^2$ that leaves the whole moduli 
space invariant and satisfies $\hat{K}_\mathrm{S}^2=\mathrm C$, becoming a symmetry generator 
already included in the traditional flavor group. Therefore, we find that the stabilizer subgroup 
is $H_{\vev T}=\Z2\x\Z2^{\CP}$. Note that the stabilizer elements that can enhance the 
traditional flavor group include the \CP-like generator $\hat{K}_*$ in all cases.
Hence, there is no flavor enhancement without \CP-like transformations at a generic point on 
the boundary of the fundamental domain in $T$ moduli space. That is, ignoring the possibility 
of \CP, the traditional flavor symmetry $\Delta'(54,2,1)\cong\Delta(54)\cup\Z9^R$ is not enhanced 
at the boundaries of the fundamental domain.

In order to figure out the flavor enhancements that the \CP-like stabilizer elements in 
eq.~\eqref{eq:CPtrafosOnBoundary} induce, we must consider i) the matrix representations of these 
stabilizer elements when acting on the twisted matter fields $\Phi_{\nicefrac{-2}{3}}$ (and their 
\CP-conjugate $\bar\Phi_{\nicefrac{-2}{3}}$), and ii) the automorphy factors associated with these 
transformations. The representations are obtained as products of $\rho(\hat K_{\mathrm S})$ and 
$\rho(\hat K_{\mathrm T})$ given in eq.~\eqref{eq:ModularKTrafoOfTwistedStringsRep}, as well as 
$\rho(\hat{K}_*)$ defined in eq.~\eqref{eq:CPTrafoOnTwistedStringsRep}. Further, by using the 
$2\x2$ matrices that represent the $\GL{2,\Z{}}_T$ generators $\hat K_{\mathrm S}$, 
$\hat K_{\mathrm T}$ (see eq.~\eqref{eq:SL2ZGeneratorsSandT}) and $\hat{K}_*$ 
(see eq.~\eqref{eq:Kstar}), we find that the automorphy factors are trivial for 
$\re\vev{T}=0,\nicefrac{-1}{2}$, i.e.
\begin{subequations}
\begin{eqnarray}
  \label{eq:TrafoK*OnPhi}
  \Phi_{\nicefrac{-2}{3}} & \xrightarrow{\hat{K}_*}                   & \bar\Phi_{\nicefrac{-2}{3}}\;,\\
  \label{eq:TrafoK*KTOnPhi}
  \Phi_{\nicefrac{-2}{3}} & \xrightarrow{\hat{K}_*\hat{K}_\mathrm{T}} & \rho(\hat{K}_\mathrm{T})^*\,\bar\Phi_{\nicefrac{-2}{3}}\;.
\end{eqnarray}
\end{subequations}
At the points \vev{T} satisfying $|\vev T|=1$, we have 
\begin{equation}
  \Phi_{\nicefrac{-2}{3}} ~\xrightarrow{\hat{K}_\mathrm{S}\hat{K}_*}~ \bar\Phi'_{\nicefrac{-2}{3}}
                          ~=~ (\vev{\bar{T}})^{\nicefrac{-2}{3}} \rho(\hat{K}_\mathrm{S})\,\bar\Phi_{\nicefrac{-2}{3}}\,.
\end{equation}
In all three cases, one can show easily that $H_{\vev T}=\Z2^{\CP}$ acts on the matter fields as 
a \Z2 symmetry.\footnote{To verify this, one must use 
$\rho(\hat K_{\mathrm T})\rho(\hat K_{\mathrm T})^*=\rho(\hat K_{\mathrm S})\rho(\hat K_{\mathrm S})^*=\Id$.} 
Thus, the order of the traditional flavor symmetry is enhanced by a factor of two, resulting
in the unified flavor symmetry
\begin{equation}
G_{\mathrm{uf},\vev{T}\in\mathrm{boundary}} ~\cong~ \Delta'(54,2,1)\cup\Z2^{\CP} ~\cong~ [324,125]
\end{equation}
at a generic point on the boundary of the fundamental domain of the $T$ moduli space.

\subsection[Flavor symmetry at <T>=i in moduli space of T2/Z3]{\boldmath Flavor symmetry at $\vev T=\I$ in moduli space of $\mathbbm T^2/\Z3$\unboldmath}
\label{sec:vevT=i}

As one can infer from figure~\ref{fig:Z3KaehlerModuliSpace}, at $\vev T=\I$ the stabilizer
subgroup includes the $\SL{2,\Z{}}_T$ transformation $\hat{K}_\mathrm{S}$,
\begin{equation}
\vev T ~\stackrel{\hat K_\mathrm{S}}{\longrightarrow}~ -\frac{1}{\vev T} ~=~ \vev T \quad\mathrm{at}\quad \vev T ~=~ \I\;,
\end{equation}
and the \CP-like generator $\hat{K}_*$ because $\vev{T}=\I$ satisfies eq.~\eqref{eq:CPOnReT=0}.
Since $\hat K_\mathrm{S}^4=\Id$, the full stabilizer subgroup is $H_{\I} \cong \Z4\x\Z2^{\CP}$, 
where the $\Z2\subset\Z4$ subgroup associated with $\hat K_\mathrm{S}^2=\mathrm{C}$  (see eq.~\eqref{eq:Delta54rhoC})
already belongs to the traditional flavor symmetry $\Delta(54)\subset\Delta'(54,2,1)$. 

As the vev $\vev T=\I$ of the modulus is invariant under $H_{\I}$, the coupling strengths 
$\hat{Y}_1(\vev T)$ and $\hat{Y}_2(\vev T)$ are also invariant. Using the general modular 
transformations eq.~\eqref{eq:YModularTrafo}, this implies particularly that the 
$\hat K_\mathrm{S}$ transformation on the coupling strengths leads to
\begin{equation}\label{eq:YEigenvectorOfS}
\hat{Y}^{(1)}_{\rep2''}(\I) ~=~
\begin{pmatrix} \hat{Y}_1(\I) \\ \hat{Y}_2(\I) \end{pmatrix} ~\stackrel{\hat K_\mathrm{S}}{\longrightarrow}~ 
    \begin{pmatrix} \hat{Y}_1\left(-\frac{1}{\I}\right) \\ \hat{Y}_2\left(-\frac{1}{\I}\right) \end{pmatrix} 
    ~=~ -\frac{1}{\sqrt{3}}\, \begin{pmatrix} 1 & \sqrt{2} \\ \sqrt{2} & -1 \end{pmatrix} \begin{pmatrix} \hat{Y}_1(\I) \\ \hat{Y}_2(\I) \end{pmatrix} 
    ~=~ \begin{pmatrix} \hat{Y}_1(\I) \\ \hat{Y}_2(\I) \end{pmatrix}\;.
\end{equation}
Consequently, the doublet of couplings $\hat{Y}^{(1)}_{\rep2''}(\I)$ is an eigenvector of the 
matrix $\rho_{\rep2''}(\hat K_{\mathrm S})$ with eigenvalue $\I$. Thus, 
eq.~\eqref{eq:YEigenvectorOfS} amounts to the vev alignment
\begin{equation}
\label{eq:T=IvevAlignment}
\hat{Y}^{(1)}_{\rep2''}(\I) ~=~ \hat{Y}_2(\I) \left(\begin{array}{c} -\frac{\sqrt{2}}{1+\sqrt{3}} \\1 \end{array}\right)\;,
\end{equation}
which provides a nontrivial relation between $\hat{Y}_1(\I)$ and $\hat{Y}_2(\I)$.
As a side remark, because of the vev alignment eq.~\eqref{eq:T=IvevAlignment}, the $T'$ modular 
forms $\hat Y^{(4)}_{\rep1}(T)$, $\hat Y^{(4)}_{\rep1'}(T)$ and $\hat Y^{(4)}_{\rep3}(T)$ 
of weight $4$ given in ref.~\cite{Nilles:2020kgo} get aligned at $\vev T=\I$ as
\begin{equation}
\hat Y^{(4)}_{\rep1}(\I) ~=~ Y^{(4)}_{\rep1'}(\I) ~=~ 3(3-2\sqrt3)\,\hat{Y}_2(\I)^4 \quad\mathrm{and}\quad \hat Y^{(4)}_{\rep3}(\I) ~=~ 3(2-\sqrt3)\,\hat{Y}_2(\I)^4 \left(\begin{array}{c} 1 \\ 1\\-1 \end{array}\right)\;.
\end{equation}
Furthermore, both modular forms, $\hat{Y}_1(\I)$ and $\hat{Y}_2(\I)$, are real at 
$\vev T=\I$ (as one can also infer directly from eq.~\eqref{eq:CPTrafoOfYukawas}) and we 
find
\begin{equation}
\hat{Y}_1(\I) ~=~ -3\sqrt{2} \frac{\eta^3(3\,\I)}{\eta(\I)} ~\approx~ -0.5234 \quad\mathrm{and}\quad \hat{Y}_2(\I) ~\approx~ 1.0112\;.
\end{equation}

We can apply the vev alignment eq.~\eqref{eq:T=IvevAlignment} also to the trilinear superpotential 
eq.~\eqref{eq:Z3superpotential} of twisted matter fields $\Phi_{\nicefrac{-2}{3}}^i = (X_i, Y_i, Z_i)^\mathrm{T}$,
$i=1,2,3$. At $\vev T=\I$, it turns out that
\begin{subequations}\label{eq:Z3superpotentialAtI}
\begin{eqnarray}
\mathcal{W}(\I,X_i,Y_i,Z_i) & \supset & c^{(1)}\, \hat{Y}_2(\I) \Big[\big( X_1\,X_2\,X_3 + Y_1\,Y_2\,Y_3 + Z_1\,Z_2\,Z_3\big) \\
                            &         & + \frac{1}{1+\sqrt{3}} \big( X_1\,Y_2\,Z_3 + X_1\,Y_3\,Z_2 + X_2\,Y_1\,Z_3\\
                            &         & \hspace{1.55cm} + X_3\,Y_1\,Z_2 + X_2\,Y_3\,Z_1 + X_3\,Y_2\,Z_1\big)\Big]\;. \nonumber
\end{eqnarray}
\end{subequations}
Under a modular $\hat K_\mathrm{S}$ transformation with its representation given 
by eq.~\eqref{eq:ModularKTrafoOfTwistedStringsRep}, twisted matter fields transform, 
according to eq.~\eqref{eq:UnbrokenModularTrafoOnPhi}, as
\begin{equation}\label{eq:AtITrafoOfTwistedFields}
\left(\begin{array}{c} X_i \\ Y_i \\ Z_i \end{array}\right) ~\stackrel{\hat K_\mathrm{S}}{\longrightarrow}~ 
\left(\begin{array}{c} X_i' \\ Y_i' \\ Z_i' \end{array}\right) ~=~ \underbrace{\frac{\I(-\I)^{\nicefrac{-2}{3}}}{\sqrt{3}}  
\left(\begin{array}{ccc} 1 & 1      & 1 \\ 
                         1 & \omega^2 &\omega \\ 
                         1 & \omega & \omega^2 
\end{array}\right)}_{=:~\rho_{\rep{3}_2,\mathrm i}(\hat K_\mathrm{S})} \left(\begin{array}{c} X_i \\ Y_i \\ Z_i \end{array}\right)\;,
\end{equation}
where the multivalued factor $(-\I)^{\nicefrac{-2}{3}}$ can be fixed to 
$-\omega^2=\exp(\nicefrac{\pi\I}{3})$ by the \Z3 point group symmetry, and the representation
$\rep{t}=\rep3_2$ following the notation of eq.~\eqref{eq:UnbrokenModularTrafoOnPhi} corresponds 
to the triplet that the matter fields build in the resulting unified flavor symmetry at $\vev T=\I$. 
It then follows that $\hat K_\mathrm{S}$ acts on the superpotential evaluated at $\vev T=\I$ as 
an $R$-symmetry,
\begin{equation}
\mathcal{W}(\I,X_i,Y_i,Z_i) ~\stackrel{\hat K_\mathrm{S}}{\longrightarrow}~ \mathcal{W}(\I,X_i',Y_i',Z_i') ~=~ \I\, \mathcal{W}(\I,X_i,Y_i,Z_i)\;,
\end{equation}
as one can also see directly from eq.~\eqref{eq:WgammaTrafoT}. Moreover, due to $\rho_{\rep{3}_2,\mathrm i}(\hat K_\mathrm{S})$,
the modular $\hat K_\mathrm{S}$ transformation at $\vev T=\I$ generates a $\Z{12}\cong\Z{4}\times\Z{3}$ 
traditional flavor symmetry when acting on matter fields, cf.\ eq.~\eqref{eq:AtITrafoOfTwistedFields}, 
where $\Z{4}\times\Z{3}$ can be generated by
\begin{equation}
\label{eq:Z4xZ3generatorsT=i}
\left(\rho_{\rep{3}_2,\mathrm i}(\hat K_\mathrm{S})\right)^3 ~=~ \frac{\I}{\sqrt{3}} \left(\begin{array}{ccc} 1 & 1        & 1 \\ 
                                                                     1 & \omega & \omega^2 \\ 
                                                                     1 & \omega^2 & \omega 
                                                  \end{array}\right) 
\qquad\mathrm{and}\qquad \left(\rho_{\rep{3}_2,\mathrm i}(\hat K_\mathrm{S})\right)^4 ~=~ \omega^2 \Id\;,
\end{equation}
respectively. It is easy to see that the $\Z{4}$ factor generated by 
$(\rho_{\rep{3}_2,\mathrm i}(\hat K_\mathrm{S}))^3$ is an $R$-symmetry, while the $\Z{3}$ factor generated 
by $(\rho_{\rep{3}_2,\mathrm i}(\hat K_\mathrm{S}))^4$ is equivalent to the point group 
symmetry~\eqref{eq:pointgroupselectionrule} of the orbifold sector.\footnote{The transformation 
$\hat K_\mathrm{S}^3$ at $\vev T=\I$ acts as a \Z4 on bosons and on $\mathcal W$. However, due 
to eq.~\eqref{eq:thetaRtrafo}, it is actually an \Z8 $R$-symmetry when acting on fermions. Since 
discrete $R$-symmetries are relevant for phenomenology~\cite{Kappl:2008ie,Lee:2010gv}, it may be 
interesting to explore the consequences of this symmetry.} Using that 
$(\rho_{\rep{3}_2,\mathrm i}(\hat K_\mathrm{S}))^6 = \rho_{\rep{3}_2}\!(\mathrm{C})$, without
including \CP-like transformations the traditional flavor symmetry is enhanced from
\begin{equation}
\label{eq:nonCPenhancementAtT=i}
\Delta'(54,2,1) \qquad\text{to}\qquad \Xi(2,2) ~\cong~ [324,111]\qquad\text{at }\vev T=\I\;.
\end{equation}

This is the largest traditional flavor group without \CP that can be obtained in the 
$\mathbbm T^2/\Z3$ orbifold sector embedded in a higher-dimensional orbifold, and contains the 
group $\Sigma(36 \times 3) \cong [108,15]$ that results when the discrete $\Z9^R$ $R$-symmetry from 
$\SL{2,\Z{}}_U$ is not considered~\cite{Baur:2019kwi,Baur:2019iai,Nilles:2020kgo}. Similarly to 
that case, the $\Xi(2,2)$ group can be generated by the matrices $\rho(\mathrm{A})$, 
$\rho(\mathrm{B})$ given in eq.~\eqref{eq:Delta54rhoAandB}, the $\Z9^R$ generator 
$\rho(\hat{\mathrm R}_1)$ eq.~\eqref{eq:R1sublatticeRotOnPhi}, and 
$\rho_{\rep{3}_2,\mathrm i}(\hat K_\mathrm{S})$ in the faithful triplet representation $\rep3_2$ of 
$\Xi(2,2)$.

As a side remark, the relation 
$\rho_{\rep{3}_2,\mathrm i}(\hat K_\mathrm{S}^4)=(\rho_{\rep{3}_2,\mathrm i}(\hat K_\mathrm{S}))^4=\omega^2 \Id$ 
in eq.~\eqref{eq:Z4xZ3generatorsT=i} is a consistent representation of $\hat K_{\mathrm S}^4=\Id$ (see 
eq.~\eqref{eq:DefiningRelationsOfSL2Z}) because the latter indicates that $\hat K_{\mathrm S}^4$ must act trivially on the 
fields up to the action of the traditional flavor symmetry.

So far, we have not yet considered the $\Z2^{\CP}$ \CP-like transformation $\hat{K}_*$ at 
$\vev T=\I$. Using the \CP-like transformation~\eqref{eq:TrafoK*OnPhi} of matter fields, one can 
explicitly show that 
\begin{equation}
\mathcal{W}(\I,X_i,Y_i,Z_i) ~\stackrel{\hat{K}_*}{\longrightarrow}~ 
\mathcal{W}(\I,\bar{X}_i,\bar{Y}_i,\bar{Z}_i) ~=~ \big(\mathcal{W}(\I,X_i,Y_i,Z_i)\big)^*\;,
\end{equation}
using the superpotential eq.~\eqref{eq:Z3superpotentialAtI} with $c^{(1)}\, \hat{Y}_2(\I) \in\mathbbm{R}$ 
and the discussion on \CP from section~\ref{sec:CP}. Consequently, if we stabilize the K\"ahler 
modulus $T$ at $\vev T = \I$, the trilinear superpotential eq.~\eqref{eq:Z3superpotentialAtI} 
respects \CP. Including \CP, the unified flavor group at $\vev T=\I$ is given by 
\begin{equation}
G_{\mathrm{uf},\vev{T}=\I} ~\cong~ [648,548] ~\cong~ \Delta'(54,2,1)\cup\Z4\cup\Z2^{\CP}\;, 
\end{equation}
where $\Xi(2,2)\cong \Delta'(54,2,1)\cup\Z4$ is the maximal flavor subgroup of 
$G_{\mathrm{uf},\vev{T}=\I}$ which does not act \CP-like.

\subsection[Flavor symmetry at <T>=omega in moduli space of T2/Z3]{\boldmath Flavor symmetry at $\vev T=\omega$ in moduli space of $\mathbbm T^2/\Z3$\unboldmath}
\label{sec:vevT=omega}

Let us now assume that the K\"ahler modulus is stabilized at $\vev T=\omega$ in moduli space. 
There, as illustrated in figure~\ref{fig:Z3KaehlerModuliSpace}, the unbroken $\mathrm{GL}(2,\Z{})_T$
symmetries can be generated by the $\Z2^{\CP}$ \CP-like transformation $\hat{K}_*\hat K_\mathrm{T}$,
and the $\Z{3}$ modular transformation $\hat K_\mathrm{S}\hat K_\mathrm{T}$, 
\begin{equation}
\vev T ~\stackrel{\hat K_\mathrm{S}}{\longrightarrow}~ -\frac{1}{\vev T} 
                       ~\stackrel{\hat K_\mathrm{T}}{\longrightarrow}~ -\frac{1}{\vev T+1} ~=~ \vev T\;.
\end{equation}
In addition, we have the \Z2 transformation $\hat K_\mathrm{S}^2 = \mathrm{C}$ that is already 
included in the traditional flavor group. The resulting stabilizer subgroup reads 
$H_{\omega}\cong\Z3\x\Z2\x\Z2^{\CP}$.

At this point in moduli space, the coupling strengths governed by the modular forms 
$\hat{Y}_1(\vev T)$ and $\hat{Y}_2(\vev T)$ with modular weight $n_Y=1$ get aligned as
\begin{equation}\label{eq:YukawaAlignmentAtOmega}
\hat{Y}^{(1)}_{\rep2''}(\omega) ~=~ \hat{Y}_2(\omega) \left(\begin{array}{c} \frac{1}{\sqrt{2}}\omega \\1 \end{array}\right)\,, \quad\mathrm{where}\quad \hat{Y}_2(\omega) ~\approx~ 0.97399899 \;.
\end{equation}
Incidentally, the vev alignment given by eq.~\eqref{eq:YukawaAlignmentAtOmega} implies that the 
$T'$ modular forms $\hat Y^{(4)}_{\rep1}(T)$, $\hat Y^{(4)}_{\rep1'}(T)$ and 
$\hat Y^{(4)}_{\rep3}(T)$ of weight $4$ given in ref.~\cite{Nilles:2020kgo} are aligned at 
$\vev T=\omega$ according to  
\begin{equation}
\hat Y^{(4)}_{\rep1}(\omega) ~=~ 0\quad,\quad \hat Y^{(4)}_{\rep1'}(\omega) ~=~ \frac{9}{4} \omega \hat{Y}_2(\omega)^4 \quad\mathrm{and}\quad \hat Y^{(4)}_{\rep3}(\omega) ~=~ \frac{3}{2}\,\hat{Y}_2(\omega)^4 \left(\begin{array}{c} 1 \\ -\frac{1}{2}\omega\\-\omega^2 \end{array}\right)\;.
\end{equation}
On the other hand, in this case the trilinear superpotential eq.~\eqref{eq:Z3superpotential} 
simplifies to        
\begin{subequations}\label{eq:Z3superpotentialAtOmega}
\begin{eqnarray}
\mathcal{W}(\omega,X_i,Y_i,Z_i) & \supset & c^{(1)}\, \hat{Y}_2(\omega) \Big[X_1\,X_2\,X_3 + Y_1\,Y_2\,Y_3 + Z_1\,Z_2\,Z_3 \\
                                &         & \hspace{16mm} - \frac{\omega}{2} \big( X_1\,Y_2\,Z_3 + X_1\,Y_3\,Z_2 + X_2\,Y_1\,Z_3 \label{eq:Z3superpotentialAtOmegaLine2}\\
                                &         & \hspace{2cm} +\, X_3\,Y_1\,Z_2 + X_2\,Y_3\,Z_1 + X_3\,Y_2\,Z_1\big)\Big]\;.\nonumber
\end{eqnarray}
\end{subequations}
Under a modular $\hat K_\mathrm{S}\,\hat K_\mathrm{T}$ transformation at $\vev T=\omega$, the three 
triplets of twisted matter fields $\Phi_{\nicefrac{-2}{3}}^i$ transform as
\begin{equation}
\label{eq:STtrafoOnPhiT=omega}
\left(\begin{array}{c} X_i  \\ Y_i  \\ Z_i \end{array}\right) ~\xrightarrow{\hat K_\mathrm{S}\,\hat K_\mathrm{T}}~ 
\left(\begin{array}{c} X_i' \\ Y_i' \\ Z_i' \end{array}\right) =
\underbrace{\frac{\I\left(\omega^2\right)^{\nicefrac{-2}{3}}}{\sqrt{3}} 
\left(\begin{array}{ccc} \omega^2 & 1        & 1 \\ 
                         \omega^2 & \omega^2 & \omega \\ 
                         \omega^2 & \omega   & \omega^2
\end{array}\right)}_{=:~\rho_{\rep{3}_2,\omega}(\hat K_\mathrm{S}\,\hat K_\mathrm{T})} 
\left(\begin{array}{c} X_i  \\ Y_i  \\ Z_i \end{array}\right)\;,
\end{equation}    
where $\rep{t}=\rep3_2$ denotes the triplet representation of the twisted matter fields 
$\Phi_{\nicefrac{-2}{3}}^i$ under the unified flavor symmetry at $\vev T=\omega$. We used 
eq.~\eqref{eq:ModularKTrafoOfTwistedStringsRep} for $\rho(\hat K_\mathrm{S})$ and 
$\rho(\hat K_\mathrm{T})$, and the automorphy factor 
$(-\vev T-1)^{\nicefrac{-2}{3}}=(\omega^2)^{\nicefrac{-2}{3}}$ for $\gamma_T=\hat K_\mathrm{S}\,\hat K_\mathrm{T}$ 
evaluated at $\vev T=\omega$ in moduli space, see eq.~\eqref{eq:UnbrokenModularTrafoOnPhi}. Note 
that the multivalued factor $\left(\omega^2\right)^{\nicefrac{-2}{3}}$ can be fixed to 
$\exp(\nicefrac{2\pi\I\,2}{9})$ by the $\Z3^{(PG)}$ point group symmetry of the $\mathbbm T^2/\Z3$ 
orbifold sector, see eq.~\eqref{eq:pointgroupselectionrule}. Due to this phase, 
$\rho_{\rep3_2,\omega}(\hat K_\mathrm{S}\,\hat K_\mathrm{T})$ generates a \Z9 symmetry that 
contains the $\Z{3}^{(PG)}$ point group symmetry as a subgroup, 
\begin{equation}
(\rho_{\rep3_2,\omega}(\hat K_\mathrm{S}\,\hat K_\mathrm{T}))^3 ~=~ \omega^2\Id\;.
\end{equation}
Thus, even though $\rho_{\rep3_2,\omega}(\hat K_\mathrm{S}\,\hat K_\mathrm{T})$ has order 9, the 
traditional flavor symmetry (without \CP) is only enhanced from 
\begin{equation}
\label{eq:nonCPenhancementAtT=omega}
\Delta'(54,2,1)\cong[162,44] \qquad\mathrm{to}\qquad H(3,2,1)\cong[486, 125]\qquad\text{at }\vev T=\omega\;,
\end{equation}
i.e.\ the order increases by a factor of three (see ref.~\cite{Jurciukonis:2017mjp} for the 
nomenclature). 

It follows that the $\Z{9}\subset H(3,2,1)$ transformation eq.~\eqref{eq:STtrafoOnPhiT=omega} acts 
on the superpotential eq.~\eqref{eq:Z3superpotentialAtOmega} as
\begin{equation}
\mathcal{W}(\omega,X_i,Y_i,Z_i) ~\xrightarrow{\hat K_\mathrm{S}\,\hat K_\mathrm{T}}~ \mathcal{W}(\omega,X_i',Y_i',Z_i') 
    ~=~ \omega\,\mathcal{W}(\omega,X_i,Y_i,Z_i)\;.
\end{equation}
This is expected due to the automorphy factor $(-\vev T-1)^{-1} = \omega$ of the superpotential 
$\mathcal{W}$ for $\gamma_T=\hat K_\mathrm{S}\,\hat K_\mathrm{T}$ evaluated at the point 
$\vev T=\omega$ in moduli space. Thus, the \Z9 symmetry enhancement yields a discrete $R$-symmetry. 

Let us now consider also the \CP-like transformation $\hat K_*\hat K_\mathrm{T}$ included in 
$H_\omega$. Under $\hat{K}_*\hat K_{\mathrm T}$, twisted matter fields $\Phi_{\nicefrac{-2}{3}}^i$ 
transform according to eq.~\eqref{eq:TrafoK*KTOnPhi}, which in terms of the component fields
reads $(X_i,Y_i,Z_i)^\mathrm{T} \xrightarrow{\hat K_*\hat K_\mathrm{T}}(\omega \bar X_i,\bar Y_i,\bar Z_i)^\mathrm{T}$.
Then, we find that the superpotential at $\vev T=\omega$, eq.~\eqref{eq:Z3superpotentialAtOmega},
respects \CP-like transformations, i.e.
\begin{equation}
\mathcal{W}(\omega,X_i,Y_i,Z_i) ~\xrightarrow{\hat{K}_*\hat K_{\mathrm T}}~ 
\mathcal{W}(\omega,\omega\bar{X}_i,\bar{Y}_i,\bar{Z}_i) ~=~ \big(\mathcal{W}(\omega,X_i,Y_i,Z_i)\big)^*\;.
\end{equation}
This can be easily confirmed by applying the identities
\begin{equation}
\hat{Y}_1(\omega) \omega ~=~ \left(\hat{Y}_1(\omega)\right)^* \quad\mathrm{and}\quad \hat{Y}_2(\omega) ~=~ \left(\hat{Y}_2(\omega)\right)^*\;,
\end{equation}
which follow from eqs.~\eqref{eq:YModularTrafo} and~\eqref{eq:CPTrafoOfYukawas}.
Finally, we find that this \CP enhancement leads to the unified flavor symmetry 
given by
\begin{equation}
G_{\mathrm{uf},\vev{T}=\omega} ~\cong~ [972,469] ~\cong~ \Delta'(54,2,1)\cup\Z9\cup\Z2^{\CP}\;, 
\end{equation}
where $H(3,2,1)\cong\Delta'(54,2,1)\cup\Z9$ is the maximal flavor subgroup of 
$G_{\mathrm{uf},\vev{T}=\omega}$ which does not act \CP-like.

\subsection[Gauge symmetry enhancement at <T>=omega]{\boldmath Gauge symmetry enhancement at $\vev T=\omega$\unboldmath}
\label{sec:accidentalU1s}

Let us analyze the ``accidental'' continuous symmetries of the superpotential 
eq.~\eqref{eq:Z3superpotentialAtOmega} at $\vev T=\omega$ in moduli space. We assume that the 
twisted matter fields $\Phi_{\nicefrac{-2}{3}}^i=(X_i, Y_i, Z_i)^\mathrm{T}$ transform identically 
for $i\in\{1,2,3\}$ (typically the three copies of twisted matter fields 
$\Phi_{\nicefrac{-2}{3}}^i$ differ in some other gauge charges, such that there is no flavor 
symmetry that mixes $i\in\{1,2,3\}$). To identify continuous symmetries, we define a general 
infinitesimal $\U{3}$ transformation
\begin{equation}
\begin{pmatrix}X_i\\Y_i\\Z_i\end{pmatrix} ~\stackrel{\U{3}}{\longrightarrow}~ \left(\Id_3 + \I\,\alpha_a\, \mathrm{T}_a\right)\,\begin{pmatrix}X_i\\Y_i\\Z_i\end{pmatrix}\;,
\end{equation}
where summation over $a=1,\ldots,9$ is implied and the $3\times 3$ matrices $\mathrm{T}_a$ denote 
the Hermitian generators of the $\U{3}$ Lie algebra. By demanding invariance of the superpotential 
eq.~\eqref{eq:Z3superpotentialAtOmega} at leading order in the parameters $\alpha_a$, we obtain two 
linear independent generators $\mathrm{t}_a$ that we choose as 
\begin{equation}\label{eq:TOmegaU1GeneratorsOrigBasis}
\mathrm{t}_1 ~:=~ \frac{1}{3\sqrt{2}}
\begin{pmatrix}
0      & \omega^2 & \omega^2\\
\omega &        0 &  1 \\ 
\omega &        1 &   0
\end{pmatrix}\quad\mathrm{and}\quad 
\mathrm{t}_2 ~:=~ \frac{\I}{3\sqrt{2}}
\begin{pmatrix}
0       & \omega^2 & -\omega^2\\
-\omega &        0 &  1 \\ 
 \omega &       -1 &  0
\end{pmatrix}\;.
\end{equation}
Note that $\mathrm{t}_1\, \mathrm{t}_2 = \mathrm{t}_2\, \mathrm{t}_1$. Hence, we found a 
$\U{1}\times\U{1}$ symmetry of the superpotential eq.~\eqref{eq:Z3superpotentialAtOmega} at 
$\vev T=\omega$. In a full string discussion (see e.g.~\cite{Beye:2014nxa}), one can show 
that this ``accidental'' symmetry is actually an exact gauge symmetry, where the $\U{1}$ gauge 
bosons correspond to certain winding strings that become massless at the self-dual point 
$\vev T=\omega$ in moduli space.

Next, we diagonalize the $\U{1}$ generators eq.~\eqref{eq:TOmegaU1GeneratorsOrigBasis} 
simultaneously by performing a (unitary) basis change in field space, i.e.\
\begin{equation}\label{eq:BasisChangeAtOmega}
\begin{pmatrix}X^\mathrm{g}_i\\Y^\mathrm{g}_i\\Z^\mathrm{g}_i\end{pmatrix} ~:=~ M^\mathrm{g}\,\begin{pmatrix}X_i\\Y_i\\Z_i\end{pmatrix} \quad\mathrm{where}\quad M^\mathrm{g} ~:=~ \frac{1}{\sqrt{3}}
\begin{pmatrix}
 1 & \omega^2 & \omega^2 \\
 1 &        1 &   \omega \\ 
 1 &   \omega &        1
\end{pmatrix}\;,
\end{equation}
where the superscript character ``g'' labels the ``gauge'' basis, in which the $\U{1}$ 
generators are diagonal. In this basis, the $\U{1}$ generators 
$\mathrm{t}^\mathrm{g}_a := M^\mathrm{g}\, \mathrm{t}_a\, (M^\mathrm{g})^{-1}$ for $a\in \{1,2\}$ 
read
\begin{equation}\label{eq:TOmegaU1Generators}
\mathrm{t}^\mathrm{g}_1 ~=~ \frac{1}{3\sqrt{2}}
\begin{pmatrix}
 2 & 0 & 0\\
 0 &-1 & 0 \\ 
 0 & 0 &-1
\end{pmatrix} \quad\mathrm{and}\quad 
\mathrm{t}^\mathrm{g}_2 ~=~ \frac{1}{\sqrt{6}}
\begin{pmatrix}
0 & 0 & 0\\
0 & 1 & 0 \\ 
0 & 0 &-1
\end{pmatrix}\;.
\end{equation}
This reproduces the $\U{1}$ charges of $\Z{3}$ twisted strings $(M_1,M_2,M_3)$ computed using a 
$\Z{3}$ shift orbifold, see ref.~\cite{Beye:2014nxa} and especially table 2 therein. Next, we 
change the basis for the generators $\rho(\mathrm{A})$, $\rho(\mathrm{B})$ and $\rho(\mathrm{C})$ 
of the traditional flavor group $\Delta(54)$ given in eqs.~\eqref{eq:Delta54rhoAandB} 
and~\eqref{eq:Delta54rhoC}. We obtain
\begin{equation}\label{eq:Delta54Generators}
\rho^\mathrm{g}(\mathrm{A}) ~:=~  \begin{pmatrix} 0 & 0 & \omega^2 \\ \omega & 0 & 0 \\ 0 & 1 & 0 \end{pmatrix}\;, \quad 
\rho^\mathrm{g}(\mathrm{B}) ~:=~  \begin{pmatrix} 0 & 1 & 0 \\ 0 & 0 & 1 \\ 1 & 0 & 0 \end{pmatrix} \quad\mathrm{and}\quad 
\rho^\mathrm{g}(\mathrm{C}) ~:=~ -\begin{pmatrix} 1 & 0 & 0 \\ 0 & 0 & 1 \\ 0 & 1 & 0 \end{pmatrix}\;.
\end{equation}
Using this, one can show easily that the $\Z{3}\times\Z{3}$ subgroup of $\Delta(54)$, generated by 
the diagonal representation matrices
\begin{equation}
\rho^\mathrm{g}(\mathrm{A})\,\rho^\mathrm{g}(\mathrm{B}) ~=~ \begin{pmatrix} \omega^2 & 0 & 0 \\ 0 & \omega & 0 \\ 0 & 0 & 1 \end{pmatrix} \quad\mathrm{and}\quad \rho^\mathrm{g}(\mathrm{B})\,\rho^\mathrm{g}(\mathrm{A}) ~=~ \begin{pmatrix} \omega & 0 & 0 \\ 0 & 1 & 0 \\ 0 & 0 & \omega^2 \end{pmatrix}\,,
\end{equation}
is contained in the continuous $\U{1}\times\U{1}$ symmetry generated by eq.~\eqref{eq:TOmegaU1Generators},
\begin{subequations}
\begin{eqnarray}
\rho^\mathrm{g}(\mathrm{A})\,\rho^\mathrm{g}(\mathrm{B}) & = & \exp\left(2\pi\I \left(+ \sqrt{2}\,\mathrm{t}^\mathrm{g}_1-\sqrt{\frac{2}{3}}\,\mathrm{t}^\mathrm{g}_2\right)\right)\;,\\
\rho^\mathrm{g}(\mathrm{B})\,\rho^\mathrm{g}(\mathrm{A}) & = & \exp\left(2\pi\I \left(- \sqrt{2}\,\mathrm{t}^\mathrm{g}_1-\sqrt{\frac{2}{3}}\,\mathrm{t}^\mathrm{g}_2\right)\right)\;.
\end{eqnarray}
\end{subequations}
The elements $\rho^\mathrm{g}(\mathrm{B})$ and $\rho^\mathrm{g}(\mathrm{C})$ generate an $S_3$ 
permutation group, which is an $R$-symmetry since its generator $\mathrm{C}$ is defined as a 
$180^\circ$ rotation in extra dimensions. This $S_3$ does not commute with the continuous 
$\U{1}\times\U{1}$ symmetry. Hence, one can show that the traditional flavor symmetry $\Delta(54)$ 
gets enhanced at $\vev T=\omega$ in moduli space to
\begin{equation}
\big(\U{1}\times\U{1}\big)\rtimes S_3^R\;,
\end{equation}
see ref.~\cite{Beye:2014nxa}. This is further enhanced by $\Z{9}^R$ (if the orbifold sector is 
embedded in a higher-dimensional orbifold) and by the unbroken $\SL{2,\Z{}}_T$ modular 
transformation $\hat K_\mathrm{S}\,\hat K_\mathrm{T}$ as discussed in section~\ref{sec:vevT=omega}. In the new 
field basis, eq.~\eqref{eq:STtrafoOnPhiT=omega} yields
\begin{equation}
\rho^\mathrm{g}_{\rep{3}_2,\omega}(\hat K_\mathrm{S}\,\hat K_\mathrm{T}) 
~:=~ M^\mathrm{g}\, \rho_{\rep{3}_2,\omega}(\hat K_\mathrm{S}\,\hat K_\mathrm{T})\, (M^\mathrm{g})^{-1} 
 ~=~ \exp(\nicefrac{2\pi\I\,2}{9}) \begin{pmatrix} \omega^2 & 0 & 0 \\ 0 & 1 & 0 \\ 0 & 0 & 1 \end{pmatrix}\;.
\end{equation}

Finally, we can perform the basis change in field space eq.~\eqref{eq:BasisChangeAtOmega} for the 
superpotential eq.~\eqref{eq:Z3superpotential} and obtain
\begin{align}\label{eq:WatTinU1xU1Basis}
&\mathcal{W}(T,X^\mathrm{g}_i,Y^\mathrm{g}_i,Z^\mathrm{g}_i) ~=~ \frac{c^{(1)}}{\sqrt{3}}\,\Big[
\left(\hat{Y}_2(T) - \sqrt{2}\,\omega^2\,\hat{Y}_1(T)\right)\, 
\left( X^\mathrm{g}_1 X^\mathrm{g}_2 X^\mathrm{g}_3 + Y^\mathrm{g}_1 Y^\mathrm{g}_2 Y^\mathrm{g}_3 + Z^\mathrm{g}_1 Z^\mathrm{g}_2 Z^\mathrm{g}_3\right)\\
& + \left(\hat{Y}_2(T) + \frac{\omega^2}{\sqrt{2}}\,\hat{Y}_1(T)\right)\, 
\left( X^\mathrm{g}_1 (Y^\mathrm{g}_2 Z^\mathrm{g}_3 + Y^\mathrm{g}_3 Z^\mathrm{g}_2) 
     + X^\mathrm{g}_2 (Y^\mathrm{g}_1 Z^\mathrm{g}_3 + Y^\mathrm{g}_3 Z^\mathrm{g}_1) 
     + X^\mathrm{g}_3 (Y^\mathrm{g}_1 Z^\mathrm{g}_2 + Y^\mathrm{g}_2 Z^\mathrm{g}_1)\right)\Big]\;.\nonumber
\end{align}
By using the $\U1\x\U1$ generators defined in eq.~\eqref{eq:TOmegaU1Generators}, we see that the 
terms from the first line of eq.~\eqref{eq:WatTinU1xU1Basis}, such as 
$Y^\mathrm{g}_1 Y^\mathrm{g}_2 Y^\mathrm{g}_3$, break this symmetry, whereas the terms from the 
second line preserve it. On the other hand, at the point $\vev T=\omega$ in moduli space, the 
couplings align as
\begin{equation}
\hat{Y}_1(\omega) ~=~ \frac{\omega}{\sqrt{2}}\,\hat{Y}_2(\omega)\;,
\end{equation}
see eq.~\eqref{eq:YukawaAlignmentAtOmega}. This alignment leads to a vanishing coefficient in the 
first line of eq.~\eqref{eq:WatTinU1xU1Basis} and we are left with
\begin{equation}\label{eq:WatOmegainU1xU1Basis}
\mathcal{W}(\omega,X^\mathrm{g}_i,Y^\mathrm{g}_i,Z^\mathrm{g}_i) ~=~ c\,\Big( 
  X^\mathrm{g}_1 (Y^\mathrm{g}_2 Z^\mathrm{g}_3 + Y^\mathrm{g}_3 Z^\mathrm{g}_2) + 
  X^\mathrm{g}_2 (Y^\mathrm{g}_1 Z^\mathrm{g}_3 + Y^\mathrm{g}_3 Z^\mathrm{g}_1) + 
  X^\mathrm{g}_3 (Y^\mathrm{g}_1 Z^\mathrm{g}_2 + Y^\mathrm{g}_2 Z^\mathrm{g}_1)\Big)\;,
\end{equation}
where $c := \nicefrac{\sqrt{3}}{2}\,c^{(1)}\, \hat{Y}_2(\omega)$. This superpotential at 
$\vev T=\omega$ is clearly invariant under the $\U{1}\times\U{1}$ transformations 
generated by eq.~\eqref{eq:TOmegaU1Generators}.

\subsection{Summary}
\label{sec:unifiedflavorsummary}

Let us now highlight the main results of this section. Unified flavor symmetries 
$G_{\mathrm{uf},\vev{T}}$ are enhancements of the traditional flavor symmetry that occur at 
different self-dual points \vev{T} and loci in moduli space of $\mathbbm T^2/\Z{K_i}$ orbifold 
sectors, where some modular transformations are linearly realized. The new flavor symmetries 
include $\SL{2,\Z{}}_T$ (and in the case of $K_i=2$ $\SL{2,\Z{}}_U$) transformations as well as 
\CP-like symmetries. Thus, in this scenario $G_{\mathrm{uf},\vev{T}}$ (and \CP) breakdown is 
triggered by moving away from the self-dual loci in moduli space. We have illustrated this in the 
$\mathbbm T^2/\Z3$ orbifold sector. At a generic point in moduli space, this orbifold exhibits a 
$\Delta'(54,2,1)\cong\Delta(54)\cup\Z{9}^R$ traditional flavor symmetry, when embedded in a 
higher-dimensional orbifold. At the points depicted in figure~\ref{fig:Z3KaehlerModuliSpace}, we 
find the following unified flavor symmetries:
\begin{subequations}
\begin{eqnarray}
  \Delta'(54,2,1) &\xrightarrow{\hat K_*\mathrm{\ or \ }\hat K_*\hat K_\mathrm{T}\mathrm{\ or \ }\hat K_\mathrm{S}\hat K_*}& G_{\mathrm{uf},\vev{T}\in\mathrm{boundary}}\cong [324,125]\,,\quad\mathrm{\small see\ section~\ref{sec:vevT@boundary}} \nonumber\\
  \Delta'(54,2,1) &\xrightarrow{\hat K_\mathrm{S}} \hspace{7.5mm}\Xi(2,2)\hspace{7.5mm} \xrightarrow{\hat K_*}& G_{\mathrm{uf},\vev{T}=\I} \cong [648,548]\,,\hspace{15mm}\mathrm{\small see\ section~\ref{sec:vevT=i}}  \nonumber\\
  \Delta'(54,2,1) &\xrightarrow{\hat K_\mathrm{S}\hat K_\mathrm{T}}~ H(3,2,1) ~\xrightarrow{\hat K_*\hat K_\mathrm{T}}& G_{\mathrm{uf},\vev{T}=\omega} \cong [972,469]\,,\hspace{14mm}\mathrm{\small see\ section~\ref{sec:vevT=omega}}  \nonumber
\end{eqnarray}
\end{subequations}
where we indicate the unbroken modular transformations that are linearly realized. In these cases, 
we have studied how the value of \vev{T} induces an alignment mechanism on the couplings, which in 
turn sets the structure of the superpotential of the theory. Finally, in 
section~\ref{sec:accidentalU1s} we observe that the traditional flavor symmetry is further enhanced 
to (continuous) gauge symmetries at self-dual points of the theory. In particular, by inspecting 
the (accidental) symmetries of the superpotential at $\vev{T}=\omega$, we have shown that the 
$\Delta(54)$ flavor symmetry is enhanced to $(\U1\x\U1)\rtimes S_3^R$, which is broken 
spontaneously elsewhere in moduli space.

\section{Conclusions and outlook}
\label{sec:conclusions}

In this paper we have presented a new step of the eclectic approach towards ten-dimensional string 
theory. This discussion goes beyond previous studies in $D=2$ that concentrated on the modular 
transformations $\SL{2,\Z{}}_T$ of a single K\"ahler modulus $T$. We now include a detailed 
analysis of $\SL{2,\Z{}}_U$ of the complex structure modulus $U$. In the orbifolds 
$\mathbbm T^2/\Z{K}$ with $K=3,4,6$ the $U$-modulus is fixed because of the orbifold twist. We show 
that even in the case of frozen $U$ we encounter remnant symmetries of $\SL{2,\Z{}}_U$ in form of 
$R$-symmetries that extend the eclectic group. When embedded in $D=6$ compact dimensions these 
symmetries are relevant as sublattice rotations. We illustrate the full eclectic picture in 
the case of $\mathbbm T^2/\Z{3}$, where we obtain $G_{\mathrm{ecl}} = \Delta(54)\cup T'\cup\Z{9}^R\cup\Z2^{\CP}$,
a discrete symmetry group of order 3,\,888.

In a next step towards $D=6$ we shall have to consider more than one unconstrained modulus beyond 
$T$. The simplest case with two unconstrained moduli ($T$ and $U$) is the $\mathbbm T^2/\Z{2}$ 
orbifold that has already been discussed in ref.~\cite{Baur:2020jwc}. There, it was shown that the 
remnants of the full modular group $O_{\hat\eta}(2,2,\Z{})$ (that contains 
$\SL{2,\Z{}}_T\times\SL{2,\Z{}}_U$) lead to the finite modular group 
$((S_3\times S_3)\rtimes \Z{4})\times\Z{2}^\CP$, where mirror symmetry acts as a $\Z{4}$ on the 
level of matter fields but only as a $\Z{2}$ permutation of $T$ and $U$, and $\Z{2}^\CP$ is a 
\CP-like transformation~\cite{Baur:2020jwc}. Together with the traditional flavor group 
$(D_8\times D_8)/\Z{2}$ this leads to an eclectic group with $4,608$ elements. The full technical 
details of the analysis of ref.~\cite{Baur:2020jwc} will be presented in ref.~\cite{Baur:2021mtl}. 
This includes a discussion of modular forms with more than one unconstrained modulus, such as 
e.g.\ Siegel modular forms. Incidentally, in a recent work of Ding, Feruglio and 
Liu~\cite{Ding:2020zxw} such a situation has been discussed from the bottom-up perspective, 
including the aforementioned finite modular group $(S_3\times S_3)\rtimes \Z{2}$ (on the level of 
the moduli and without \CP). It will be interesting to see how our top-down approach from string 
theory can make contact with the bottom-up discussion of ref.~\cite{Ding:2020zxw}.

After having control of the $\mathbbm T^2/\Z{2}$ case, all of the relevant building blocks for the 
$D=6$ case are at our disposal. A discussion of the generic case with $D=6$ will be too difficult 
to be analyzed in detail. Simpler cases can be found in models of elliptic fibrations of $D=6$ 
compact space. Sublattice rotations of the various $D=2$ sectors would be relevant for the eclectic 
group. Finally, the next step would require the selection of specific string models that could 
successfully describe the flavor structure of quarks and leptons to make connections to the 
bottom-up constructions (see e.g.~\cite{Feruglio:2019ktm} and references therein).

\section*{Acknowledgments}

The work of S.R.-S.\ was partly supported by CONACyT grants F-252167 and 278017.
The work of P.V. is supported by the Deutsche Forschungsgemeinschaft (SFB1258).


\providecommand{\bysame}{\leavevmode\hbox to3em{\hrulefill}\thinspace}

\end{document}